\documentclass[a4paper,11pt]{article}
\pdfoutput=1 

\usepackage{jheppub} 

\usepackage[T1]{fontenc}
\usepackage[utf8]{inputenc}

\usepackage{hyperref}
\usepackage[capitalise]{cleveref}
\usepackage{orcidlink} 

\usepackage{amsmath,amsfonts,amssymb,amsthm,amsxtra,bbm,mathrsfs,mathtools}
\usepackage{slashed,cancel}
\usepackage{bm} 
\usepackage{bbold} 
\usepackage{pifont} 

\usepackage{graphicx}
\usepackage{epsfig}
\usepackage{subcaption} 
\usepackage{overpic}
\usepackage{color,xcolor}

\usepackage{tikz}
\usepackage{tikz-feynman}
\usepackage[force]{feynmp-auto}
\usepackage{booktabs}
\usepackage{multirow}
\usepackage{enumitem}
\usepackage{lipsum}

\usepackage[normalem]{ulem} 
\usepackage{comment}
\usepackage{eqparbox}
\usepackage{braket}

\newcommand{\hide}[1]{\iffalse #1 \fi}

\newcommand{\be}{\begin{equation}}
\newcommand{\ee}{\end{equation}}
\newcommand{\bea}{\begin{eqnarray}}
\newcommand{\eea}{\end{eqnarray}}
\newcommand{\ba}{\begin{aligned}}
\newcommand{\ea}{\end{aligned}}

\newcommand{\equaref}[1]{Eq.~(\ref{#1})}

\newcommand{\figref}[1]{Fig.~\ref{#1}}

\newcommand{\secref}[1]{Section~\ref{#1}}

\newcommand{\tabref}[1]{Tab.~\ref{#1}}

\newcommand{\mybox}[1]{\boxed{\phantom{{}^a_1}\!\! #1}}

\preprint{IPPP/25/80}

\title{Domain Walls in $A_4$ Flavour Models}

\author[a,b]{Bowen Fu~\orcidlink{0000-0003-2270-8352}\,,}
\affiliation[a]{Key Laboratory of Cosmology and Astrophysics (Liaoning) \& College of Sciences, Northeastern University, Shenyang 110819, China}
\affiliation[b]{Foshan Graduate School of Innovation, Northeastern University, Foshan 528312, China}
\author[c]{Stephen F. King~\orcidlink{0000-0002-4351-7507}\,,}
\affiliation[c]{School of Physics and Astronomy, University of Southampton, Southampton, SO17 1BJ, U.K.}
\author[d]{Luca Marsili~\orcidlink{0000-0002-9085-8160}\,,}
\affiliation[d]{Instituto de F\`isica Corpuscular (IFIC), Universitat de Val\`encia, Parc Cientific UV, C/ Catedratico Jose Beltran 2, E-46980 Paterna, Spain}
\author[e]{Jessica Turner~\orcidlink{0000-0002-9679-5252}\,,}
\affiliation[e]{Institute for Particle Physics Phenomenology, Department of Physics, Durham University, Durham DH1 3LE, U.K.}
\author[f]{and Ye-Ling Zhou~\orcidlink{0000-0002-3664-9472}}
\affiliation[f]{School of Fundamental Physics and Mathematical Sciences, Hangzhou Institute for Advanced
Study, UCAS, Hangzhou 310024, China}

\emailAdd{fubowen@neu.edu.cn}
\emailAdd{s.f.king@soton.ac.uk}
\emailAdd{luca.marsili@ific.uv.es}
\emailAdd{jessica.turner@durham.ac.uk}
\emailAdd{zhouyeling@ucas.ac.cn}

\abstract{The spontaneous breaking of an $A_4$ flavour symmetry, often used to predict leptonic mixing, can lead to the formation of domain walls which can annihilate and generate a stochastic gravitational wave background. We study this phenomenon in three scenarios where the nature of the scalar field responsible for breaking the $A_4$ symmetry spontaneously differs: real, complex, and supersymmetric. For the real scalar, a biased potential produces metastable walls that decay into oscillating two-wall systems with important consequences for gravitational wave signals. In the complex scalar case, we discuss the interplay between domain walls and global strings and classify the types of domain walls that form in terms of the $A_4$ group symmetries. We investigate the properties of supersymmetric $A_4$ domain walls, and highlight the BPS walls. Through a detailed analysis of these models with non-Abelian symmetries, we discover new kinds of domain walls, which we denote as ``oreo''-type composite domain walls, CP-violating domain walls and SUSY non-Abelian domain walls.
Finally we show how these results may be achieved in leptonic $A_4$ flavour models, with and without supersymmetry, and discuss their distinctive gravitational wave signatures.}

\begin{document}
\maketitle
\flushbottom
\section{Introduction}
Discrete non-Abelian flavour symmetries have long provided a compelling organising principle for lepton mixing, with the tetrahedral group \(A_4\) playing a central role due to its triplet and three singlet representations (two of which are non-trivial)~\cite{Ma:2001dn,Altarelli:2005yx,King:2013eh}. In canonical constructions, charged lepton and neutrino mass matrices arise from flavon fields whose vacuum expectation values (VEVs) spontaneously break \(A_4\) to its Abelian discrete subgroups, leading to predictive mixing patterns at leading order, the most famous being the tribimaximal (TBM) mixing pattern~\cite{Harrison:2002er,Harrison:2002kp,Xing:2002sw,Harrison:2002et,Harrison:2003aw}. Subleading corrections to the flavon alignments can accommodate deviations from exact mixing patterns, giving viable fits to observed lepton mixing data~\cite{Altarelli:2005yx,Lam:2008rs}. There is a rich phenomenology associated with leptonic flavour models. In particular, collider signatures and lepton flavour violation in the context of $A_4$ symmetries have been studied extensively~\cite{Tsumura:2009yf,Berger:2014gga,Arroyo-Urena:2018mvl,Bauer:2016rxs,Heinrich:2018nip}, with constraints from $\mu\to e\gamma$~\cite{MEG:2016leq} already probing symmetry breaking scales just below the TeV level.

Beyond collider and charged lepton flavour phenomenology, leptonic flavour models can have interesting cosmological consequences. The spontaneous breaking of discrete symmetries in the early Universe generically leads to the formation of domain walls (DWs). For example, it is well known that the spontaneous breaking of a discrete \(\mathbb{Z}_2\) symmetry has important cosmological implications~\cite{Vilenkin:1984ib}. This is due to the vacua in different causally disconnected spatial regions of the early Universe randomly settling into one of the two allowed values. Later, when these regions come into causal contact, cosmological domain walls form between the different vacua according to the Kibble mechanism~\cite{Kibble:1976sj}. If such walls are long-lived, their energy density scales slowly and can come to dominate the Universe's energy density, conflicting with standard cosmology.

However, if scalar potentials are biased by a term which explicitly breaks the symmetry by a small amount, this can lead to domain wall annihilation, which in turn can source an observable stochastic gravitational wave (GW) background~\cite{Zeldovich:1974uw,Kibble:1976sj,Vilenkin:1984ib,Saikawa:2017hiv}. Recent theoretical  studies have sharpened this connection, mapping wall tensions, biases and decay temperatures to the amplitude and peak frequency of the GW signal, and explore the sensitivity of current and future GW observatories to such signals (see $e.g.$ Ref.~\cite{Saikawa:2017hiv} and references therein). These developments motivate a systematic treatment of domain walls in realistic flavour models based on {\em non-Abelian} discrete family symmetries and their associated GW signatures.
Extending these considerations of domain walls to the breaking of a {\em non-Abelian} discrete group introduces qualitatively new features compared to the simplest \(\mathbb{Z}_2\) domain wall case. The vacuum manifolds contain multiple classes of minima typically preserving different residual subgroups, which in turn admit distinct \emph{families} of walls with group-theoretic relations among their tensions. This structure, together with generic explicit bias terms, can yield richer annihilation histories and multi-peaked GW spectra~\cite{Gelmini:2020bqg,Fu:2024jhu}. A recent detailed analysis for \(S_4\) highlighted how residual \(Z_2\) and \(Z_3\) vacua lead to several topologically different wall varieties with calculable profiles, tensions, and possible decay channels and showed how explicit breaking lifts vacuum degeneracies with clear GW consequences~\cite{Fu:2024jhu,Jueid:2023cgp}. It is therefore timely to extend this programme to the widely used \(A_4\) symmetry that underlies many leptonic flavour models. In this work, we present a comprehensive study of domain walls arising from $A_4$ flavour breaking and their cosmological phenomenology, including their impact on the stochastic GW background. We analyse three theoretically and phenomenologically motivated scenarios, distinguished by the nature of the flavon sector, where, in typical flavour models, flavon VEVs will preserve either a $Z_2$ residual symmetry associated with the $\mathrm{S}$ generator, or a $Z_3$ residual symmetry associated with the $\mathrm{T}$ generator of $A_4$. We first consider an $A_4$ real scalar triplet with the most general renormalisable potential and find that the cubic $A_4$ invariant splits the eight $Z_3$ preserving vacua into two sets, giving rise to both metastable and stable pairs of domain walls. Secondly, we extend this to a complex triplet in an $A_4 \times U(1)$ setting (relevant to models with extra Abelian charges and CP violation), where the global $U(1)$ phase gives rise to global strings in addition to domain walls. We classify the possible $\mathrm{S}$- and $\mathrm{T}$-type walls, including those with non-trivial relative phases, and discuss how composite string–wall networks can form and radiate GWs. Finally, we consider a supersymmetric (SUSY) framework in which the scalar potential is $F$-term dominated and the $A_4$ breaking scale lies well above the soft SUSY-breaking scale. In this case, the potential admits first-order (BPS-like~\cite{Bogomolny:1975de,Prasad:1975kr,Chibisov:1997rc}) wall equations. The vacuum manifold includes the origin and four $\mathrm{T}$-type vacua related by $A_4$, for which we find analytic wall profiles and exact tensions. Soft terms lift the degeneracy between the trivial and non-trivial vacua, ensuring annihilation of SUSY domain walls and linking the soft-breaking scale to a GW peak. Subsequent non-SUSY walls, generated between the non-trivial vacua themselves, can then produce an additional GW contribution. Across these cases, we classify the vacuum structure, construct the corresponding wall solutions, and compute their tensions and relative stability. We provide a qualitative discussion on the GW signals produced in these scenarios. We then embed these results in explicit lepton flavour models. In particular, we show how the $\mathrm{S}$- and $\mathrm{T}$-type vacua arise in minimal real flavon $A_4$ models, in complex flavon models with an additional $U(1)_L$ (or its discrete/gauged variants) that introduce global or local strings, and in SUSY $A_4$ models with driving fields and an $R$-symmetry. In each case, we identify the resulting network of non-Abelian walls and Abelian defects and discuss the associated GW signatures, including characteristic multi-peak spectra.

This paper is structured as follows. In \secref{sec:non-susy-dws}, we review the \(A_4\) group structure, construct the most general \(A_4\)-invariant scalar potential, and analyse the resulting non-supersymmetric domain wall solutions, including their cosmological evolution and  gravitational wave emission. In \secref{sec:superfield}, we turn to the supersymmetric setting, deriving BPS-like walls, incorporating \(R\)-symmetries and driving fields, and studying how soft-term biases control wall decay. In \secref{sec:models}, we embed these constructions into realistic flavour models and follow in \secref{sec:sol_nonSUSY_DW} and \secref{sec:GW} with a qualitative discussion on the cosmological solutions to the domain wall problem and gravitational wave signals from such non-Abelian domain walls. Finally, we make concluding remarks in \secref{sec:conclusion}.

\section{Domain Walls from $A_4$ Symmetry \label{sec:non-susy-dws}}
We consider the flavour symmetry of the lepton sector to be the tetrahedral group $A_{4}$~\cite{Ma:2001dn}, which is the group of even permutations of four objects. This group is generated by two elements, $\mathrm{S}$ and $\mathrm{T}$, that satisfy the relations $S^{2} = T^{3} = (ST)^{3} = \mathbb{1}$. From these generators, the twelve group elements are constructed:
$\mathbb{1}$, $\mathrm{S}$, $\mathrm{T}$, $T^{2}$, $ST$, $TS$, $STS$, $ST^{2}$, $T^{2}S$, $TST$, $TST^{2}$, and $T^{2}ST$. $A_{4}$ is the smallest non-Abelian discrete group that has a three-dimensional irreducible representation, denoted as $\mathbf{3}$. It also includes three one-dimensional irreducible representations: the trivial singlet $\mathbf{1}$ and two non-trivial singlets $\mathbf{1'}$ and $\mathbf{1''}$. In this work, we adopt the Ma-Rajasekaran (MR) basis~\cite{Ma:2001dn}. While physical predictions are basis-independent, this choice makes a clear classification of all possible vacuum configurations of the scalar field $\phi$. In this basis, the generators are
\begin{eqnarray}
T = \left(
\begin{array}{ccc}
 0 & 0 & 1 \\
 1 & 0 & 0 \\
 0 & 1 & 0 \\
\end{array}
\right)\,,\quad
S = \left(
\begin{array}{ccc}
 1 & 0 & 0 \\
 0 & -1 & 0 \\
 0 & 0 & -1 \\
\end{array}
\right)\,.
\label{eq:generator1}
\end{eqnarray}
In realistic models, the discrete non-Abelian flavour symmetry is spontaneously broken to an Abelian residual subgroup. This process is driven by scalar fields acquiring non-zero vacuum expectation values (VEVs). These scalars, commonly referred to as \textit{flavons}, may be real, complex, or even superfields. 

\subsection{Domain Walls from Real Scalar Fields}\label{sec:realtrip}
\begin{figure}
  \centering
  \includegraphics[width=0.33\linewidth]{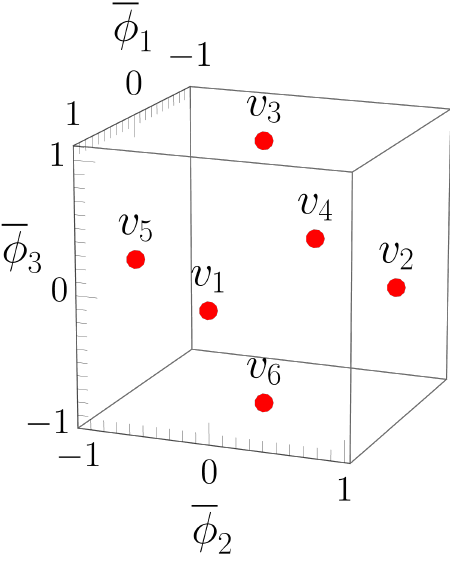}\qquad
  \includegraphics[width=0.33\linewidth]{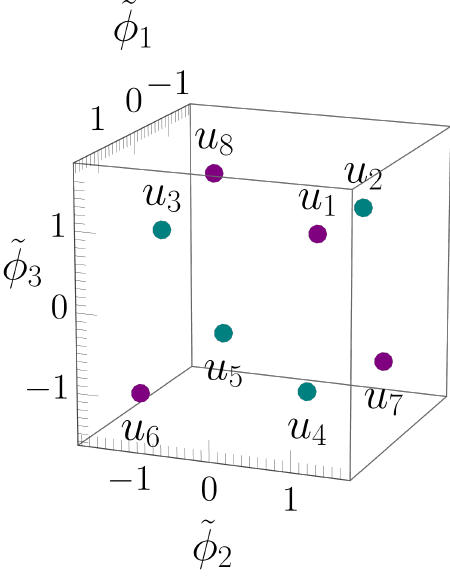}
  \caption{$\mathrm{S}$-type vacua (left panel) and $T_+$-type ($T_-$-type) vacua (right panel) in teal (purple) colour. 
  The fields are rescaled as $\tilde{\phi}_i = \phi_i \sqrt{3g_1+2g_2}/\mu$ and $a=0.1$ is chosen as a benchmark case. }
  \label{fig:u-vacua}
\end{figure}

Here we consider a real scalar triplet \(\phi=(\phi_{1},\phi_{2},\phi_{3})^{T}\) and its vacuum structure. The most general renormalisable \(A_{4}\)-invariant potential is
\begin{equation}\label{eq:potential}
V(\phi)= -\frac{\mu^{2}}{2}\,I_{1}+ \frac{g_{1}}{4}\,I_{1}^{2}+ \frac{g_{2}}{2}\,I_{2}+ A\,I_{3}\,,
\end{equation}
where the invariants are defined as
\begin{equation}\label{eq:A4_invariant}
\begin{aligned}
I_{1}&=\phi_{1}^{2}+\phi_{2}^{2}+\phi_{3}^{2}\,,\\
I_{2}&=\phi_{1}^{2}\phi_{2}^{2}+\phi_{2}^{2}\phi_{3}^{2}+\phi_{3}^{2}\phi_{1}^{2}\,,\\
I_{3}&=\phi_{1}\phi_{2}\phi_{3}\,.
\end{aligned}
\end{equation}
Vacuum stability requires that \(g_{1}>0\) and \(g_{2}>-\tfrac{3}{2}g_{1}\).
The first two terms of Eq.~\eqref{eq:potential} preserve an \(O(3)\simeq SO(3)\times\mathbb{Z}_{2}\) symmetry; adding \(I_{2}\) reduces this to \(A_{4}\times\mathbb{Z}_{2}\) and the \(I_{3}\) term breaks the \(\mathbb{Z}_{2}\).\footnote{We note that if the first two terms of the potential are large compared to the last two, an approximate $O(3)$ symmetry is released. The breaking of such a symmetry can give rise to a rich zoology of defects including walls, junctions and knots \cite{Kubotani:1991kw,Vilenkin:2000jqa}.}  Note that the absence of the last term, $i.e.$ in the limit \( A \to 0 \), reduces the $A_4$-invariant potential to an $S_4$-invariant potential which was explored in Ref.~\cite{Pascoli:2016eld}.
A necessary condition for \( \phi \) to be in a vacuum is \( \partial V(\phi)/\partial \phi_i = 0 \). From Eq.~\eqref{eq:potential}, this condition takes the form
\begin{eqnarray}\label{eq:realextrem}
\frac{\partial V(\phi)}{\partial \phi_i} = \phi_i \left[-\mu^2+ g_1 I_1 + g_2 ( \phi_j^2 + \phi_k^2) \right] + A \phi_j \phi_k = 0\,,
\end{eqnarray} 
where \( i,j,k = 1,2,3 \) with \( i \neq j \neq k\). 
Depending on whether $g_2>0$ or $g_2<0$, the system admits six {$Z_2$}-preserving vacua and eight {$Z_3$}-preserving vacua, which are respectively given by 
\be
\begin{aligned}
S:& & \Braket{\phi_S} &= \left\{
\begin{pmatrix} 1 \\ 0 \\ 0 \end{pmatrix}, 
\begin{pmatrix} 0 \\ 1 \\ 0 \end{pmatrix}, 
\begin{pmatrix} 0 \\ 0 \\ 1 \end{pmatrix}, 
\begin{pmatrix} -1 \\ 0 \\ 0 \end{pmatrix}, 
\begin{pmatrix} 0 \\ -1 \\ 0 \end{pmatrix}, 
\begin{pmatrix} 0 \\ 0 \\ -1 \end{pmatrix}
\right\}v\,,
\\[1em]
T:& & \Braket{\phi_T} &= \left\{
\begin{pmatrix} 1 \\ 1 \\ 1 \end{pmatrix}, 
\begin{pmatrix} -1 \\ 1 \\ 1 \end{pmatrix}, 
\begin{pmatrix} 1 \\ -1 \\ 1 \end{pmatrix}, 
\begin{pmatrix} 1 \\ 1 \\ -1 \end{pmatrix}, 
\begin{pmatrix} -1 \\ -1 \\ -1 \end{pmatrix}, 
\begin{pmatrix} 1 \\ -1 \\ -1 \end{pmatrix}, 
\begin{pmatrix} -1 \\ 1 \\ -1 \end{pmatrix}, 
\begin{pmatrix} -1 \\ -1 \\ 1 \end{pmatrix}
\right\}u\,,
\end{aligned}
\ee
where the vacuum expectation value magnitudes $v$ and $u$ are defined as
\begin{equation}
v = \frac{\mu}{\sqrt{g_1}} \quad \text{and} \quad u = \frac{\mu}{\sqrt{3g_1+2g_2}}\,.
\end{equation}
To avoid ambiguity, we denote residual symmetries of $A_4$ by $Z_n$ ($e.g.$\ $Z_2, Z_3$) and all other discrete cyclic symmetries by $\mathbb{Z}_N$. 
Each vacuum of the $\mathrm{S}$-type preserves a $Z_2$ symmetry. 
For example, $(1,0,0)^T$ is invariant under a $Z_2$ transformation generated by $\mathrm{S}$, and $(0,1,0)^T$ is invariant under a $Z_2$ transformation generated by $TST^{-1}$. 
Since any vacuum of this class is invariant under either the transformation of $\mathrm{S}$ or a conjugate transformation of $\mathrm{S}$, we denote these vacua as the $\mathrm{S}$-type vacua, and their normalised values are shown in the left panel of \figref{fig:u-vacua}.
Solving the scalar equations of motion (EoM), one can obtain the domain wall solutions between various vacua; see Ref.~\cite{Fu:2024jhu} for a detailed analysis. 
In the case of the $\mathrm{S}$-type domain walls, there are 15 possible ways to connect two distinct $\mathrm{S}$-type vacua. 
The classic $\mathbb{Z}_2$ domain walls form between the vacua $v_1$--$v_4$, $v_2$--$v_5$ and $v_3$--$v_6$ with wall tensions of $\sigma_{\mathrm{SI}}=\frac{2}{3} \sqrt{2 g_1} v^3=\frac{2}{3} m_1 v^2$. Following the notation in \cite{Wu:2022tpe}, we denote these walls as $\mybox{v_1}\mybox{v_4}$, $\mybox{v_2}\mybox{v_5}$ and $\mybox{v_3}\mybox{v_6}$, respectively.
The remaining 12 possible connections, for example between $v_1$--$v_2$, have associated wall tensions that are 
function of the Lagrangian parameters $g_1$ and $g_2$ \cite{Fu:2024jhu}.
Similarly, each $u_n$ in the second class of solutions preserves a $Z_3$ symmetry $e.g.$ $(1,1,1)^T$ is invariant under a $Z_3$ transformation generated by $\mathrm{T}$ and $(-1,1,1)^T$ is invariant under a $Z_3$ transformation generated by $STS$. 
We refer to these vacua as the $\mathrm{T}$-type vacua and the domain walls that can form between any two of them were studied in detail in Ref.~\cite{Fu:2024jhu}. 
The $\mathrm{T}$-type vacua are shown on the right of \figref{fig:u-vacua}. 
We observe that geometrically the $\mathrm{S}$ vacua lie on the coordinate axes, whereas the $\mathrm{T}$ vacua point toward cube diagonals.
When the \( I_3 \) term is included, the six \( S \)-type vacua remain unchanged, but the eight \( T \)-type solutions \( u_n \) split into two groups:
\begin{eqnarray}
T_-: \quad u_{1,6,7,8} &=& 
\left\{
\begin{pmatrix} 1 \\ 1 \\ 1 \end{pmatrix}, 
\begin{pmatrix} 1 \\ -1 \\ -1 \end{pmatrix},
\begin{pmatrix} -1 \\ 1 \\ -1 \end{pmatrix},
\begin{pmatrix} -1 \\ -1 \\ 1 \end{pmatrix}
\right\}u_-\,,\label{eq:uminus}\\
T_+: \quad u_{2,3,4,5} &=& 
\left\{
\begin{pmatrix} -1 \\ 1 \\ 1 \end{pmatrix},
\begin{pmatrix} 1 \\ -1 \\ 1 \end{pmatrix},
\begin{pmatrix} 1 \\ 1 \\ -1 \end{pmatrix},
\begin{pmatrix} -1 \\ -1 \\ -1 \end{pmatrix}
\right\}u_+ \label{eq:uplus}\,,
\end{eqnarray}
where 
\begin{eqnarray}
u_{\mp}=u\left( \sqrt{a^2+1} \mp a \right)\,, \quad \text{with} \quad a=\frac{A}{2\mu\sqrt{3g_1+2g_2}}.
\end{eqnarray}
We note that this splitting will lead to the formation of metastable walls. 
In the right panel of \figref{fig:u-vacua}, the purple and teal points denote the \(u_+\) and \(u_-\) vacua, respectively.
The potential at these solutions takes two distinct values:
\begin{eqnarray}
V_{\mp}=-\frac{3\mu^4}{4 (3 g_1 + 2 g_2)}
\left(1+\frac{2}{3}a^2 \mp \frac{2}{3}a\sqrt{a^2+1}\right)
\left(\sqrt{a^2+1} \mp a \right)^2\,, \label{eq:V_pm}
\end{eqnarray}
where the \(\mp\) signs correspond to the \(u_-\) and \(u_+\) vacua, respectively.
For \(A<0\) ($i.e.$\ \(a<0\)), we find \(V_-<V_+\), making \(V_-\) the global minimum.
Consequently, the \(u_-\) vacua are the true vacua, while the \(u_+\) vacua are metastable (false) vacua.
Conversely, for \(A>0\) ($i.e.$\ \(a>0\)), the situation is reversed.
In either case, the scalar mass-squared eigenvalues are
\begin{eqnarray}
&& m_{1,\mp}^2 = 2\mu^2\left(1+a^2 \mp a\sqrt{a^2+1}\right)\,,\nonumber\\
&& m_{2,\mp}^2=m_{3,\mp}^2 =
- \frac{2 g_2}{3 g_1 + 2 g_2}\,\mu^2
\left[\,1 - 2a\left(a \mp \sqrt{a^2+1}\right)\!\left(1+\frac{3 g_1}{g_2}\right)\right] \,.
\end{eqnarray}
We now consider the properties of domain walls in the real-triplet \(A_4\) case.
Starting from the Lagrangian \(\mathcal{L}=\sum_n \tfrac{1}{2}\,\partial^\mu\phi_n\,\partial_\mu\phi_n - V(\phi)\) for real scalars, the equations of motion (EoM) follow from the Euler–Lagrange equations,
\begin{equation}
\partial^\mu\!\left(\frac{\partial \mathcal{L}}{\partial(\partial^\mu\phi_i)}\right)
-\frac{\partial \mathcal{L}}{\partial\phi_i}
=\partial^\mu\partial_\mu\phi_i+\frac{\partial V}{\partial\phi_i}=0\,.
\label{eq:EoM_t}
\end{equation}
Assuming a static planar wall in the \(x\)–\(y\) plane, so that the fields vary only along the transverse coordinate \(z\), the static equations reduce to
\begin{equation}
\frac{d^2\phi_i(z)}{dz^2}=\frac{\partial V(\phi)}{\partial \phi_i}\,.
\label{eq:eom}
\end{equation}
Since the cubic term modifies the \(Z_3\) (\(\mathrm{T}\)-type) vacuum structure but leaves the \(Z_2\) (\(\mathrm{S}\)-type) vacua unchanged\footnote{The $\mathrm{S}$-type vacua have two vanishing components, so $I_3=\phi_1\phi_2\phi_3=0$ and are unchanged by the cubic term.
}, we focus below on \(Z_3\)-preserving walls (the \(\mathrm{S}\)-type walls have the same properties as in Ref.~\cite{Fu:2024jhu}).
\begin{figure}[t!]
\centering
\subfloat[][TI]{\includegraphics[width=.33\textwidth]{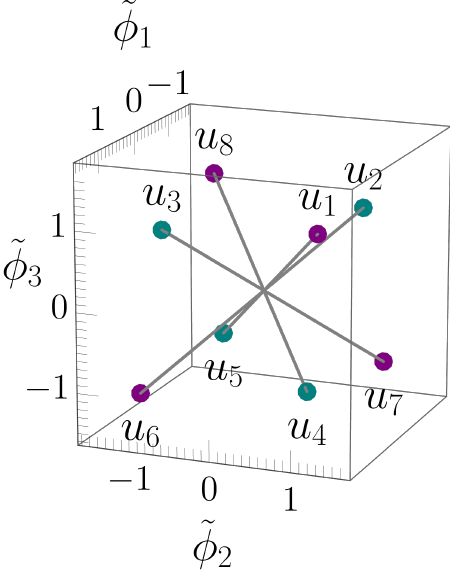}}
\subfloat[][TII]{\includegraphics[width=.33\textwidth]{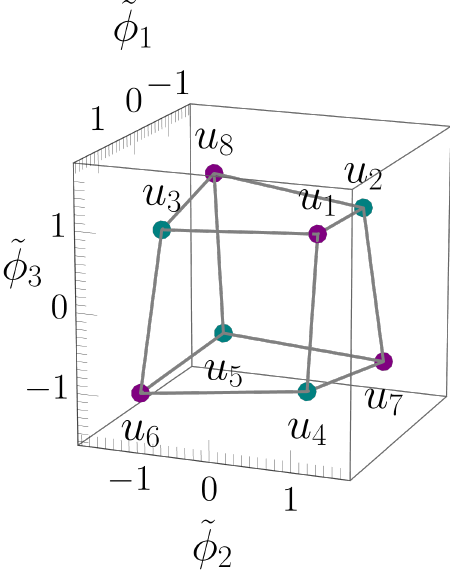}}
\subfloat[][TIII]{\includegraphics[width=.33\textwidth]{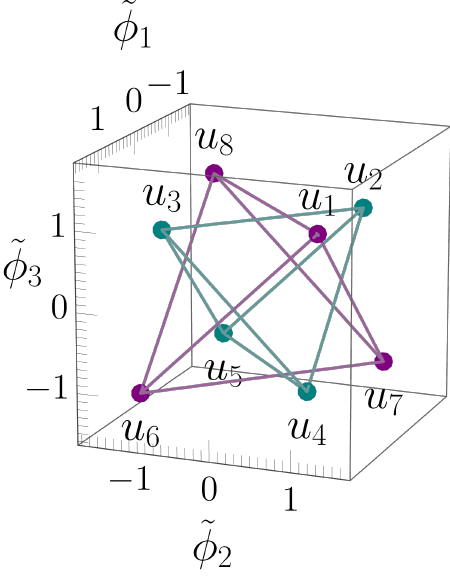}}
\caption{$Z_3$-preserving vacua of $A_4$ and three topologically different types of DWs. TI, TII and TIII DWs are given in the left, middle and right panels. The purple and teal lines of the rightmost figure show two different kinds of DWs of type TIII. 
\label{fig:DW_T}}
\end{figure}
Following Ref.~\cite{Fu:2024jhu}, we rescale the fields and the spatial coordinate as
\begin{equation}
\tilde{\phi}_i=\phi_i/u,
\qquad
\overline{z}=\mu\,z\,,
\end{equation}
so that the EoM become
\begin{equation}
\frac{d^2\tilde{\phi}_i}{d\overline{z}^2} =\tilde{\phi}_i\!\left[ -1+\frac{\tilde{\phi}_1^2+\tilde{\phi}_2^2+\tilde{\phi}_3^2}{3+2\beta} +\frac{\beta}{3+2\beta}\,(\tilde{\phi}_j^2+\tilde{\phi}_k^2) \right] +2a\,\tilde{\phi}_j\tilde{\phi}_k\,, 
\label{eq:EoM}
\end{equation}
where $i\neq j\neq k$ and $\beta\equiv g_2/g_1$.
As in the \(S_4\) case, there are three topologically distinct \(Z_3\)-preserving walls (TI, TII, and TIII) shown in \figref{fig:DW_T}, In the leftmost figure of \figref{fig:DW_T}, the grey lines represent one type of DWs, TI. In the central figure, the grey lines represent the  type TII DW. While in the rightmost figure, the teal and purple lines show two different kinds of type TIII DWs.
A key observable of domain walls is the \emph{tension} (energy/mass per unit area).
From the stress--energy tensor \(T_{\mu\nu}=\sum_i \partial_\mu\phi_i\,\partial_\nu\phi_i-\mathcal{L}\,g_{\mu\nu}\), the energy density across a static wall is
\begin{equation}
\varepsilon(z)=\frac12\sum_i\!\left[\frac{d\phi_i(z)}{dz}\right]^2+\Delta V\big(\phi(z)\big),
\qquad
\Delta V(\phi)\equiv V(\phi)-V_{\min}\,,
\label{eq:varepsilon}
\end{equation}
where the first term is the gradient energy and the second term is the potential excess above the true vacuum. 
The tension is obtained by integrating the energy density over \(z\):
\begin{equation}
\sigma=\int_{-\infty}^{+\infty}\!dz\,\varepsilon(z).
\end{equation}
Writing the magnitudes of the \(\mathrm{S}\)- and \(\mathrm{T}\)-type vacuum expectation values as
\(v=\mu/\sqrt{g_1}\) and \(u=\mu/\sqrt{3g_1+2g_2}\), we expect the parametric scalings
\be
\sigma_S\sim \mu\,v^2,\qquad
\sigma_T\sim \mu\,u^2
\ee
for the tensions of the \(\mathrm{S}\)- and \(\mathrm{T}\)-type domain walls, respectively, where there will be $\mathcal{O}(1)$ coefficients present that depend on the paths in field space.

\begin{figure}[t!]
\centering
\subfloat[$a=-0.00003$, $\beta=-0.01$, $R_\sigma = \sigma_\mathrm{1p}/\sigma_\mathrm{2p} = 1.41$.]{\includegraphics[width=.35\textwidth]{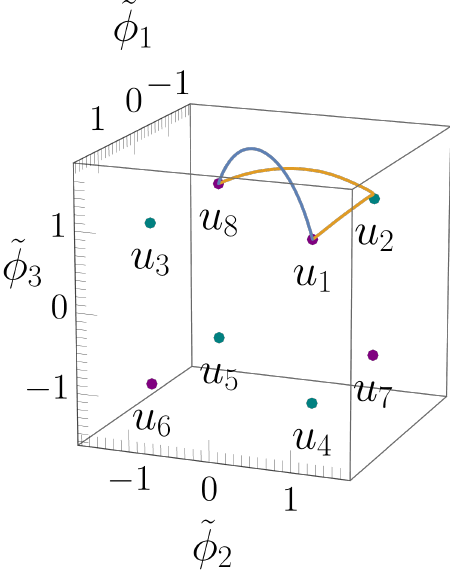}
\includegraphics[height=.35\textwidth]{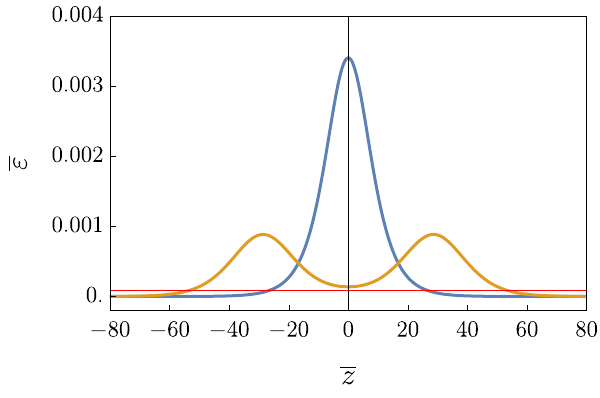}}\\
\subfloat[$a=-0.0003$, $\beta=-0.01$, $R_\sigma = \sigma_\mathrm{1p}/\sigma_\mathrm{2p} = 1.14$]{\includegraphics[width=.35\textwidth]{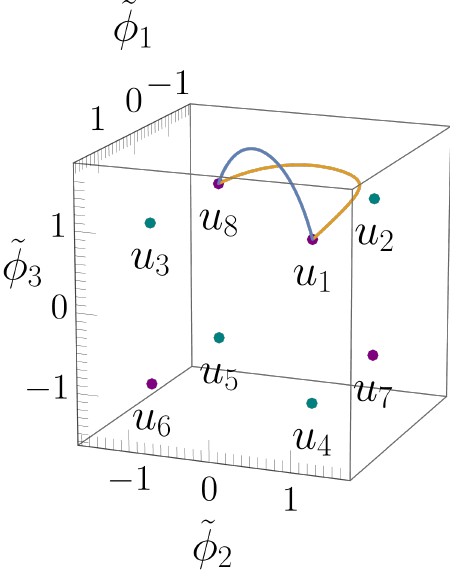}
\includegraphics[height=.35\textwidth]{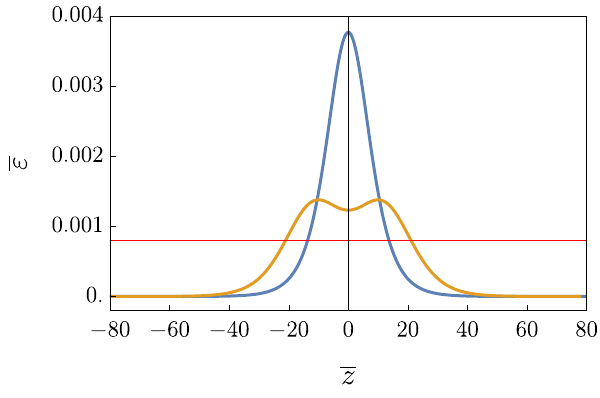}}
\caption{The left panel shows the domain wall solutions $i.e.$ solutions to EoM Eq.~\ref{eq:EoM} between the vacua $u_1$ and $u_8$. The right panel shows the normalised energy density as a function of the normalised $z$-coordinate for the different domain wall solutions. The red line marks $\Delta V$. \label{fig:path_and_en}}
\end{figure}

The TI and TII walls interpolate between the \emph{split} \(\mathrm{T}\)-vacua \(u_-\) and \(u_+\) ($e.g.$, \(u_1\!\leftrightarrow\!u_5\) for TI, \(u_1\!\leftrightarrow\!u_4\) for TII), whereas TIII walls connect vacua of the same depth. Similar to the $S_4$ model, these TIII solutions can be unstable. This phenomenon is shown in \figref{fig:path_and_en}, which plots trajectories found by solving Eq.~\eqref{eq:EoM} for TIII boundary conditions, \(\tilde{\phi}(-\infty)=u_1\) and \(\tilde{\phi}(+\infty)=u_8\), across two benchmark points (rows). We find two distinct TIII solutions.
The blue curve is the direct,  ``one-peak'' TIII solution, which connects \(u_1 \to u_8\) via a single path. The yellow curve is an indirect, ``two-peak'' TIII solution. We emphasise that this is also a TIII solution, as it begins at $u_1$ and ends at $u_8$. However, its field trajectory is different, bending \emph{toward} the \(u_2\) vacuum, as shown in the left panels.
This two-peak structure arises from the underlying physics: different from the $S_4$ case~\cite{Fu:2024jhu}, the $A \neq 0$ term lifts the degeneracy, making \(u_2\) a higher-energy false vacuum relative to \(u_1\) and \(u_8\). 
The direct one-peak path (blue) is energetically costly. 
The system prefers the indirect path (yellow), which is physically analogous to a {bound state of two TII-like walls}. 
The right panels show the corresponding energy densities \(\varepsilon(z)\): the blue profile has a single central peak, while the yellow profile has two. 
Because the yellow path only \emph{approaches} \(u_2\) but does not settle there, the energy density in the ``dip'' between the peaks remains above the \(u_{+}\) vacuum value. We dub this the ``oreo'' domain wall configuration. 
This two-peak configuration is stable because it selects an intermediate separation that balances the gradient energy (which would push the peaks apart) and the potential energy (which would pull them together)\footnote{Heuristically, one can picture the decay of a TIII wall as a force pushing the central peak down and the energy difference between $u_+$ and $u_-$ vacua as a force pushing the two peaks to the center; the resulting balance fixes both the peak separation and the intermediate energy level.}.
Defining the tensions of the one-peak and two-peak solutions as $\sigma_\mathrm{1p}$ and $\sigma_\mathrm{2p}$, respectively, we note that for both benchmark points, the ratio $\sigma_\mathrm{1p}/\sigma_\mathrm{2p}$ is sufficiently large ($> 1$). This confirms that the {direct, one-peak TIII solution is unstable and will decay} into the energetically preferred, stable two-peak TIII configuration.
\begin{figure}
    \centering
    \includegraphics[width=0.8\linewidth]{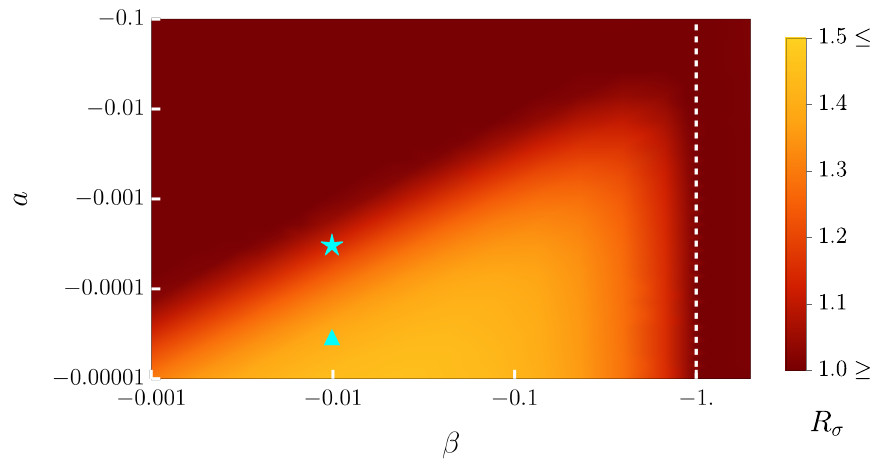}
    \caption{The ratio $R_\sigma=\sigma_\mathrm{1p}/\sigma_\mathrm{2p}$ for different $\beta$ and $a$.
    The benchmark points in \figref{fig:path_and_en} are marked out by the star and the triangle.  }
    \label{fig:Ratio_Heatmap}
\end{figure}
The value of ratio $R_\sigma=\sigma_\mathrm{1p}/\sigma_\mathrm{2p}$ in the $\beta$-$a$ plane is shown in \figref{fig:Ratio_Heatmap}. The red region indicates when $\sigma_\mathrm{1p} \leq \sigma_\mathrm{2p}$ and the one-peak defect cannot decay to the two-peak configuration. In contrast the yellowish regions show when the decay \emph{can} occur. 
On the one hand, the one-peak configuration is favoured as $\beta$ approaches $-1$ and becomes stable for $\beta\leq -1$, which is consistent with the $S_4$ limit (for negligible $|a|$ in the bottom of \figref{fig:Ratio_Heatmap}). 
On the other hand, as $a$ becomes large (roughly $0.1\beta$), the two-peak configuration is also disfavoured because the bias between $u_+$ and $u_-$ is too large. 

To analyse the dynamics of domain walls (DWs) we must compute the wall tensions. 
For the TI and TII walls this is non-trivial because they interpolate between vacua 
\(u_+\) and \(u_-\) with different energy densities. Physically, such walls are 
immediately biased: the explicit \(\mathbb{Z}_2\) breaking lifts the degeneracy across 
the wall and creates a uniform pressure difference \(\Delta V\) that accelerates the wall. 
A practical approach is to compute the tension in the \(S_4\)-symmetric limit, where 
\(u_+\) and \(u_-\) are degenerate and a static solution exists, and then treat the 
splitting as a bias term. Below we discuss the qualitative wall dynamics but do not compute the tensions of TI and TII explicitly.

If the tension of a TIII wall exceeds that of two TII walls, $
\sigma(\mathrm{TIII})>2\,\sigma(\mathrm{TII})$, 
it can decay into two TII walls, with decay rate \(\Gamma\propto\Delta\sigma\), where $
\Delta\sigma\equiv \sigma(\mathrm{TIII})-2\,\sigma(\mathrm{TII})$.
In this \(A_4\) model with \(A\neq 0\), every TI/TII wall separates vacua of \emph{unequal energy}, \(u_-\) and \(u_+\).
For \(A<0\) one has \(V_-<V_+\) and the energy gap is:
\[
\Delta V\equiv V_+-V_-=-\frac{2\mu^{4}}{3g_{1}+2g_{2}}\,a\,(1+a^{2})^{3/2}\,,
\]
as derived from Eq.~\eqref{eq:V_pm}. This energy difference provides a uniform pressure \(P=\Delta V\) that pushes the wall from the \(u_-\) side toward the \(u_+\) side (for \(A>0\) the roles of \(u_\pm\) are reversed).
Consequently, when a TIII wall decays into two TII walls, the pair are driven back toward each other by this pressure.
The instantaneous acceleration of a planar TII wall is
\[
\mathrm{acc}=\frac{\Delta V}{\sigma_{\mathrm{TII}}}\,,
\]
and, using \(\sigma_{\mathrm{TII}}\sim \mu\,u^2=\mu^{3}/(3g_{1}+2g_{2})\), we obtain the parametric estimate
\[
\mathrm{acc}\sim \mu\,a\,(1+a^{2})^{3/2}\,,
\]
in natural units. As expected, when \(|a|\ll 1\) the pressure difference \(\Delta V\), and hence the wall acceleration, is small.
In the absence of Hubble friction and scalar radiation, the two–wall system produced when a TIII wall decays into two TII walls undergoes undamped oscillations: the walls repeatedly collide and re-separate, and energy conservation prevents their annihilation. These oscillations are not expected to be an efficient GW source compared to the final collapse, but they can store energy and delay annihilation, effectively increasing \(\sigma_{\mathrm{eff}}\) and delaying annihilation. Including Hubble friction and scalar radiation introduces dissipation, damping the oscillations and driving the system to annihilation in finite time. We note that this two-wall oscillatory behaviour is unique to $A_4$ (this does not occur with the $S_4$ symmetry). A full dynamical analysis including Hubble friction will be presented elsewhere.

\subsection{Domain Walls from Complex Scalar Fields}\label{sec:complextrip} 
\begin{figure}
  \centering
  \includegraphics[width=0.6\linewidth]{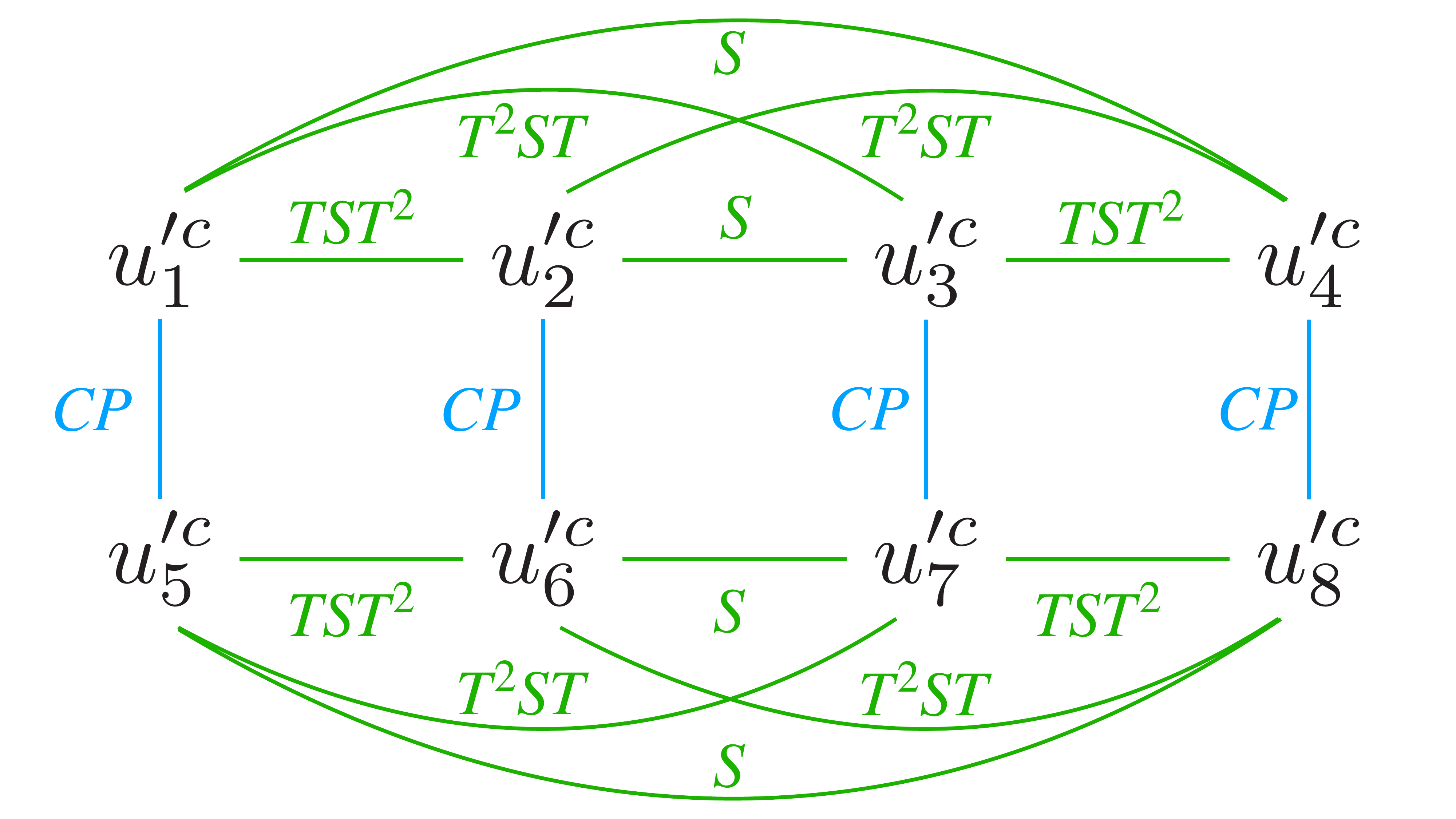}
  \caption{Relation between the vacua under $A_4$ group transformations and CP conjugation.}
  \label{fig:tcp_vacua}
\end{figure}

Complex scalar triplets are well motivated in flavour model building, as their additional phase degrees of freedom enrich the vacuum structure, allow for CP-violation and often appear in realistic flavon sectors or supersymmetric completions. 
We now consider flavour models with a complex scalar triplet, 
$\varphi=(\varphi_{1},\varphi_{2},\varphi_{3})^{T}$.
Because the fields carry complex phases, the most general renormalisable 
\(A_{4}\)-invariant potential contains more operators than in the real-triplet case. 
Up to hermitian conjugation, every renormalisable term can be expressed as a 
linear combination of the invariants listed in App.~\ref{app:complex_invariants}. 
The potential then takes the schematic form 
\be\label{eq:potential_complex}
\begin{aligned}
V(\varphi, \varphi^*)\;=\;&
-\mu_{1}^{2}\,\mathcal{I}_{11} + g_{11}\,\mathcal{I}_{11}^{2}
+2g_{21}\,\mathcal{I}_{21} + \bigl(g_{22}\,\mathcal{I}_{22}+{\rm h.c.}\bigr) \\
&+\Bigl[ -\mu_{2}^{2}\,\mathcal{I}_{12}
+ g_{12}\,\mathcal{I}_{12}^{2} + g_{13}\,\mathcal{I}_{11}\mathcal{I}_{12}
+ g_{23}\,\mathcal{I}_{23} + g_{24}\,\mathcal{I}_{24} 
\\&\qquad\qquad\qquad\qquad\qquad
+ g_{25}\,\mathcal{I}_{25}
+ A_{1}\,\mathcal{I}_{31} + A_{2}\,\mathcal{I}_{32}
+ \text{h.c.} \Bigr].
\end{aligned}
\ee
Coefficients $\mu_1^2$, $g_{11}$, $g_{21}$ are real, while all other coefficients are complex. 
This potential contains many free parameters and in realistic model building, one usually invokes additional symmetries to reduce the number of parameters. For example, assuming a conserved \(U(1)\) charge and an unbroken CP symmetry (so that all couplings are real) eliminates the bracketed terms, and the potential reduces to 
\be\label{eq:potential_c}
V(\varphi, \varphi^*)=
-\mu_{1}^{2}\,\mathcal{I}_{11}
+g_{11}\,\mathcal{I}_{11}^{2}
+2g_{21}\,\mathcal{I}_{21}
+g_{22}\bigl(\mathcal{I}_{22}+{\rm h.c.}\bigr),
\ee
which is equally valid for an \(S_{4}\times U(1)\) symmetry. 
Since the potential is independent of the overall \(U(1)\) phase, this angle is undetermined by minimisation and labels an \(S^{1}\) of degenerate vacua. The spontaneous breaking of this global \(U(1)\) produces a massless Goldstone mode corresponding to uniform shifts of the common phase, and leads to the formation of a network of global strings. This is distinguishable from the walls attached strings, which usually arises from the breaking chain $U(1) \to Z_n \to 1 $ \cite{Hiramatsu:2012sc,Wu:2022tpe}. The DWs and global strings in our case, are independent global defects. Replacing the \(U(1)\) by a discrete subgroup changes the operator content. For example, in \(A_{4}\times\mathbb{Z}_{3}\) the cubic invariant \(\mathcal{I}_{31}\) is allowed, distinguishing \(A_{4}\) from \(S_{4}\). Nonetheless, in what follows we focus on the case with the additional $U(1)$. 
It will be convenient to parametrise the three components of the complex scalar in polar form, 
\[
\varphi_i = \frac{\phi_i}{\sqrt{2}} e^{i\alpha_i}.
\] 
We distinguish three types of global phases: 
\begin{itemize}
  \item \(\alpha_i\): the phase associated with each field. 
  \item \(\alpha_{ij}=\alpha_i-\alpha_j\): physical phase differences between components that enter the potential. 
  \item \(\alpha\): the common shift \(\alpha_i\mapsto\alpha_i+\alpha\) for all \(i\). The potential is \(U(1)\)-invariant under this shift, so \(\alpha\) labels an \(S^1\) of degenerate vacua and is responsible for cosmic string formation. 
\end{itemize}
Using this polar parametrisation, the potential depends only on the relative phases \(\alpha_{12}\) and \(\alpha_{23}\), and becomes 
\be\label{eq:potential_c_polar}
V(\phi,\alpha_{ij})=
-\frac{\mu^{2}}{2}\,I_{1}
+\frac{g_{1}}{4}\,I_{1}^{2}
+\frac{g_{2}}{2}\,I_{2}
+\frac{g_{2}'}{2}\,I_{2}'\,,
\ee 
where \(I_{1}\) and \(I_{2}\) retain the forms of Eq.~\eqref{eq:A4_invariant} and 
\[
I_{2}'=
\phi_{1}^{2}\phi_{2}^{2}\cos 2\alpha_{12}
+\phi_{2}^{2}\phi_{3}^{2}\cos 2\alpha_{23}
+\phi_{3}^{2}\phi_{1}^{2}\cos 2(\alpha_{12}+\alpha_{23})\,,
\] 
with the simplified notation 
\(\{\mu_{1},g_{11},g_{21},g_{22}\}\equiv\{\mu,g_{1},g_{2},g_{2}'\}\). 
A limitation of the polar form is that if any radial field \(\phi_i=0\), the corresponding phase \(\alpha_i\) is ill-defined. 
When two of the radial fields are zero, we obtain the S-type ($Z_2$-preserving) vacua, which are 
\begin{equation}
\mathrm{S}^c:\quad 
\left\{
\begin{pmatrix} 1 \\ 0 \\ 0 \end{pmatrix}, 
\begin{pmatrix} 0 \\ 1 \\ 0 \end{pmatrix},
\begin{pmatrix} 0 \\ 0 \\ 1 \end{pmatrix}
\right\} v e^{i\alpha}, 
\qquad v=\frac{\mu}{\sqrt{g_1}}\,.
\label{eq:Sc-vacua}
\end{equation}
For each vacuum, only one individual phase is well defined (the one associated with the nonzero field). Along a domain wall interpolating between two such vacua, however, more than one phase can vary simultaneously, which we will analyse in detail later. 
When all three radial fields are nonzero, all three phases are well-defined. In this case, two independent relative phases remain physical, while the common overall $U(1)$ phase corresponds to the global degeneracy discussed above. Combining the phase and modulus conditions, one obtains the $Z_3$-preserving vacua: 
the $\mathrm{T}^c$-type branch for $g_2'<0$
\begin{equation}
\left\{u^c_1,u^c_2,u^c_3,u^c_4\right\}= 
\left\{
\begin{pmatrix} 1 \\ 1 \\ 1 \end{pmatrix},\;
\begin{pmatrix} 1 \\ -1 \\ -1 \end{pmatrix},\;
\begin{pmatrix} -1 \\ 1 \\ -1 \end{pmatrix},\;
\begin{pmatrix} -1 \\ -1 \\ 1 \end{pmatrix}
\right\} u^c e^{i\alpha},\qquad\quad
\label{eq:Tc-vacua}
\end{equation}
the $\mathrm{T}'^c_{+}$ branch for $0<g_2'<-g_2$ 
\begin{equation} 
\left\{u'^c_1,u'^c_2,u'^c_3,u'^c_4\right\}=
\left\{
\begin{pmatrix} 1 \\ \omega \\ \omega^* \end{pmatrix},\;
\begin{pmatrix} 1 \\ -\omega \\ \omega^* \end{pmatrix},\;
\begin{pmatrix} \omega^* \\ 1 \\ -\omega \end{pmatrix},\;
\begin{pmatrix} -\omega \\ \omega^* \\ 1 \end{pmatrix}
\right\} u'^c\,e^{i\alpha}\,,\quad
\label{eq:Tcp-vacua}
\end{equation}
and its CP-conjugated branch $\mathrm{T}'^c_{-}$
\begin{equation}
\left\{u'^c_5,u'^c_6,u'^c_7,u'^c_8\right\}=
\left\{
\begin{pmatrix} 1 \\ \omega^* \\ \omega \end{pmatrix},\;
\begin{pmatrix} 1 \\ -\omega^* \\ \omega \end{pmatrix},\;
\begin{pmatrix} \omega \\ 1 \\ -\omega^* \end{pmatrix},\;
\begin{pmatrix} -\omega^* \\ \omega \\ 1 \end{pmatrix}
\right\} u'^c\,e^{i\alpha},
\end{equation}
where $\omega = e^{i2\pi/3}$ ($\omega^2=\omega^*$).
Within each branch of $\mathrm{T}'^c_{+}$ and $\mathrm{T}'^c_{-}$, the four vacua are related by order-2 $A_4$ transformations, as shown in \figref{fig:tcp_vacua}.
Besides, each vacuum is invariant under an order-3 transformation up to a global phase: $u'^c_1$, $u'^c_2$, $u'^c_3$ and $u'^c_4$ are invariant under $\mathrm{T}$, $ST$, $TS$ and $STS$, respectively. 
And their CP-conjugated vacua are invariant under the same transformations.
The two branches together give eight $Z_3$-preserving vacua when $0<g_2'<-g_2$.
The factor $e^{i\alpha}$ is the global $U(1)$ phase (a degenerate $S^1$ vacuum manifold), while the entries with $\omega,\omega^*$ encode physical relative phases. 
The corresponding vacuum magnitudes are 
\begin{equation}
u^c = \frac{\mu}{\sqrt{3g_1+2g_2+2g_2'}}\,,
\qquad
u'^c = \frac{\mu}{\sqrt{3g_1+2g_2-g_2'}}\,.
\end{equation}
The overall phase $\alpha$ is free and cannot be determined by minimisation of the potential. This phase is a manifestation of the global $U(1)$ symmetry and, upon spontaneous breaking, leads to the formation of a cosmic string network. In addition to strings, domain walls can also form between non-identical vacua. For example, one can consider the wall interpolating between the first two vacua of $\mathrm{T}^c$ in Eq.~\eqref{eq:Tc-vacua}. While the vacua for $g_2'<0$ are essentially extensions of the real scalar case with only a global phase, the vacua for $g_2'>0$ introduce nontrivial relative phases between the components. 
For a set of complex scalars $\varphi_i$, the one-dimensional static equations of motion along the $z$-direction are 
\begin{equation}
\frac{d^2\varphi_i(z)}{dz^2} = \frac{\partial V(\varphi, \varphi^*)}{\partial \varphi_i^*}\,,
\qquad \text{for each complex scalar } \varphi_i \,.
\label{eq:eom_complex}
\end{equation}
The energy density is given by 
\begin{equation}
\varepsilon(z) = \sum_i \frac{d\varphi_i(z)}{dz}\frac{d\varphi_i^*(z)}{dz} 
+ \Delta V\bigl(\varphi(z),\varphi^*(z)\bigr)\,.
\label{eq:varepsilon_complex}
\end{equation}
To simplify the analysis, we focus on $A_4\times U(1)$ models with the reduced potential of Eq.~\eqref{eq:potential_c}. 
Moving to the parameterisation 
\begin{equation}
\varphi_i = \frac{1}{\sqrt{2}}\,(h_i+i a_i)\,,
\end{equation}
the equations of motion for the real and imaginary components become 
\begin{align}
\frac{d^2 h_i}{dz^2} &= h_i\Big[-\mu^2 + g_1 \sum_j(h_j^2+a_j^2) 
+ \sum_{j\neq i}\Big(g_2(h_j^2+a_j^2)+g_2'(h_j^2-a_j^2)\Big)\Big]
+2g_2' a_i \sum_{j\neq i} h_j a_j, \\
\frac{d^2 a_i}{dz^2} &= a_i\Big[-\mu^2 + g_1 \sum_j(h_j^2+a_j^2) 
+ \sum_{j\neq i}\Big(g_2(h_j^2+a_j^2)-g_2'(h_j^2-a_j^2)\Big)\Big]
+2g_2' h_i \sum_{j\neq i} h_j a_j.
\end{align}

\subsubsection{$\mathrm{S}^c$-type domain walls}
We now focus on walls between $\mathrm{S}^c$-type vacua, such as the one interpolating between 
$(1,0,0)^{T} v e^{i\alpha_1}$ and $(0,1,0)^{T} v e^{i\alpha_2}$. 
As the fields transition between these vacua, a relative phase 
$\alpha_{12}\equiv\alpha_1-\alpha_2$ can arise. 
At spatial infinity, where the configuration must settle into a true vacuum, 
the potential is minimised only for discrete values of this phase:
\[
\alpha_{12}=n\frac{\pi}{2},\qquad n\in\mathbb{Z}.
\]
For a minimal tension static wall, this relative phase must be 
constant across the wall, $i.e.$\ $\alpha_{12}'(z)=0$. 
This configuration simultaneously minimises both the kinetic energy term and the phase-dependent potential energy. As detailed in Appendix~\ref{app:Sdomain}, an analysis of the conserved current associated with the $U(1)$ symmetry leads to the same conclusion.

The crucial insight is that for a stable wall, the relative phase $\alpha_{12}$ must be constant. This reduces the complex equations of motion into a form that is mathematically identical to a well-understood real scalar model with an $S_4$ symmetry~\cite{Fu:2024jhu}. This resemblance is significant, as it allows us to directly apply the known results from that simpler case to our analysis.
In this simplified picture, the influence of the constant phase gets absorbed into a new effective coupling parameter, $\mathbf{B}$:
\[
\mathbf{B} \equiv \beta+\beta'\cos 2\alpha_{12}\,.
\]
The equations of motion for the field magnitudes then take on the familiar $S_4$ form:
\[
\overline{\phi}_1''=\overline{\phi}_1\!\left[-1+\overline{\phi}_1^2+\overline{\phi}_2^2+\mathbf{B}\,\overline{\phi}_2^2\right]\,,\quad
\overline{\phi}_2''=\overline{\phi}_2\!\left[-1+\overline{\phi}_1^2+\overline{\phi}_2^2+\mathbf{B}\,\overline{\phi}_1^2\right]\,.
\]
From the analysis in Ref.~\cite{Fu:2024jhu}, the tension of the domain wall increases as $\mathbf{B}$ increases. The stable domain wall will be the one corresponding to the phase $\alpha_{12}$ that {minimises the value of $\mathbf{B}$}. The system has two families of choices for the phase: $\alpha_{12}=0,\pi$ (making $\cos 2\alpha_{12}=+1$) or $\alpha_{12}=\pm\pi/2$ (making $\cos 2\alpha_{12}=-1$). The preferred choice depends on the sign of $\beta'$:
\begin{itemize}
  \item If \textbf{$\beta' > 0$}, the system chooses $\alpha_{12}=\pm\pi/2$ to make the second term negative, resulting in $\mathbf{B}=\beta-\beta'$. 
  \item If \textbf{$\beta' < 0$}, the system chooses $\alpha_{12}=0,\pi$ to make the second term negative (since $\beta'$ itself is negative), resulting in $\mathbf{B}=\beta+\beta'$.
\end{itemize}
In both cases, the minimal tension branch achieves the lowest possible effective coupling of $\mathbf{B}_{\text{min}}=\beta-|\beta'|$.

\begin{figure}[t!]
  \centering
  \includegraphics[width=0.35\linewidth]{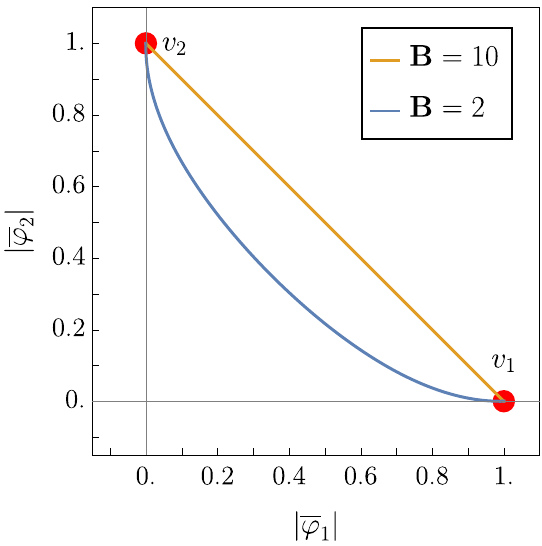}
  \label{fig:SIIc_path_bc1}
  \includegraphics[width=0.5\linewidth]{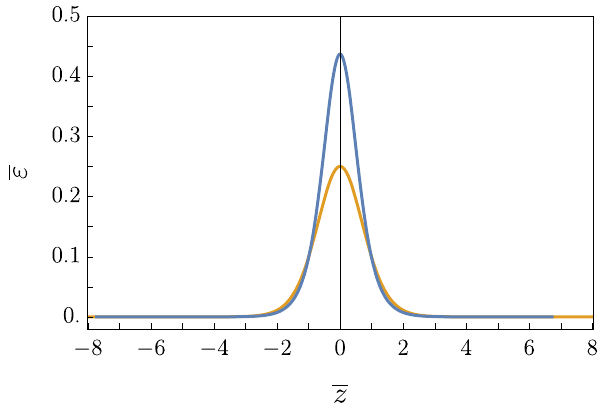}
  \caption{Field–space paths between the vacua (red points) $v_1$ and $v_2$. The orange path corresponds to $\mathbf{B}=2$ and the blue path to $\mathbf{B}=10$. The stable, minimal-tension wall always follows the path with the smaller $\mathbf{B}$ value.}
  \label{fig:SIIc_path}
\end{figure}
\figref{fig:SIIc_path} provides an explicit example. 
The two paths in the left panel correspond to the two values of $\mathbf{B}$: the orange path for $\mathbf{B}=2$ and the blue path for $\mathbf{B}=10$. 
The right panel shows the dimensionless energy density, defined as 
\begin{eqnarray}
\overline{\varepsilon} = \int_{-\infty}^{+\infty} d\overline{z} \, \left\{ \frac{1}{2} \big[\overline{\phi}^{\prime 2}_1(\overline{z}) + \overline{\phi}^{\prime 2}_2(\overline{z}) + \overline{\phi}^{\prime 2}_3(\overline{z}) \big] + \Delta \overline{V}(\overline{\phi}(\overline{z})) \right\}\,,
\end{eqnarray}
along the $z$-axis, and it is clear that the orange path ($\mathbf{B}=2$) has lower energy. 
Those values of $\mathbf{B}$ can be realised in two distinctive cases: 
\begin{itemize}
  \item When $\beta=6$ and $\beta'=4$, $\mathbf{B}=2$ is realised for $\alpha_{12}=\pm\pi/2$ while $\mathbf{B}=10$ is realised for $\alpha_{12}=0,\,\pi$;
  \item When $\beta=6$ and $\beta'=-4$, $\mathbf{B}=2$ is realised for $\alpha_{12}=0,\,\pi$ while $\mathbf{B}=10$ is realised for $\alpha_{12}=\pm\pi/2$.
\end{itemize}
It is therefore sufficient to analyse two representative boundary conditions:
\begin{align*}
  &\text{BP1}: z=-\infty:\ (1,0)^{T}\,v,\quad z=+\infty:\ (0,1)^{T}\,v,\qquad \beta=6\,, \beta'=-4 \\
  &\text{BP2}: z=-\infty:\ (1,0)^{T}\,v,\quad z=+\infty:\ (0,i)^{T}\,v,\qquad \beta=6\,, \beta'=4 
\end{align*}
corresponding to $\alpha_{12}=0$ and $\alpha_{12}=\pi/2$, respectively. 
For each benchmark, we determine wall profiles between these vacua using a shooting method akin to 
\texttt{CosmoTransitions} \cite{Wainwright:2011kj}. 
\begin{figure}[t!]
  \centering
  \subfloat[BP1]{\includegraphics[width=0.45\linewidth]{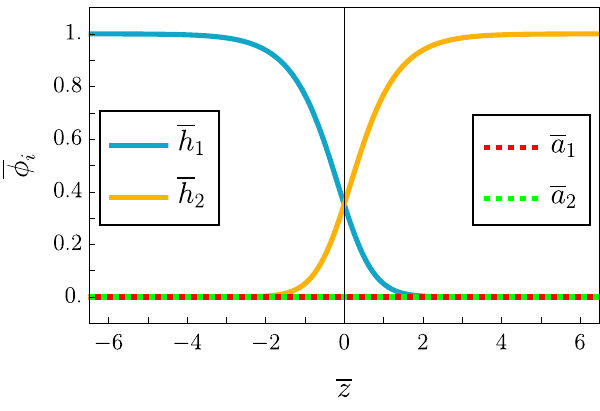}
  \label{fig:BP1sols}}
  \subfloat[BP2]{\includegraphics[width=0.45\linewidth]{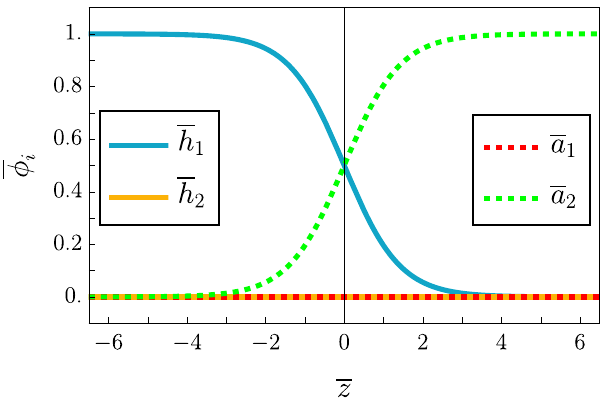}
  \label{fig:BP2sols}}
  \caption{(a) BP1 wall interpolating between $(1,0)^T v$ and $(0,1)^T v$ with $\beta=6$ and $\beta'=-4$. 
  (b) BP2 wall interpolating between $(1,0)^T v$ and $(0,i)^T v$ with $\beta=6$ and $\beta'=4$.}
\end{figure}
The BP1 solutions are shown in \figref{fig:BP1sols}: $h_1$, the real part of 
$\varphi_1$, interpolates from $1\to0$ and $h_2$, the real part of $\varphi_2$, interpolates from $0\to1$. 
In this case the relative phase is $\alpha_{12}=0$, and the imaginary parts of the fields, $a_{1,2}$, remain zero. 
While $h_{1,2}$ interpolate smoothly, $a_{1,2}$ vanish, reducing to the real scalar case with $\alpha_{12}=0$ and vanishing $U(1)$ current. 
We note that in this case, $\mathbf{B}=2$ and the DW solution is stable. 
The BP2 solutions in \figref{fig:BP2sols} show $h_1$ and $a_2$ interpolating in $z$ while $h_2$ and $a_1$ vanishing, corresponding to a constant relative phase $\alpha_{12}=\pi/2$ and vanishing $U(1)$ current in the $z$-direction.
In both benchmark cases, the constant phase condition ensures that the Noether current is conserved and vanishes, and that the wall profiles minimise the tension and the domain walls are stable. 

\subsubsection{$\mathrm{T}^c$-type domain walls}

\begin{figure}[t!]
  \centering
  \includegraphics[width=0.45\linewidth]{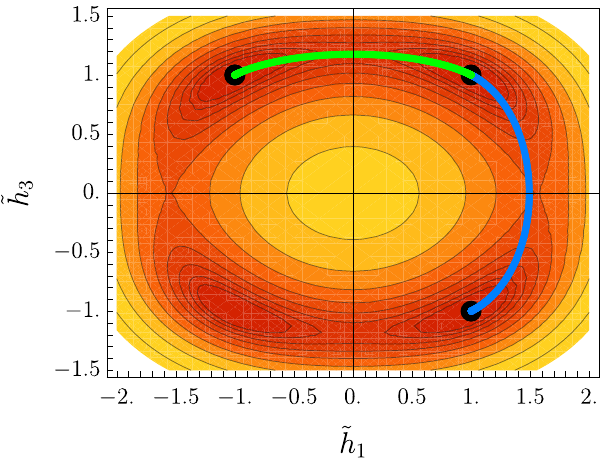}
  \includegraphics[width=0.5\linewidth]{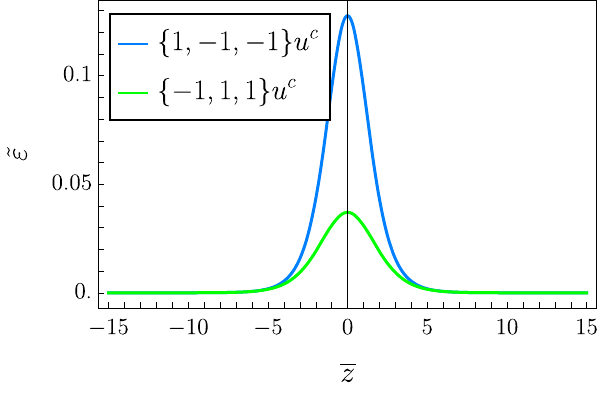}  
  \caption{Left panel: the path in $\tilde{h}_1-\tilde{h}_3$ field space for domain wall solutions $(1,1,1)u^c \;\to\; (-1,1,1)u^c$ (the green path) and $(1,1,1)u^c \;\to\; (1,-1,-1)u^c$ (the blue path). Right panel: the energy density of the domain walls for both paths in the $\overline{z}$ direction. }
  \label{fig:Tc}
\end{figure}

The $\mathrm{T}^c$-type vacua are listed in \equaref{eq:Tc-vacua} and to investigate
their associated domain walls we consider the interpolation between $(1,1,1)u^c$ and $(1,-1,-1)u^c e^{i\alpha}$ as an example. 
Without loss of generality, we fix the overall $U(1)$ phase of $(1,1,1)u^c$, the left-vacuum at $z=-\infty$, to be unity. 
The relative phase of the right-vacuum, at $z=+\infty$, can then be taken as either $\alpha=0$ or $\alpha=\pi$, which correspond to two distinct wall solutions:
\begin{enumerate}
  \item[(i)] $(1,1,1)u^c \;\to\; (-1,1,1)u^c$ \quad ($\alpha=\pi$), 
  \item[(ii)] $(1,1,1)u^c \;\to\; (1,-1,-1)u^c$ \quad ($\alpha=0$).
\end{enumerate}
In both cases, the solutions to the EoMs are paths in field space that lie fully in the real subspace, as shown in the left panel of \figref{fig:Tc} by the blue and green paths, respectively
\footnote{At any point in field space there exists a flat U(1) direction associated with the global symmetry. 
Variation along this direction corresponds to a string solution, requiring an extra spatial dimension to fully describe.}. 
In such a case, the EoMs can be reduced to \footnote{It is convenient to work in the algebraic form of the fields rather than polar coordinates: the polar form suffers from singularities when $\phi_i=0$ (where $\alpha_i$ is ill defined), and the radial equations contain additional phase-kinetic terms $\phi_i^2(\partial_\mu \alpha_i)^2$ that complicate the analysis. }
\begin{eqnarray}
  \frac{d^2h_i(z)}{dz^2} &=& 
  h_i\Big[-\mu^2 + g_1 \sum_{j} h_j^2 + (g_2+g'_2)\sum_{j\neq i} h_j^2\Big]\,, 
\end{eqnarray}
which, after rescaling $\tilde{h}_i = h_i/u^c$ and $\overline{z}=\mu z$, become
\begin{eqnarray}
  \frac{d^2\tilde{h}_1}{d\overline{z}^2} &=& 
  \tilde{h}_1\Big[-1 + \frac{1}{3+2\mathbf{B}}\,\tilde{h}_1^2 
  + \frac{2+2\mathbf{B}}{3+2\mathbf{B}}\,\tilde{h}_3^2\Big]\,, \\
  \frac{d^2\tilde{h}_3}{d\overline{z}^2} &=& 
  \tilde{h}_3\Big[-1 + \frac{2+\mathbf{B}}{3+2\mathbf{B}}\,\tilde{h}_3^2 
  + \frac{1+\mathbf{B}}{3+2\mathbf{B}}\,\tilde{h}_1^2\Big]\,, 
\end{eqnarray}
with $\mathbf{B}=\beta+\beta'$. We note that there is a permutation symmetry between $\varphi_2$ and $\varphi_3$ and so the EoM for $\tilde{h}_2$ is not necessary to solve ($\tilde{h}_2=\tilde{h}_3$). 
These are precisely the equations of motion of the $S_4$ symmetric real scalar theory \cite{Fu:2024jhu}. 

For the domain wall associated to (i), the asymptotic vacua of $\varphi_1$ change from $1$ to $-1$, which can be read from the green curve in the left panel of \figref{fig:Tc}. 
While the field value of $\tilde{h}_3$ (and also $h_2$) remain the same at $z=\pm\infty$, there is non-trivial change in its field value away from the asymptotic vacua. 
In contrast, the path in the field space for solution (ii), both the right-vacua of $\varphi_2$ and $\varphi_3$ change. 
This manifests in a higher energy density for solution (ii) compared to (i) as shown on the right panel of \figref{fig:Tc}.

\subsubsection{$\mathrm{T}'^c$-type domain walls}
\begin{figure}[t!]
  \centering
  \includegraphics[width=0.6\linewidth]{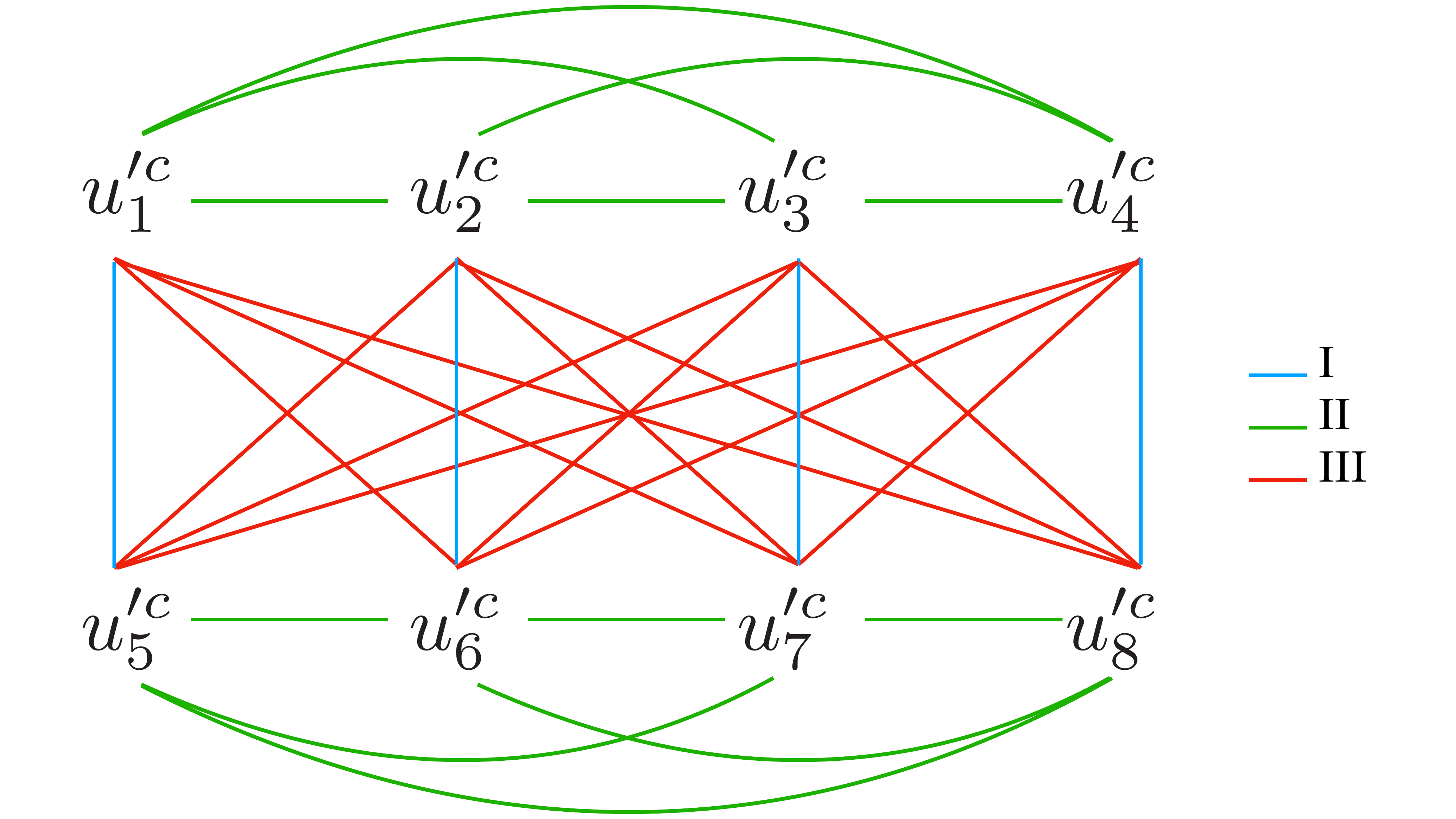}
  \caption{Diagram showing different types of domain walls between the 8 $\mathrm{T}'^c$-type vacua.}
  \label{fig:DW_sum}
\end{figure}

\begin{figure}
  \centering
  \includegraphics[width=0.8\linewidth]{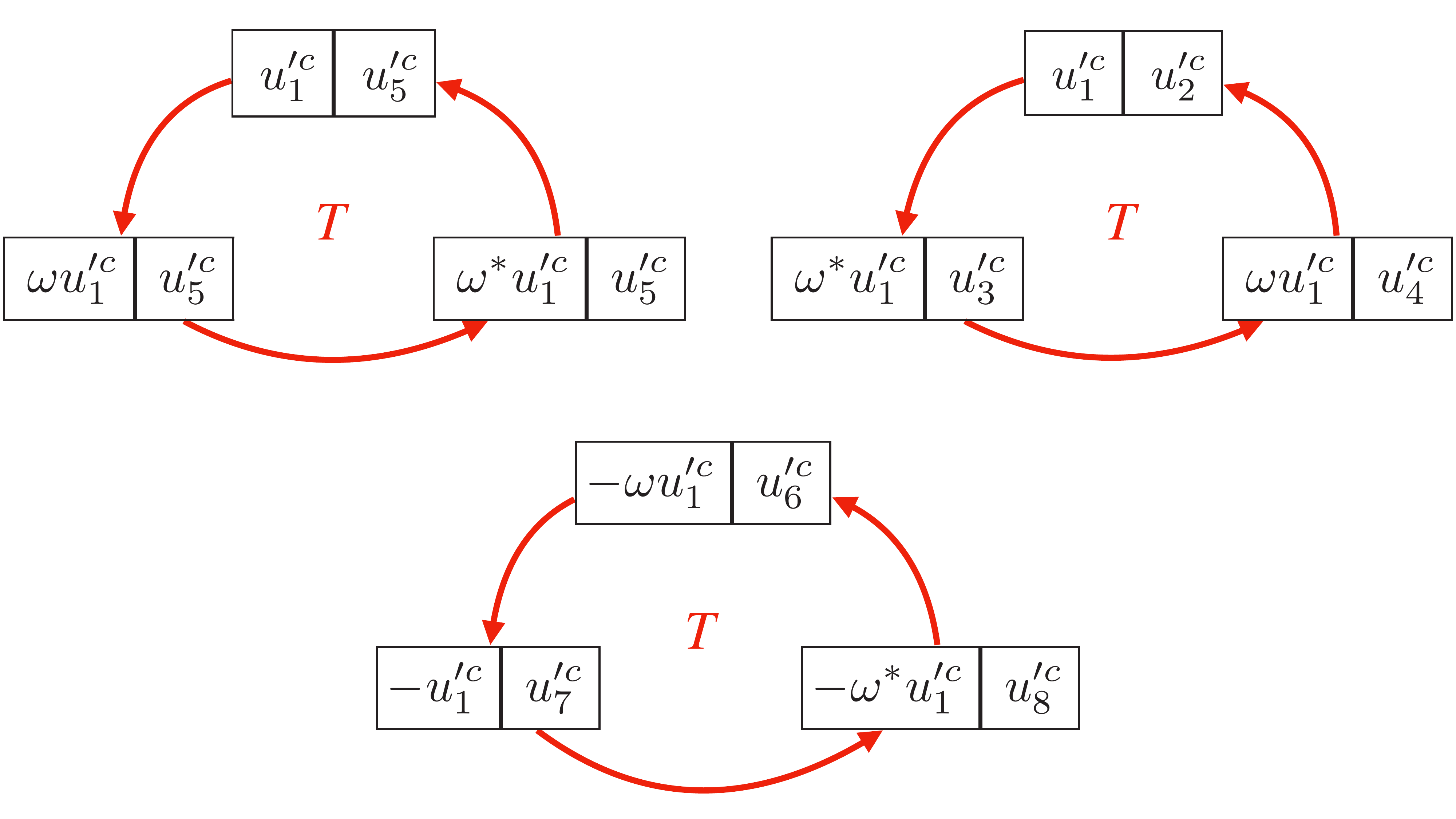}
  \caption{Behaviour of DWs under the $\mathrm{T}$ transformation}
  \label{fig:wall_trans}
\end{figure}

In contrast to the restricted set of possibilities of $\mathrm{T}^c$ domain walls, the $\mathrm{T}'^c$-type vacua admit a richer spectrum of DWs. 
There are three distinctive types of DWs between the $\mathrm{T}'^c$-type vacua:
\begin{itemize}
  \item $\mathrm{T}'^c$I between two CP-conjugated vacua, $e.g.$ $u^{\prime c}_1$ and $u^{\prime c}_5$ (with relative phase $0$ or $\pm 2\pi/3$);
  \item $\mathrm{T}'^c$II between two vacua within the same branch of either $\mathrm{T}'^c_+$ or $\mathrm{T}'^c_-$, $e.g.$ $u^{\prime c}_1$ and $u^{\prime c}_2$ (with relative phase $0$) or $u^{\prime c}_3$ (with relative phase $0$) or $u^{\prime c}_4$ (with relative phase $\pi$);
  \item $\mathrm{T}'^c$III between two vacua in two branches that are not CP conjugates of each other, $e.g.$ $u^{\prime c}_1$ and $u^{\prime c}_6$ (with relative phase $\pi/3$) or $ u^{\prime c}_7$ (with relative phase $-\pi/3$) or $u^{\prime c}_8$ (with relative phase $0$).
\end{itemize}
All the possible DWs between all the $\mathrm{T}'^c$-type vacua are shown in \figref{fig:DW_sum}. 
Within each type, the DWs are related by the group transformations.
Here we take the DWs with $u'^c_1$ on one side as concrete examples. 
As the $u'^c_1$ is invariant under the $\mathrm{T}$ transformation up to a global phase $-2 \pi/3$, one can use the $\mathrm{T}$ transformation to turn one DW into another:
\begin{itemize}
  \item For the $\mathrm{T}'^c$I-type DWs, $\mybox{u'^c_1}\mybox{u'^c_5}$ can be turned into $\mybox{\omega u'^c_1}\mybox{u'^c_5}$ by $\mathrm{T}$ transformation, and similarly into $\mybox{\omega^* u'^c_1}\mybox{u'^c_5}$ by $T^2$ transformation.
  \item For the $\mathrm{T}'^c$II-type DWs, $\mybox{u'^c_1}\mybox{u'^c_2}$ can be turned into $\mybox{\omega^* u'^c_1} \mybox{u'^c_3}$ by $\mathrm{T}$ transformation, and into $\mybox{\omega u'^c_1}\mybox{u'^c_4}$ by $T^2$ transformation.
  \item For the $\mathrm{T}'^c$III-type DWs, $\mybox{\hspace{-5pt}-\hspace{-2pt}\omega u'^c_1}\mybox{u'^c_6}$ can be turned to $\mybox{\hspace{-5pt}-\hspace{-2pt}u'^c_1}\mybox{u'^c_7}$ by $\mathrm{T}$ transformation, and similarly to $\mybox{\hspace{-5pt}-\hspace{-2pt}\omega^* u'^c_1}\mybox{u'^c_8}$ by $T^2$ transformation.
\end{itemize}
The property of walls under group transformation not only proves the equal tension of walls in each class, but also explains the relative phases between the two vacua of the walls.
The behaviour of DWs (with $u'^c_1$ on one side) under the $\mathrm{T}$ transformation is summarised in \figref{fig:wall_trans}.
\begin{figure}
  \centering  \includegraphics[width=0.7\linewidth]{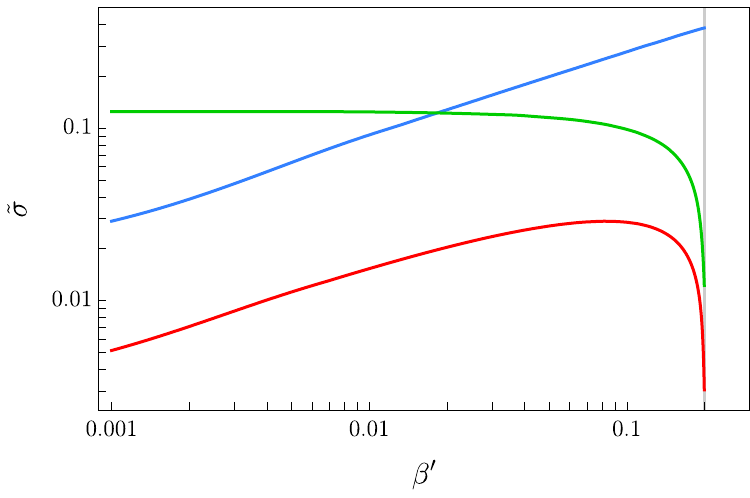}
  \caption{The dimensionless tension of DWs between $\mathrm{T}'^c$-type vacua when $\beta=-0.2$. The blue, green, and red curves correspond to the $\mathrm{T}'^c$I, $\mathrm{T}'^c$II, and $\mathrm{T}'^c$III walls, respectively.}
  \label{fig:Tcp_tension}
\end{figure}
To discuss the properties of each type of DWs, we define the dimensionless tension $\tilde{\sigma}$ as
\begin{eqnarray}
\tilde{\sigma} = \int_{-\infty}^{+\infty} d\overline{z}\, \left\{ \sum_i \frac{d\tilde{\varphi}_i(\overline{z})}{d\overline{z}}\frac{d\tilde{\varphi}_i^*(\overline{z})}{d\overline{z}} 
+ \Delta V\bigl(\tilde{\varphi}(\overline{z}),\tilde{\varphi}^*(\overline{z})\bigr)
\right\} \,,
\end{eqnarray}
which can be related to the DW tension by 
$\sigma = \mu (u'^c)^2 \tilde{\sigma}$.
The values of the dimensionless tension for different types of DWs between are shown in \figref{fig:Tcp_tension} where the blue corresponds to $\mathrm{T}'^c$I, green  to $\mathrm{T}'^c$II and  red to $\mathrm{T}'^c$III.
As an example, we choose $\beta=-0.2$ as a benchmark value and discuss the dependence of the dimensionless tension $\tilde{\sigma}$ on $\beta'$. 
The hierarchy of tensions determines the stability of the domain wall network, which varies significantly with the parameter $\beta'$. In the regime where $\beta'$ is small ($\beta' \ll -\beta$), as shown on the left side of \figref{fig:Tcp_tension}, the tension hierarchy follows $\tilde{\sigma}_{\mathrm{T}'^c\text{II}} > \tilde{\sigma}_{\mathrm{T}'^c\text{I}} > \tilde{\sigma}_{\mathrm{T}'^c\text{III}}$. Crucially, we find that $\tilde{\sigma}_{\mathrm{T}'^c\text{II}} > \tilde{\sigma}_{\mathrm{T}'^c\text{I}} + \tilde{\sigma}_{\mathrm{T}'^c\text{III}}$ in this region, indicating that $\mathrm{T}'^c$II walls are unstable and will decay into a composite configuration of $\mathrm{T}'^c$I and $\mathrm{T}'^c$III walls. Conversely, in the large $\beta'$ regime ($\beta' \lesssim -\beta$), the tension of $\mathrm{T}'^c$I walls (blue) rises while that of $\mathrm{T}'^c$II (green) and $\mathrm{T}'^c$III (red) falls, shifting the hierarchy to $\tilde{\sigma}_{\mathrm{T}'^c\text{I}} > \tilde{\sigma}_{\mathrm{T}'^c\text{II}} > \tilde{\sigma}_{\mathrm{T}'^c\text{III}}$. Eventually, the condition $\tilde{\sigma}_{\mathrm{T}'^c\text{I}} > \tilde{\sigma}_{\mathrm{T}'^c\text{II}} + \tilde{\sigma}_{\mathrm{T}'^c\text{III}}$ is satisfied, rendering the $\mathrm{T}'^c$I walls unstable against decay into $\mathrm{T}'^c$II and $\mathrm{T}'^c$III walls.

\section{Supersymmetric $A_4$ Domain Walls}\label{sec:superfield}
In $\mathcal{N}=1$ supersymmetry (SUSY), the interactions of the theory are primarily determined by the Kähler potential, $K$, and the holomorphic superpotential, $W$. The fundamental objects are chiral superfields, denoted by $\hat{\varphi}$. Each chiral superfield is a function on superspace and can be expanded in the Grassmann coordinate $\theta$ as
$ \hat{\varphi}(x,\theta) \;=\; \varphi(x) + \sqrt{2}\,\theta\,\chi(x) + \theta^2 F(x)$
where $\varphi(x)$ is a complex scalar, $\chi(x)$ is a chiral fermion, and $F(x)$ is a non-dynamical auxiliary field. This structure ensures that SUSY is manifest.
The superpotential, $W(\hat{\varphi})$, is a holomorphic function of the superfields, which constrains the allowed interactions in the theory.
The full SUSY action is constructed from these components:
\begin{equation}
  S = \int d^4x d^4\theta \, K(\hat{\varphi}, \hat{\varphi}^*) + \left( \int d^4x d^2\theta \, W(\hat{\varphi}) + \text{h.c.} \right) \,.
\end{equation}
Our analysis will focus on the consequences of the superpotential $W$.
Within a flavoured SUSY theory, the superpotential must also respect the flavour symmetry, which in our case is $A_4$. 
For an $A_4$-triplet chiral superfield $\hat{\varphi}$, the most general renormalisable superpotential consistent with $A_4$ and holomorphy is
\begin{equation} \label{eq:w_A4}
  W(\hat{\varphi}) \;=\; \frac{1}{2}\,\mu \big(\hat{\varphi}_1^2 + \hat{\varphi}_2^2 + \hat{\varphi}_3^2\big)
  \;-\; {\rm g}\, \hat{\varphi}_1 \hat{\varphi}_2 \hat{\varphi}_3\,,
\end{equation}
where $\mu$ is a parameter with mass dimension one and ${\rm g}$ is a dimensionless coupling. 
By suitable phase redefinitions of the superfields and superspace coordinates, both $\mu$ and ${\rm g}$ can be chosen real and positive.
In general, the scalar potential for the complex scalars $\varphi_n$ is obtained from the superpotential via the $F$-terms,
\begin{equation}
  V(\varphi,\varphi^*) \;=\; \sum_n F_n(\varphi)\,F_n^*(\varphi^*) \;+\; V_{\rm soft} \;+\; \cdots \,,
\end{equation}
where
\begin{equation}
  F_n(\varphi) \;=\; \left.\frac{\partial W(\hat{\varphi})}{\partial \hat{\varphi}_n}\right|_{\theta=0},
  \qquad
  F_n^*(\varphi^*) \;=\; \left.\frac{\partial W^*(\hat{\varphi}^\dagger)}{\partial \hat{\varphi}_n^\dagger}\right|_{\bar{\theta}=0}\,,
\end{equation}
are the $F$-terms and their complex conjugates. 
The index $n$ runs over all scalar components in the theory. 
The term $V_{\rm soft}$ collects the soft SUSY-breaking contributions, while the ellipsis denotes the $D$-term contribution and possible higher-order corrections.

Given the superpotential in Eq.~\eqref{eq:w_A4}, the scalar potential $V(\varphi,\varphi^*)$ can be written explicitly as 
\begin{eqnarray} \label{eq:potential_SUSY}
V(\varphi, \varphi^*) &=& |\mu \varphi_1 - {\rm g} \varphi_2\varphi_3 |^2 + |\mu \varphi_2 - {\rm g} \varphi_3\varphi_1 |^2 + |\mu \varphi_3 - {\rm g} \varphi_1\varphi_2 |^2 \nonumber\\
&+& m_\varphi^2 (|\varphi_1|^2 + |\varphi_2|^2 + |\varphi_3|^2) \,,
\end{eqnarray}
where the quadratic soft mass term is chosen to be invariant under the $A_4$ symmetry. 
Compared with the general complex-scalar case in Eq.~\eqref{eq:potential_c}, the SUSY potential does not contain the cubic invariant
\begin{eqnarray}
\mathcal{I}_{32} = \frac{1}{3}(\varphi_1 \varphi_2 \varphi_3^*+ \varphi_1 \varphi_2^* \varphi_3 + \varphi_1^* \varphi_2 \varphi_3)\,,
\end{eqnarray}
which leads to important differences in the vacuum structure. 

The vacuum-alignment problem in the SUSY framework can be simplified if the soft term is negligibly small. 
In the exact SUSY limit, the minimum of the potential is obtained by imposing
\begin{eqnarray}
  F_n (\varphi) = 0 \,,
\end{eqnarray}
$i.e.$ along the so-called $F$-flat directions. 
The $F$-terms in this case are
\begin{eqnarray} \label{eq:F_0}
  F_1 &=& \mu \varphi_1 - {\rm g} \varphi_2\varphi_3 \,, \nonumber\\
  F_2 &=& \mu \varphi_2 - {\rm g} \varphi_3\varphi_1 \,, \nonumber\\
  F_3 &=& \mu \varphi_3 - {\rm g} \varphi_1\varphi_2 \,.
\end{eqnarray}
The solutions include the trivial vacuum 
\begin{eqnarray}
  \varphi = 0 \,,\label{eq:0-vacuum}
\end{eqnarray}
and the $\mathrm{T}^s$-type vacua
\begin{eqnarray}
  \mathrm{T}^s: w_{1,2,3,4} =
    \left\{
    \begin{pmatrix} 1 \\ 1 \\ 1 \end{pmatrix}, 
    \begin{pmatrix} 1 \\ -1 \\ -1 \end{pmatrix},
    \begin{pmatrix} -1 \\ 1 \\ -1 \end{pmatrix},
    \begin{pmatrix} -1 \\ -1 \\ 1 \end{pmatrix}
    \right\}w\,,
    \label{eq:Ts-vacua}
\end{eqnarray}
where $w = \mu/{\rm g}$. 
Although the flavon is complex in both cases, the SUSY vacuum structure differs significantly from the general non-SUSY complex-scalar case in two respects:
\begin{itemize}
  \item SUSY automatically allows a trivial vacuum that is degenerate with the non-trivial ones.
  \item The non-trivial vacua are discrete: there is no continuous $U(1)$ degeneracy (unlike in the non-SUSY complex-scalar case).
\end{itemize} 
Although the degeneracy between the trivial and non-trivial vacua can be lifted by including SUSY-breaking effects such as soft mass terms, this lifting is not an intrinsic property of supersymmetric domain walls and should instead be regarded as a feature of the SUSY breaking rather than of the SUSY theory itself.In addition to the $\mathrm{T}$-type vacua, the $\mathrm{S}$-type vacua are also required in many flavour models. 
They do not arise directly from Eq.~\eqref{eq:w_A4}, but can be generated if the quadratic term in the superpotential is forbidden (equivalently, if $\mu=0$). 
This can be achieved by imposing an additional $\mathbb{Z}_3$ symmetry. 
In this case, the $F$-term conditions admit solutions of the form 
\begin{eqnarray}
  \mathrm{S}^s: 
    \left\{
    \begin{pmatrix} 1 \\ 0 \\ 0 \end{pmatrix}, 
    \begin{pmatrix} 0 \\ 1 \\ 0 \end{pmatrix},
    \begin{pmatrix} 0 \\ 0 \\ 1 \end{pmatrix}
    \right\} s\,,
    \label{eq:Ss-vacua}
\end{eqnarray}
where $\mathrm{S}$ is a free parameter. 
In the exact SUSY limit, the $\mathrm{S}^s$ vacua correspond to the flat directions: the alignment is fixed, but the absolute value of the non-zero component is undetermined. 
Such a situation arises in many SUSY model-building realisations, $e.g.$, \cite{Altarelli:2005yx, Bazzocchi:2007na, Ding:2013bpa}. 
One solution to this problem is to introduce extra fields mass dimensions in their couplings, with an special example presented later in \secref{sec:SUSYdwA4}.
Alternatively, the minimal extension is by including a soft SUSY-breaking mass term. Then, this flatness is removed and the magnitude $\mathrm{S}$ is fixed by the soft mass term.
With the soft SUSY-breaking term included, the potential becomes
\begin{eqnarray} \label{eq:potential_SUSY_S}
V(\varphi, \varphi^*) &=& |{\rm g}|^2 (|\varphi_2\varphi_3 |^2 + |\varphi_3\varphi_1 |^2 + |\varphi_1\varphi_2 |^2 )
+ m_\varphi^2 (|\varphi_1|^2 + |\varphi_2|^2 + |\varphi_3|^2) \,,
\end{eqnarray}
This is a special case of the non-SUSY potential in Eq.~\eqref{eq:potential_c}. 
In such case, $\mathrm{S}$ is determined to be $s = \sqrt{{-m_\varphi^2}/{|{\rm g}|^2}} e^{i \alpha}$ with $\alpha$ an arbitrary overall phase.

In the rest of this section, we discuss the main properties of non-Abelian SUSY domain walls. 
We distinguish SUSY domain walls from non-SUSY ones by the following feature: the flavon acquires a non-trivial VEV and the non-Abelian discrete symmetry is spontaneously broken at a scale much higher than the soft SUSY-breaking scale. 
We use domain walls in SUSY $A_4$ as a concrete example. 
In \secref{sec:superfield}, we study SUSY domain walls arising from the superpotential of a single $A_4$ triplet flavon superfield, and in \secref{sec:SUSY_R} we extend this setup by including a triplet driving superfield with an $R$-symmetry. 
To provide a simpler illustration, Appendix~\ref{app:Z2SUSY} presents toy models of SUSY $Z_2$ domain walls with and without an $R$-symmetry.

\subsection{Simplest SUSY Domain Walls}

\begin{figure}[t!]
  \centering
  \includegraphics[width=0.45\linewidth]{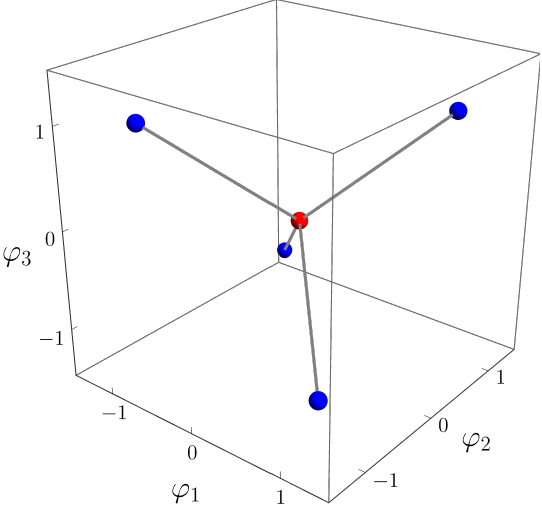}
  \caption{Arrangement of SUSY vacua in the field space of $({\varphi}_1, {\varphi}_2, {\varphi}_3)$ and the corresponding topologies of SUSY DWs. 
  The trivial vacuum is located at the origin ($\varphi_i = 0$), while the four non-trivial $\mathrm{T}$-type vacua sit at the vertices of a regular tetrahedron. 
  This tetrahedral geometry is a direct consequence of the model's $A_4$ flavour symmetry. 
  The plot is shown for a vacuum expectation value $w = \mu/g = 1$, with the overall phase rotated away.
  The relative phases between the vacua are also indicated.}
  \label{fig:SUSYvacua}
\end{figure}

We start from the superpotential in Eq.~\eqref{eq:w_A4} for a triplet flavon superfield. 
The $F$-flat directions yield only the $\mathrm{T}^s$-type vacua in Eq.~\eqref{eq:Ts-vacua} and the trivial vacuum at the origin, as illustrated in \figref{fig:SUSYvacua}.
For the $\mathrm{T}^s$-type vacua, the VEV scale $w$ is much larger than the soft mass parameter $m_\varphi$, and these VEVs are generated purely by SUSY dynamics. 
Following the discussion above, domain walls interpolating between the trivial vacuum and the $\mathrm{T}^s$-type vacua are therefore regarded as SUSY DWs. All vacua on both sides of the wall respect SUSY.
By contrast, domain walls associated with soft SUSY breaking, $e.g.$, the $\mathrm{S}^s$-type vacua derived from soft mass term, should be regarded as non-SUSY domain walls, since the VEV scale $\mathrm{S}$ is fixed in vacua where SUSY is broken. 
In the exact SUSY limit, $i.e.$ at energy scales much higher than the SUSY-breaking scale, the $\mathrm{S}^s$ vacua form a flat direction and do not give rise to topologically stable DWs.

In the following, we focus on walls in the SUSY limit and neglect soft-breaking terms. 
This is a good approximation when the flavour-breaking scale is much higher than the SUSY-breaking scale. 
We will see that the properties of SUSY domain walls differ qualitatively from those in the non-SUSY case.
The scalar potential in the SUSY limit is purely $F$-term:
\begin{equation}
  V(\varphi, \varphi^*) = \sum_n F_n(\varphi)\, F^*_n(\varphi^*) \,.
\end{equation}
The $F$-flatness conditions ($F_n=0$) admit five vacua: the trivial solution $\varphi_i=0$ and the four $\mathrm{T}^s$ vacua, arranged as shown in \figref{fig:SUSYvacua}. 
The trivial vacuum is unavoidable in this setup, since $F_n(\varphi=0)=0$. 
Consequently, the lowest-tension domain walls connect the trivial vacuum to each $\mathrm{T}^s$ vacuum. 
A direct wall between two $\mathrm{T}^s$ vacua is energetically disfavoured: the configuration naturally relaxes through the degenerate vacuum at the origin, effectively splitting into two separate walls.

As an explicit example, we focus on the wall connecting the trivial vacuum to $(1,1,1)^T w$. 
The solutions to the second-order Euler–Lagrange equation,
\begin{equation}
  \frac{d^2\varphi_i}{dz^2} = \frac{\partial V}{\partial \varphi_i^*}\,,
\end{equation}
can be obtained by first solving a simpler first-order equation
\begin{equation}
  \frac{d\varphi_i}{dz} = e^{i\theta} F_i^*(\varphi^*) \,.
  \label{eq:eom_susy}
\end{equation}
Here $\theta$ is a constant phase to be fixed. 
Any solution of this BPS equation automatically solves the full second-order equation, which can be seen by differentiating Eq.~\eqref{eq:eom_susy} with respect to $z$ and apply the chain rule (treating $\varphi_i$ and $\varphi_i^*$ as independent variables):
\begin{align}
  \frac{d^2\varphi_i}{dz^2} &= e^{i\theta} \sum_n \frac{\partial F_i^*}{\partial \varphi_n^*} \frac{d\varphi_n^*}{dz} = e^{i\theta} \sum_n \frac{\partial F_i^*}{\partial \varphi_n^*} \left(e^{-i\theta} F_n\right) \nonumber\\
  &= \sum_n F_n \frac{\partial F_i^*}{\partial \varphi_n^*}
  \;=\; \frac{\partial}{\partial \varphi_i^*} \left( \sum_n |F_n|^2 \right)
  \;=\; \frac{\partial V}{\partial \varphi_i^*} \,.
\end{align}
This reproduces the correct Euler–Lagrange equation.

The wall tension can be found by rewriting the energy integral and completing the square:
\begin{align}
  \sigma &= \int_{-\infty}^{+\infty} dz \left( \left|\frac{d\varphi_i}{dz}\right|^2 + |F_i|^2 \right) \nonumber\\
  &= \int dz \left| \frac{d\varphi_i}{dz} - e^{i\theta} F_i^* \right|^2 
  + \int dz \left( e^{i\theta} F_i^* \frac{d\varphi_i^*}{dz} + e^{-i\theta} F_i \frac{d\varphi_i}{dz} \right) \,.
\end{align}
The first term is manifestly non-negative and vanishes if Eq.~\eqref{eq:eom_susy} is satisfied. 
Using $F_i = \partial W / \partial \varphi_i$, the second term becomes
\begin{align}
  \sigma &= \int_{\varphi(-\infty)}^{\varphi(+\infty)} \left( e^{i\theta} dW^* + e^{-i\theta} dW \right)
  \;=\; e^{i\theta} \Delta W^* + e^{-i\theta} \Delta W
  \;=\; 2\,\text{Re}\!\left(e^{-i\theta} \Delta W\right) \,,
\end{align}
where $\Delta W = W(\varphi_{+\infty}) - W(\varphi_{-\infty})$ and $\theta$ is the phase of $\Delta W$.
Then it yields
\begin{equation}
  \sigma = 2|\Delta W|\,.
\end{equation}
For the superpotential in Eq.~\eqref{eq:w_A4} and the wall connecting $\varphi=0$ to $\varphi=w_1\equiv (1,1,1)^T w$, $i.e.$, $\mybox{0}\mybox{w_1}$, this evaluates to $\sigma = \mu w^2$. The equation of motion \equaref{eq:eom_susy} can be solved analytically, yielding the familiar kink profile:
\begin{equation} \label{eq:DW_solution_SUSY}
  \varphi(z) = \begin{pmatrix} 1 \\ 1 \\ 1 \end{pmatrix} \frac{w}{2}\left(1+\tanh\frac{\mu z}{2}\right)\,.
\end{equation}
While the first-order condition provides a powerful method for obtaining the simplest, lowest-tension walls, it is a sufficient but not necessary condition for solutions of the second-order equation. From \equaref{eq:DW_solution_SUSY}, it is straightforward to construct the scalar profile for any wall $\mybox{0}\mybox{w_i}$ by replacing the vector direction $(1,1,1)^T$ with the direction corresponding to the chosen vacuum $w_i$.

\subsection{Domain Walls with R-symmetry \label{sec:SUSY_R}}

A $U(1)_R$ symmetry is widely used in SUSY model building. 
In SUSY flavour models, driving superfields are standard tools for realising particular flavon VEV alignments and for suppressing unwanted cross couplings between different flavons. 
These superfields carry $R$-charge $2$ and therefore can only appear linearly in the superpotential. 
In this subsection, we discuss how the vacuum structure and domain wall properties are modified once the driving superfields are introduced. 

Given a triplet driving superfield $\hat{\xi}$ and a triplet flavon superfield $\hat{\varphi}$, the $A_4$-invariant superpotential can be written as
\begin{eqnarray}
W(\hat{\varphi},\hat{\xi}) &=& \mu \left( \hat{\varphi}_1 \hat{\xi}_1 +
\hat{\varphi}_2 \hat{\xi}_2 + \hat{\varphi}_3 \hat{\xi}_3 \right)
- {\rm g} \left(\hat{\varphi}_1 \hat{\varphi}_2 \hat{\xi}_3
+ \hat{\varphi}_2 \hat{\varphi}_3 \hat{\xi}_1 
+ \hat{\varphi}_3 \hat{\varphi}_1 \hat{\xi}_2 \right)\,.
\end{eqnarray}
Since the superspace coordinate $\theta$ carries $R$-charge $+1$, the superpotential transforms with $R$-charge $+2$ under $U(1)_R$ rather than being invariant. 
The flavon VEVs are determined by the $F$-flatness conditions with respect to the driving fields $\xi$,
\begin{equation}
  F_{\xi_i} \;=\; \left.\frac{\partial W(\hat{\varphi},\hat{\xi})}{\partial \hat{\xi}_i}\right|_{\theta=0}\!,
\end{equation}
which are explicitly
\begin{eqnarray}
  F_{\xi_1} (\varphi) &=& \mu \varphi_1 - {\rm g} \varphi_2 \varphi_3\,, \nonumber\\
  F_{\xi_2} (\varphi) &=& \mu \varphi_2 - {\rm g} \varphi_3 \varphi_1\,, \nonumber\\
  F_{\xi_3} (\varphi) &=& \mu \varphi_3 - {\rm g} \varphi_1 \varphi_2 \,.
\end{eqnarray}
For comparison, the $F$-terms with respect to the flavon $\varphi$ are
\begin{eqnarray}
  F_{\varphi_1} (\varphi, \xi) &=& \mu \xi_1 - {\rm g} \left(\varphi_2 \xi_3 + \varphi_3 \xi_2\right)\,, \nonumber\\
  F_{\varphi_2} (\varphi, \xi) &=& \mu \xi_2 - {\rm g} \left(\varphi_3 \xi_1 + \varphi_1 \xi_3\right)\,, \nonumber\\
  F_{\varphi_3} (\varphi, \xi) &=& \mu \xi_3 - {\rm g} \left(\varphi_1 \xi_2 + \varphi_2 \xi_1\right)\,.
\end{eqnarray}
Imposing $F_{\xi_i}=0$ and $F_{\varphi_i}=0$ leads to the same flavon VEVs as in Eqs.~\eqref{eq:0-vacuum} (trivial) and Eqs.~\eqref{eq:Ts-vacua} ($\mathrm{T}^s$-type), together with vanishing VEVs for the driving fields, $\langle \xi_i\rangle = 0$. 

The scalar potential is constructed following the same procedure as in the previous subsection. 
In general, it receives contributions from both sets of $F$-terms
\begin{equation}
  V(\varphi,\varphi^*,\xi,\xi^*) \;=\; \sum_i F_{\xi_i} (\varphi)\, F^*_{\xi_i} (\varphi^*) 
  + \sum_i F_{\varphi_i} (\varphi, \xi)\, F^*_{\varphi_i} (\varphi^*, \xi^*) \,.
\end{equation}
The second term is proportional to $\xi \xi^*$ and does not affect the flavon VEV alignment. 
Taking the limit $\xi \to 0$, such contribution disappears and we obtain the effective potential relevant for the flavon alignment
\begin{eqnarray}
  V(\varphi,\varphi^*) \;=\; \sum_i F_{\xi_i} (\varphi)\, F^*_{\xi_i} (\varphi^*)\,.
\end{eqnarray}
This potential has exactly the same polynomial dependence on $\varphi$ and $\varphi^*$ as \equaref{eq:potential_SUSY}. 
As before, domain walls form between the trivial vacuum and each of the $\mathrm{T}^s$-type vacua, but not directly between two $\mathrm{T}^s$-type vacua. 
Solving the second-order equation of motion with the boundary conditions $\varphi(-\infty)=0$ and $\varphi(+\infty)=(1,1,1)^T w$, we can obtain the domain wall solution analytically: it coincides with the profile in \equaref{eq:DW_solution_SUSY}. 
Along this solution, the driving fields remain zero for all $z$.

We now ask whether the first-order equation \equaref{eq:eom_susy} remains valid once the driving fields are included. 
In the present case, the natural generalisation of \equaref{eq:eom_susy} to the full field space is
\begin{eqnarray}
  \frac{d\varphi_i(z)}{dz} &=& F^*_{\varphi_i}(\varphi^*,\xi^*) \,, \nonumber\\
  \frac{d\xi_i(z)}{dz} &=& F^*_{\xi_i}(\varphi^*)\,,
\label{eq:eom_susy_R}
\end{eqnarray}
where we have fixed the overall phase on the right-hand side to zero, since there is no relative phase between the boundary vacua. 
It is straightforward to verify that the domain wall solution in \equaref{eq:DW_solution_SUSY} does not satisfy Eq.~\eqref{eq:eom_susy_R}. 
Instead, it solves the alternative first-order equation
\begin{eqnarray}
  \frac{d\varphi_i(z)}{dz} &=& F^*_{\xi_i}(\varphi^*)\,,
\label{eq:eom_susy_Rp}
\end{eqnarray}
with $\xi_i(z)=0$ along the wall. This example illustrates that the first-order condition \equaref{eq:eom_susy} is a sufficient, but not necessary, condition for obtaining a domain wall solution.

\section{Domain Walls from $A_4$ Lepton Flavour Models}\label{sec:models}
In realistic flavour models, multiple triplet scalar fields are typically present.
The observed mixing between neutrino flavours can naturally emerge from the misalignment of scalar VEVs that preserve different residual symmetries in the charged lepton and neutrino sectors.
A common example is the case where a scalar in the charged lepton sector acquires a 
$\mathrm{T}$-type VEV, while another scalar in the neutrino sector acquires an 
$\mathrm{S}$-type VEV. This framework is widely employed in flavour model building and provides a compelling explanation for the origin of large lepton mixing angles.
Explicit examples of such models will be presented in the following section.

We address how the vacuum alignment of each scalar is affected in detail below.
Consider two real scalars, \(\phi^S\) and \(\phi^T\), with \(\phi^S\) acquiring an \(\mathrm{S}\)-type VEV (\(e.g.\), \(v_1\)) and \(\phi^T\) a \(\mathrm{T}\)-type VEV (\(e.g.\), \(u_1\)). 
As discussed in \secref{sec:realtrip}, such VEVs can be obtained from their respective potentials, \(V(\phi^S)\) and \(V(\phi^T)\), through appropriate quartic couplings. 
However, the full potential 
\begin{eqnarray}
  V(\phi^S, \phi^T) = V(\phi^S) + V(\phi^T) + V_{\rm cross}(\phi^S, \phi^T)\,,
\end{eqnarray}
also contains cross-coupling terms. 
These couplings typically fall into two categories. 
The first type, \(e.g.\), \(((\phi^S_1)^2 + (\phi^S_2)^2 + (\phi^S_3)^2)((\phi^T_1)^2 + (\phi^T_2)^2 + (\phi^T_3)^2)\), does not alter the VEV direction but changes its magnitude. 
The second type, \(e.g.\), \(\phi^S_2\phi^S_3 \phi^T_2 \phi^T_3 + \phi^S_3 \phi^S_1 \phi^T_3 \phi^T_1 + \phi^S_1 \phi^S_2 \phi^T_1 \phi^T_2\), can rotate the VEV direction away from the original \(Z_2\)- or \(Z_3\)-preserving alignment. 
Such terms are critical for VEV alignment: small rotations can yield small corrections to the mixing angles, potentially improving agreement with data, whereas large cross couplings can destroy the correlation between the mixing and residual symmetries. 

Additional symmetries can reduce, but generally cannot eliminate, terms of the second type.  
These must therefore be assumed to be phenomenologically small or suppressed by embedding the model in a different framework, such as SUSY or extra dimensions. 
In what follows, we will work in the decoupling limit, $i.e.$, under the assumption that cross couplings are negligible compared to the individual scalar potentials and neglect their impact on domain wall properties. 

We briefly review DWs arising in realistic lepton flavour model building. 
In most $A_4$ models, the three SM lepton doublets $\ell=(\ell_1, \ell_2, \ell_3)^T$ (with $\ell_i = (\nu_{iL}, l_{iL})$) are arranged as a triplet of $A_4$, and right-handed charged leptons $l_R = (e_R, \mu_R, \tau_R)^T$ are taken as singlets $(\mathbf{1}, \mathbf{1}', \mathbf{1}'')$, respectively. 
The three copies of right-handed neutrinos $\nu_R=(\nu_{1R}, \nu_{2R},\nu_{3R})^T$ are also arranged as a triplet of $A_4$. 
In addition, we introduce triplet vector-like charged leptons $E = (E_1, E_2, E_3)^T$ (with $E_i = E_{iL}+E_{iR}$) for the UV-completion of the model. 

To achieve the flavour misalignment between neutrinos and charged leptons, often two triplet flavons are introduced: one in the charged lepton sector and the other in the neutrino sector. 
In the minimal case, these flavons are assumed to be real and denoted as $\phi^T = (\phi^T_1, \phi^T_2, \phi^T_3)$ and $\phi^S = (\phi^S_1, \phi^S_2, \phi^S_3)$. 
The labels $\mathrm{T}$ and $\mathrm{S}$ indicate that these flavons will acquire the $\mathrm{T}$- and $\mathrm{S}$-type VEVs, respectively. 
However, if the flavon is covariant along with additional symmetries, $e.g.$, $\mathbb{Z}_N$ ($N \geqslant 3$) or $U(1)$, the internal field space requires the flavon to be complex. 
In this case, we denote the corresponding flavons as $\varphi^T = (\varphi^T_1, \varphi^T_2, \varphi^T_3)$ and $\varphi^S = (\varphi^S_1, \varphi^S_2, \varphi^S_3)$. 
Furthermore, the flavons must be complex in the SUSY framework. 
In this section, we will briefly review three types of lepton flavour models:
\begin{itemize}
\item[1)] Real flavon $A_4$ models. 
All scalars are assumed to be real, which introduces the minimal physical degrees of freedom of flavons.
This approach has advantages from the phenomenological point of view. 

\item[2)] Complex flavon $A_4$ models. 
Enforcing additional Abelian asymmetries helps to forbid some unnecessary couplings in the Yukawa sector, but requires flavons to be complex. 
Here, in particular, we will focus on $A_4\times U(1)$. 
Discussion of this approach can be generalised to $A_4 \times \mathbb{Z}_N$ (for $N>2$), which will not be repeated. 
However, it is worth mentioning that spontaneous breaking of these extra Abelian symmetries might lead to additional topological defects, which requires a case-by-case study. 

\item[3)] SUSY flavon $A_4$ models. 
A large set of flavour models have been constructed in the SUSY framework, providing the advantage of VEV alignment and UV completion in the high-scale SUSY.

\end{itemize}
 In the following subsections we present representative models in each class. Our aim is not to construct fully realistic models that fit all current data, but to outline the model structures and their correlation with non-Abelian domain walls in each framework.

\subsection{Real Flavon $A_4$ Models}
We consider two real flavons, which implies that $(\phi^S_i)^* = \phi^S_i$ and $(\phi^T_i)^* = \phi^T_i$ in the Ma-Rajasekaran (MR) basis \cite{Ma:2001dn}.\footnote{This condition depends upon the representation basis. For example, in the Altarelli-Feruglio (AF) basis \cite{Altarelli:2005yx}, which is widely used in flavour model building, the condition becomes $(\phi^S_1)^* = \phi^S_1$ and $(\phi^S_2)^* = \phi^S_3$ \cite{Pascoli:2016eld}.}
In addition, we assume the $A_4$ symmetry is accompanied by a $\mathbb{Z}_4$ symmetry to forbid unwanted couplings. 
The field assignments in such an $A_4 \times \mathbb{Z}_4$ model are given in \tabref{tab:model_1}. 
Based on the field content, the most general renormalisable $A_4$-invariant Yukawa terms for leptons can be written as
\begin{eqnarray}
-\mathcal{L}_{l,\nu} &=& 
y_\nu \bar{\ell}_i \tilde{H} \nu_{iR} + 
\frac{y_1}{2} \bar{\nu}_{iR}^c \nu_{jR} \, \phi^S_k + \frac{y_2}{2} \bar{\nu}_{iR}^c \nu_{iR} \, \eta \nonumber\\
&+& 
y_E \bar{\ell}_i H E_i + M_E \bar{E}_i E_i + \phi^T_i \bar{E}_i ( y_e e_R + 
 \omega^{1-i} y_\mu \mu_R + 
 \omega^{i-1} y_\tau \tau_R) +  
\text{h.c.} \,, 
\label{eq:Yukawa_1}
\end{eqnarray}
where $i,j,k$ are generational indices, running from $1$ to $3$ with $i \neq j \neq k \neq i$, and $\omega = e^{i 2\pi/3}$. $y_{\nu,E,1,2,e,\mu,\tau}$ are dimensionless Yukawa coefficients and $M_E$ is the mass parameter for the vector-like charged lepton $E$. 
In addition, we introduce a real singlet scalar $\eta$, which follows the same transformation property as $\phi^S$ under $\mathbb{Z}_4$ needed to generate the mass splitting among the neutrinos. 
The flavon $\phi^S$ is covariant under the subgroup $\mathbb{Z}_2 \subset \mathbb{Z}_4$, which ensures the flavon is real. 
Allowing complex components would require an enlarged field space and additional phase‐alignment conditions. 
For further simplicity, we assume an unbroken CP symmetry, which forces all Yukawa couplings to be real.

\begin{table}[t]
\begin{center}
\begin{tabular}{c| c c c c| c c c c}
\hline\hline
Fields & $\ell$ & $(e_R,\mu_R,\tau_R)$ & $\nu_R$ & $E$ & $H$ & $\phi^T$
& $\phi^S$ & $\eta$  \\ \hline

$A_4$ & $\mathbf{3}$ & $(\;\mathbf{1},\;\mathbf{1}',\;\mathbf{1}''\;)$ & $\mathbf{3}$ & $\mathbf{3}$ & $\mathbf{1}$ & $\mathbf{3}$ & $\mathbf{3}$ & $\mathbf{1}$ \\

$\mathbb{Z}_4$ & $i$ & $i$ & $i$ & $i$ & $1$ & $1$ & $-1$ & $-1$ \\

\hline
\hline
\end{tabular}
\caption{\label{tab:model_1} Field assignment of Model I in $A_4 \times \mathbb{Z}_4$.}
\end{center}
\end{table}

The most general potential of flavons is given by 
\begin{eqnarray}
V = V(\phi^T) + V(\phi^S) + V(\phi^T, \phi^S) + \cdots\,,
\label{eq:potentialA4real}
\end{eqnarray}
where $V(\phi^T)$ and $V(\phi^S)$ include only self-couplings of $\phi^T$ and $\phi^S$, respectively.
$V(\phi^T, \phi^S)$ denotes their cross-couplings, and the dots denote other terms, $e.g.$, couplings with $\eta$, which are not crucial for the alignment of vacuum directions in the flavour space. 
$V(\phi^T)$ takes the same form as the general $A_4$-invariant potential in \equaref{eq:potential}; $V(\phi^S)$ has a simple form with the cubic terms forbidden by the parity symmetry induced by $\mathbb{Z}_4$. 
Ignoring the cross-couplings, $\phi^T$ and $\phi^S$ acquire their $Z_3$ and $Z_2$-invariant VEVs separately, depending on the sign of the corresponding coefficient $g_2$. 

Flavons spontaneously gain VEVs and break $A_4 \times \mathbb{Z}_4$. As outlined in the \secref{sec:non-susy-dws}, $\phi^S$ and $\phi^T$ can take $\mathrm{S}$- and $\mathrm{T}$-type
VEVs, respectively, if signs of the corresponding coefficients in the potential are arranged appropriately. 
Here, without loss of generality, we choose $\langle \phi^S \rangle$ and $\langle \phi^T \rangle$ as
\begin{eqnarray} \label{eq:model_1_vev}
\langle \phi^S \rangle = \begin{pmatrix} 1 \\ 0 \\ 0 \end{pmatrix} v_S\,, \quad
\langle \phi^T \rangle = \begin{pmatrix} 1 \\ 1 \\ 1 \end{pmatrix} v_T \,,
\end{eqnarray} 
where $v_S$ and $v_T$ take similar form as $v$ and $u$ in \equaref{eq:uminus} for given $A_4$-invariant potentials, respectively. Taken together, these VEVs totally break the $A_4$ symmetry. However, the individual flavons $\phi^S$ and $\phi^T$ take $Z_2$- and $Z_3$-preserving VEVs separately.
The $\phi^S$ VEV leads to the right-handed Majorana neutrino mass matrix of the form
\begin{eqnarray} \label{eq:mass_R} 
M_R = \begin{pmatrix} 
y_2 v_\eta & 0 & 0 \\
0 & y_2 v_\eta & y_1 v_S \\
0 & y_1 v_S & y_2 v_\eta
\end{pmatrix}\,. 
\end{eqnarray} 
Due to the $Z_2$-preserving VEV of $\phi^S$, the mass matrix $M_R$ is also invariant under the transformation of $\mathrm{S}$: $S M_R S^T = M_R$. 
In the charged lepton sector, the Yukawa coupling $Y_l \bar{\ell} H l_R$ is generated after the heavy lepton $E$ decouples and $\varphi$ gains the VEV, where the $3\times 3$ Yukawa coupling matrix is given by 
\begin{eqnarray} \label{eq:Yukawa_l} 
Y_l = \frac{y_E v_T}{M_E}\begin{pmatrix} 
y_e & y_\mu & y_\tau \\
y_e & \omega^2 y_\mu & \omega y_\tau \\
y_e & \omega y_\mu & \omega^2 y_\tau
\end{pmatrix}\,,
\end{eqnarray}
with $\omega = e^{i 2\pi/3}$. After the SM Higgs gains its VEV, $\langle H \rangle = v_H = 174$~GeV, the mass matrix for charged leptons $M_l = Y_l v_H$ and 
the Dirac mass matrix for neutrinos,
\be
M_D = y_{\nu} \begin{pmatrix}
  1 & 0 & 0 \\ 0 & 1 & 0 \\ 0 & 0 & 1
\end{pmatrix} v_H
\ee
is generated. 
The Majorana mass matrix for light neutrinos, via the seesaw formula $M_\nu = M_D M_R^{-1} M_D^T$, can be explicitly written as
\begin{eqnarray}
  M_\nu = \begin{pmatrix}
    b-\frac{a^2}{b} & 0 & 0 \\ 0 & b & a \\ 0 & a & b 
  \end{pmatrix}\,,
\end{eqnarray}
where 
\begin{equation}
a = \frac{y_1 v_S y_\nu^2 v_H^2 }{(y_1 v_S)^2- (y_2 v_\eta)^2}\,,\quad b = \frac{y_2 v_\eta y_\nu^2 v_H^2}{(y_2 v_\eta)^2 - (y_1 v_S)^2}\,. 
\end{equation}
The charged lepton mass matrix is diagonalised via $U_l^\dag M_l M_l^\dag U_l = {\rm diag} \{ m_e, m_\mu, m_\tau \}$, and the light neutrino mass matrix via $U_\nu^\dag M_\nu U_\nu^* = {\rm diag} \{ m_1, m_2, m_3 \}$, where $U_l$ and $U_\nu$ are unitary matrices. 
Their product $U \equiv U_l^\dag U_\nu$ enters in the Standard Model charged current Lagrangian and is the leptonic mixing matrix, whose values can be measured by neutrino oscillation experiments.
With the right-handed neutrino mass matrix given in \equaref{eq:mass_R} and the charged lepton Yukawa terms given in \equaref{eq:Yukawa_l}, we find
\begin{eqnarray}
U_l = U_\omega \equiv \frac{1}{\sqrt{3}}
\begin{pmatrix}
  1 & 1 & 1 \\
 1 & \omega ^2 & \omega \\
 1 & \omega & \omega ^2
\end{pmatrix} \,, \quad
U_\nu = \begin{pmatrix}
  0 & 1 & 0 \\ 
  \frac{1}{\sqrt{2}} & 0 & \frac{-1}{\sqrt{2}} \\
  \frac{1}{\sqrt{2}} & 0 &\frac{1}{\sqrt{2}}
\end{pmatrix} \,,
\end{eqnarray}
up to some undetermined phases and $m_1 = |b-a^2/b|$, $m_2 = |b-a|$ and $m_3 = |b+ a|$. We further obtain tribimaximal (TBM) mixing, with absolute values of each entry given by
\begin{eqnarray} \label{eq:TBM}
|U| = \left(
\begin{array}{ccc}
 \frac{2}{\sqrt{6}} & \frac{1}{\sqrt{3}} & 0 \\
 \frac{1}{\sqrt{6}}& \frac{1}{\sqrt{3}} & \frac{1}{\sqrt{2}} \\
 \frac{1}{\sqrt{6}}& \frac{1}{\sqrt{3}} & \frac{1}{\sqrt{2}} \\
\end{array}
\right) \,.
\end{eqnarray} 
The TBM mixing is well-known for its prediction of mixing angles $\theta_{12}\simeq 35.3^\circ$ and $\theta_{23}=45^\circ$~\cite{Harrison:2002er,Harrison:2002kp,Xing:2002sw}, which are still consistent with neutrino oscillation data within $3\sigma$ deviation.
The prediction of TBM in $A_4$ models is not accidental but largely fixed by the symmetry. 
It is a consequence of a residual $Z_3$ symmetry preserved in the charged lepton sector, and a $Z_2\times \mathbb{Z}_2^{\mu\tau}$ in the neutrino sector \cite{Lam:2008rs}. 
Here, $\mathbb{Z}_2^{\mu\tau}$ refers to the $\nu_\mu$-$\nu_\tau$ permutation symmetry. 
It is not a sub-symmetry of $A_4$, but is accidentally preserved in the most general Lagrangian compatible with $A_4$ and the specified representation content for the flavons \cite{Altarelli:2010gt}. However, as the vanishing $\theta_{13}$ predicted by TBM has been ruled out by neutrino oscillation data, corrections have to be introduced.

Before discussing the corrections to the TBM, we will comment on the Altarelli-Feruglio (AF) basis, a three-dimensional irreducible basis widely used in flavour model building. Generations of $\mathrm{S}$ and $\mathrm{T}$ in the AF basis obtained through a basis transformation of \equaref{eq:generator1}: 
\begin{eqnarray}
  S_{\rm AF} = U_\omega S U_\omega^\dag = 
  \frac{1}{3} \left(
\begin{array}{ccc}
 -1 & 2 & 2 \\
 2 & -1 & 2 \\
 2 & 2 & -1 \\
\end{array}
\right)\,,\quad
  T_{\rm AF} = U_\omega T U_\omega^\dag = 
  \left(
\begin{array}{ccc}
 1 & 0 & 0 \\
 0 & \omega ^2 & 0 \\
 0 & 0 & \omega \\
\end{array}
\right)\,.
\end{eqnarray} 
As a comparison, the representation basis presented in \equaref{eq:generator1} is regarded as the MR basis. 
In the AF basis, vacua in \equaref{eq:model_1_vev} are transformed to
\begin{eqnarray} 
\langle \phi^S \rangle_{\rm AF} = \begin{pmatrix} 1 \\ 1 \\ 1 \end{pmatrix} \frac{v_S}{\sqrt{3}}\,, \quad
\langle \phi^T \rangle_{\rm AF} = \begin{pmatrix} 1 \\ 0 \\ 0 \end{pmatrix} \sqrt{3} v_T \,.
\end{eqnarray} 
The AF basis is widely applied in the literature since the leading-order charged lepton Yukawa matrix invariant under $T_{\rm AF}$ is diagonal in this basis, 
\begin{eqnarray}
  M_{l,{\rm AF}} = U_\omega^\dag M_{l} = \frac{y_E v_T}{M_E}\begin{pmatrix} 
y_e & 0 & 0 \\
0 & y_\mu & 0 \\
0 & 0 & y_\tau
\end{pmatrix} v_H\,.
\end{eqnarray}
The Dirac neutrino mass matrix in the AF basis is given by $M_{D,{\rm AF}} = U_\omega^\dag M_D U_\omega = M_D$.
The Majorana mass matrix for heavy neutrinos is given by
\begin{eqnarray}
&&M_{R,{\rm AF}} = 
\left(
\begin{array}{ccc} 
y_2v_\eta+ \frac{2}{3}y_1v_S & - \frac{1}{3} y_1v_S & - \frac{1}{3} y_1v_S \\
 - \frac{1}{3} y_1v_S & \frac{2}{3}y_1v_S & y_2v_\eta- \frac{1}{3} y_1v_S \\
 - \frac{1}{3} y_1v_S & y_2v_\eta- \frac{1}{3} y_1v_S & \frac{2}{3}y_1v_S \\
\end{array}
\right)\,.
\label{eq:neutrino_mass}
\end{eqnarray}
The light neutrino mass matrix is straightforwardly written out accordingly after applying the seesaw formula, which will not be repeated again. 
In the AF basis, the mixing matrix is determined via the diagonalisation of $M_\nu$ at leading order. 
Compared with the AF basis, the MR basis treats the three components symmetrically under permutations, which is more convenient for presenting the domain wall field configurations.
Thus, we will work in the MR basis in the following study.


The economical model presented above includes the basic feature of most $A_4$ models, namely the realisation of the TBM mixing at leading order. 
Although the vanishing $\theta_{13}$ does not fit the current neutrino data any more, several different types of effects have been considered in the literature to pull the model consistent with data. Very recently, JUNO released their first measurement of $\theta_{12}$ and $\Delta m^2_{21}$ with best-fit values consistent with global data of past experiments and $1\sigma$ errors less than $1.6\%$ and $2.8\%$ \cite{JUNO:2025gmd}. The increasingly precise data for oscillation parameters should be taken into account in realistic model building. A common way to include deviations from the leading-order mixing pattern, in this case TBM, is to include cross-couplings between different flavours. 
 These cross-couplings explicitly break the residual Abelian symmetries and can be included in two ways, via the lepton Yukawa couplings or flavon potential. 
 We discuss each scenario below:
\begin{itemize}
\item Cross couplings between different flavours can contribute to the lepton Yukawa/mass matrices directly, modifying flavour textures and further generating deviation of mixing from its leading-order result. 
As an example, we consider the non-renormalisable operators, $\frac{1}{\Lambda}\bar{\nu}_{R}^c \nu_{R} (\phi^S \phi^T)_{3_{S}}$ and $\frac{1}{\Lambda}\bar{\nu}_{R}^c \nu_{R} (\phi^S \phi^T)_{3_{A}}$ where $(\phi^S \phi^T)_{3_{S}}$ and $(\phi^S \phi^T)_{3_{A}}$ are symmetric and antisymmetric triplet decompositions of the product of $\phi^T$ and $\phi^S$ (see the appendix). 
These two fields acquire the following vacuum alignments, $(0,1,1)^T$ and $(0,1,-1)^T$, respectively. 
They induce the following corrections to the right-handed neutrino mass matrix:
\begin{eqnarray}
  \epsilon_1 v_S \begin{pmatrix}
    0 & 1 & 1 \\ 1 & 0 & 0 \\ 1 & 0 & 0
  \end{pmatrix} + 
  \epsilon_2 v_S \begin{pmatrix}
    0 & -1 & 1 \\ -1 & 0 & 0 \\ 1 & 0 & 0
  \end{pmatrix}\,,
\end{eqnarray}
where $\epsilon_1$ and $\epsilon_2$ are small dimensionless parameters of order $v_\phi^T/ \Lambda$.
Since the VEV of $\phi^T$ only preserves a $Z_3$ symmetry, the triplet contraction $(\phi^S \phi^T)_{3_{S,A}}$ breaks the $Z_2$ residual symmetry in the neutrino sector, leading to a deviation from \equaref{eq:mass_R} in the mass matrix $M_R$.
As a result, the property $S M_R S^T = M_R$ does not hold any more and the $Z_2$ symmetry in the neutrino sector is explicitly broken.

\item Cross-couplings can also be included in the flavon potential via either renormalisable terms (if it is not forbidden by symmetry) or higher dimensional operators. 
A general flavon potential should include both self-couplings for each flavon, $i.e.$, $V(\phi^S)$ and $V(\phi^T)$ and their cross couplings, which we denote as $V(\phi^S, \phi^T)$. 
These terms contribute to the vacuum alignment via the minimisation of the potential. 
As a consequence, the VEVs of $\phi^S$ and $\phi^T$ are shifted from their original $Z_2$- and $Z_3$-preserving ones. 
Both flavons $\phi^T$ and $\phi^S$ modify the directions of $\mathrm{S}$-type and $\mathrm{T}$-type vacua, leading to a non-vanishing $\theta_{13}$ proportional to the size of cross couplings \cite{Pascoli:2016eld}. 
\end{itemize}
We comment on domain walls generated in this model. 
In the neutrino sector, the S-type DWs are generated as $\phi^S$ takes $Z_2$-preserving VEVs, with a tension of order $\mu_S v_S^2$. 
In the charged lepton sector, the T-type DWs are generated as $\phi^T$ takes $Z_3$-preserving VEVs, with tension of order $\mu_T v_T^2$. 
Since the $\mathrm{S}$-type and $\mathrm{T}$-type walls are associated with independent flavons, they correspond to distinct symmetry-breaking patterns, $A_4 \to Z_2$ and $A_4 \to Z_3$, respectively. 
In the absence of sizeable cross couplings in the scalar potential, the two wall networks coexist and evolve essentially independently in the Universe, each approaching its own scaling regime and contributing additively to the total domain wall energy density. 
If cross couplings between $\phi^S$ and $\phi^T$ or additional explicit breaking terms are present, interactions between the two networks and biases among the vacua can arise, potentially leading to the partial annihilation of one network against the other. In the present discussion we focus on the decoupling limit, where the dynamics of the $\mathrm{S}$-type and $\mathrm{T}$-type walls can be treated separately.

\subsection{Complex flavon $A_4$ models}\label{sec:comDWA4}
Complex flavons arise naturally in flavour model building when the scalar sector is charged under additional Abelian symmetries beyond $A_4$. 
Such extensions are well motivated: they can forbid unwanted operators in the Yukawa sector, provide mechanisms for spontaneous CP violation and offer additional handles for controlling the vacuum alignment. 
Here we focus on a global $U(1)_L$ lepton number symmetry, though the discussion generalises straightforwardly to discrete $\mathbb{Z}_N$ subgroups (with $N > 2$) or a gauged $U(1)_{B-L}$.
Different choices for the additional symmetry lead to qualitatively different topological defects upon spontaneous breaking: a discrete $\mathbb{Z}_N$ produces Abelian domain walls, a global $U(1)_L$ gives rise to global cosmic strings and a gauged $U(1)_{B-L}$ yields local  cosmic strings.

We assign $U(1)_L$ charges as follows: all lepton fields carry charge $+1$, while the flavons $\varphi^S$ and $\eta$ responsible for generating the right-handed neutrino Majorana mass must carry charge $-2$ to form $U(1)_L$-invariant bilinears $\bar{\nu}_R^c \nu_R \varphi^S$.
This charge assignment forces $\varphi^S$ and $\eta$ to be complex.
For simplicity, we retain $\phi^T$ as a real field (uncharged under $U(1)_L$).
The complete field assignments for this Model II are summarised in \tabref{tab:model_2}.

The Yukawa Lagrangian takes the same form as in \equaref{eq:Yukawa_1} apart from $\phi^S$ and $\eta$ being complex.
Provided $\phi^T$ and $\varphi^S$ acquire the same VEV directions as in \equaref{eq:model_1_vev}, we find that
\begin{eqnarray}
\langle \varphi^S \rangle = \begin{pmatrix} 1 \\ 0 \\ 0 \end{pmatrix} v_S e^{i\alpha_S}\,, \quad
\langle \phi^T \rangle = \begin{pmatrix} 1 \\ 1 \\ 1 \end{pmatrix} v_T \,,
\end{eqnarray}
where $\alpha_S$ is an overall $U(1)_L$ phase. This model reproduces the TBM mixing pattern at leading order. 
The phase $\alpha_S$ is unphysical at the level of the mass matrices and can be absorbed by a field redefinition. However, it plays an important role in the vacuum manifold and the formation of topological defects.
The scalar potential for the complex flavon $\varphi^S$ takes the form discussed in \secref{sec:complextrip}. 
As a result, the vacuum manifold for $\varphi^S$ contains a continuous $S^1$ degeneracy parametrised by $\alpha_S$, in addition to the discrete $A_4$ degeneracy among the three $\mathrm{S}$-type directions.
Spontaneous breaking of this $U(1)_L$ produces a network of global cosmic strings, with string tension $\mu_{\rm str} \sim v_S^2$.
The interplay between the discrete $A_4$ domain walls and the continuous $U(1)_L$ strings leads to a composite defect network which we discussed in \secref{sec:complextrip}.

The global $U(1)_L$ symmetry can be generalised to other Abelian symmetry groups such as $\mathbb{Z}_N$ ($N \geq 2$) and a gauged $U(1)_{B-L}$. In the former, 
replacing $U(1)_L$ with a discrete $\mathbb{Z}_N$ subgroup restricts the allowed operators. After the spontaneous breaking of $\mathbb{Z}_N$, Abelian domain walls are produced.
These walls, together with the non-Abelian $A_4$ walls, form a multi-component network whose dynamics depends on the relative tensions and the structure of explicit breaking terms. Promoting $U(1)_L$ to a gauged $U(1)_{B-L}$ eliminates global strings in favour of local strings.
The gauge symmetry is broken by the flavon VEV, producing local strings with tension $\mu_{\rm str} \sim v_S^2$ that are cosmologically long-lived.
Domain walls in this scenario can end on local strings, and the string-wall network evolves differently from the global case due to the absence of long-range Goldstone interactions.

In summary, complex flavon $A_4$ models naturally accommodate additional Abelian symmetries that enrich both the flavour structure and the topological defect spectrum. The interplay between non-Abelian domain walls from $A_4$ breaking and strings or walls from the Abelian sector provides new handles for model building and distinctive gravitational wave signatures that depend on the symmetry breaking pattern.
\begin{table}[t]
\begin{center}
\begin{tabular}{c| c c c c| c c c c}
\hline\hline
Fields & $\ell$ & $(e_R,\mu_R,\tau_R)$ & $\nu_R$ & $E$ & $H$ & $\phi^T$
& $\varphi^S$ & $\eta$  \\ \hline

$A_4$ & $\mathbf{3}$ & $(\;\mathbf{1},\;\mathbf{1}',\;\mathbf{1}''\;)$ & $\mathbf{3}$ & $\mathbf{3}$ & $\mathbf{1}$ & $\mathbf{3}$ & $\mathbf{3}$ & $\mathbf{1}$ \\

$U(1)_L$ & $1$ & $1$ & $1$ & $1$ & $0$ & $0$ & $-2$ & $-2$ \\

\hline
\hline
\end{tabular}
\caption{\label{tab:model_2} Field assignments of Model II in $A_4 \times U(1)_L$.}
\end{center}
\end{table}
\\

\subsection{SUSY flavon $A_4$ models}\label{sec:SUSYdwA4}
In the SUSY framework, all leptons and flavons are promoted to chiral superfields. 
For the superfields containing left-handed fermions and their scalar partners, we follow the same notation as in the non-SUSY case, but add a hat to denote a chiral superfield. 
For right-handed superfields, we use a superscript \(^{c}\), e.g. \((\hat{e}^c, \hat{\mu}^c, \hat{\tau}^c)\) for \((e_R, \mu_R, \tau_R)\). 
The superfield assignments in the SUSY \(A_4\) model are summarised in \tabref{tab:model_3}.

There is a continuous \(U(1)_R\) symmetry which contains the usual \(R\)-parity as a subgroup. 
The \(R\)-charge splits the superfields into three sectors: matter superfields with \(R\)-charge~1, Higgs and flavon superfields with \(R\)-charge~0 and the so-called driving superfields, to be introduced below, with \(R\)-charge~2. 
An additional \(\mathbb{Z}_3\) symmetry is imposed to forbid unwanted couplings among superfields. 
This setup follows the original Altarelli–Feruglio model of \cite{Altarelli:2005yx}, but we consider its UV completion by including a heavy lepton superfield pair \((\hat{E},\hat{E}^c)\) and the right-handed neutrino superfield \(\hat{\nu}^c\). 
The most general renormalisable superpotential for the Yukawa couplings is then given by
\begin{eqnarray}
W_{l,\nu} &=& 
y_\nu^* \hat{\ell}_i \hat{h}_u \hat{\nu}_i^c + 
\frac{y_1^*}{2} \hat{\nu}_i^c \hat{\nu}_j^c \hat{\varphi}^S_k 
+ \frac{y_2^*}{2} \hat{\nu}_i^c \hat{\nu}_i^c \hat{\eta} \nonumber\\
&+& 
y_E^* \hat{\ell}_i \hat{h}_d \hat{E}_i^c 
+ M_E^* \hat{E}_i \hat{E}_i^c 
+ \hat{\varphi}^T_i \hat{E}_i 
\left( y_e^* \hat{e}^c 
+ \omega^{1-i} y_\mu^* \hat{\mu}^c 
+ \omega^{i-1} y_\tau^* \hat{\tau}^c \right) \,,
\label{eq:Yukawa_3}
\end{eqnarray}
where a complex conjugate sign is included for all coefficients such that same mass matrices as in the non-SUSY case are reproduced.
The SUSY approach of \cite{Altarelli:2005yx} provides a systematic and technically natural way to separate the vacuum alignments of the flavons in the charged-lepton and neutrino sectors. 
The key idea is to introduce driving superfields, denoted by \(\hat{\xi}^{T}\), \(\hat{\xi}^{S}\) and \(\hat{\xi}_0\) in \tabref{tab:model_3}. 
These driving superfields carry \(R\)-charge~2 and therefore enter the superpotential only linearly. 
All renormalisable terms in the superpotential that involve the driving superfields and flavons are
\begin{eqnarray}
W_{\rm d} &=& \hat{\xi}^{T}_i \left(\mu \hat{\varphi}^T_i + {\rm g} \hat{\varphi}^T_j \hat{\varphi}^T_k \right) \nonumber\\
&+& {\rm g}_1 \hat{\xi}^{S}_i \hat{\varphi}^S_j \hat{\varphi}^S_k 
+ \xi_0 \left({\rm g}_2 \hat{\varphi}^S_i \hat{\varphi}^S_i + {\rm g}_3 \hat{\eta}^2 \right) 
+ \xi'_0 \left({\rm g}_4 \frac{\hat{\eta}^3}{\Lambda_1} + \Lambda_2^2 \right)\,.
\end{eqnarray}
There are no couplings of the driving superfields to the Higgs superfields, which is consistent with the assumption that the flavour symmetry breaking scale is naturally much higher than the electroweak scale. 

As reviewed in \secref{sec:superfield}, the effective scalar potential for the flavons is obtained from the \(F\)-terms, to which we add \(A_4\)-invariant soft SUSY-breaking mass terms. 
Writing \(V = V_l + V_\nu\), we have
\begin{eqnarray} \label{eq:potentialSUSY}
V_l &=& \left|\mu \varphi^T_1 + {\rm g} \varphi^T_2 \varphi^T_3 \right|^2 
   + \left|\mu \varphi^T_2 + {\rm g} \varphi^T_3 \varphi^T_1 \right|^2 
   + \left|\mu \varphi^T_3 + {\rm g} \varphi^T_1 \varphi^T_2 \right|^2 \nonumber\\
&+& m_T^2 \left(|\varphi^T_1|^2 + |\varphi^T_2|^2 + |\varphi^T_3|^2 \right) \,,\nonumber\\[1ex]
V_\nu &=& |{\rm g}_1|^2 \left(|\varphi^S_2 \varphi^S_3|^2 + |\varphi^S_3 \varphi^S_1|^2 + |\varphi^S_1 \varphi^S_2|^2 \right) \nonumber\\
&+& \left|{\rm g}_2 \left[(\varphi^S_1)^2 + (\varphi^S_2)^2 + (\varphi^S_3)^2\right] + {\rm g}_3 \eta^2 \right|^2 
+ \left|{\rm g}_4 \frac{\eta^3}{\Lambda_1} + \Lambda_2^2 \right|^2 \nonumber\\
&+& m_S^2 \left(|\varphi^S_1|^2 + |\varphi^S_2|^2 + |\varphi^S_3|^2 \right) \,,
\end{eqnarray}
where \(m_T^2\) and \(m_S^2\) are soft mass parameters. 
Cross couplings between \(\varphi^T\) and \(\varphi^S\) are forbidden at the renormalisable level by their charge assignments under \(\mathbb{Z}_3 \times U(1)_R\). 

A straightforward derivation shows that \(V_l\) has almost the same structure as \equaref{eq:potential_c}, up to one additional cubic invariant,
\begin{eqnarray}
\mathcal{I}_{32} = \frac{1}{3}\left(\varphi_1 \varphi_2 \varphi_3^*
+ \varphi_1 \varphi_2^* \varphi_3 
+ \varphi_1^* \varphi_2 \varphi_3 \right)\,,
\end{eqnarray}
and its complex conjugate. 
By treating \(\eta\) as a background number (irrespective of whether \(\eta\) vanishes or not), \(V_\nu\) takes exactly the same form as \equaref{eq:potential_c}. 

\begin{table}[t]
\begin{center}
\begin{tabular}{c| c c c c | c c c c | c c c c}
\hline\hline
Superfields & $\hat{\ell}$ & $(\hat{e}^c,\hat{\mu}^c,\hat{\tau}^c)$ & $\hat{\nu}^c$ & $(\hat{E},\hat{E}^c)$ & 
$\hat{h}_{u,d}$ & $\hat{\varphi}^T$ & $\hat{\varphi}^S$ & $\hat{\eta}$ & 
$\hat{\xi}^{T}$ & $\hat{\xi}^{S}$ & $\hat{\xi}_0$ & $\hat{\xi}'_0$ \\ \hline
$A_4$ & $\mathbf{3}$ & $(\;\mathbf{1},\;\mathbf{1}',\;\mathbf{1}''\;)$ & $\mathbf{3}$ & $\mathbf{3}$ & 
$\mathbf{1}$ & $\mathbf{3}$ & $\mathbf{3}$ & $\mathbf{1}$ & 
$\mathbf{3}$ & $\mathbf{3}$ & $\mathbf{1}$ & $\mathbf{1}$ \\
$\mathbb{Z}_3$ & $\omega$ & $\omega^2$ & $\omega^2$ & $(\omega, \omega^2)$ & 
$1$ & $1$ & $\omega^2$ & $\omega^2$ & 
$1$ & $\omega^2$ & $\omega^2$ & $1$ \\
$U(1)_R$ & $1$ & $1$ & $1$ & $1$ & 
$0$ & $0$ & $0$ & $0$ &
$2$ & $2$ & $2$ & $2$ \\
\hline\hline
\end{tabular}
\caption{\label{tab:model_3} Superfield assignments of Model~III in $A_4 \times \mathbb{Z}_3 \times U(1)_R$.}
\end{center}
\end{table}
Given that the soft mass terms are much smaller than the flavour symmetry breaking scale, the flavon VEVs are determined along the approximately flat directions. 
The flatness conditions associated with \(\hat{\xi}^{T}\) coincide with \equaref{eq:F_0}, and therefore fix the \(\mathrm{T}\)-type vacuum alignments. SUSY domain walls are then expected to form after \(A_4\) breaking. 
Similarly, the flatness conditions associated with \(\hat{\xi}^{S}\), as encoded in \equaref{eq:potential}, give the \(\mathrm{S}\)-type vacua. 
The non-renormalisable term involving \(\xi'_0\) fixes the magnitudes of the VEVs via
\begin{equation}
s^2 = -\frac{{\rm g}_3}{{\rm g}_2}\,\eta^2 \,, 
\qquad 
\eta^3 = -{\rm g}_4 \Lambda_1 \Lambda_2^2 \,,
\end{equation}
which completes the specification of the SUSY flavon vacuum structure.

\section{Cosmological Solutions to the non-Abelian Domain Wall Problem}\label{sec:sol_nonSUSY_DW}
It is well established that domain walls arising from spontaneous discrete symmetry breaking lead to a serious cosmological problem. If generated during the radiation era, their energy density scales more slowly than that of radiation, eventually coming to dominate the energy density of the Universe and leading to ``overclosure''~\cite{Zeldovich:1974uw, Kibble:1976sj, Vilenkin:1984ib}.  In this section, we review several mechanisms to solve this problem. We first outline standard non-supersymmetric solutions below, before discussing specific solutions for SUSY domain walls in \secref{sec:SUSY_solution}.

\subsection{Non-SUSY Solutions}
One standard resolution is to assume that the $A_4$-breaking phase transition occurs before or during inflation. If the symmetry is broken during inflation and never restored after reheating, the domain wall network is inflated away and effectively diluted~\cite{Antusch:2008gw}. Alternatively, the production of domain walls can be avoided entirely in explicit $A_4$ models where the waterfall field responsible for the transition receives a small shift during inflation, selecting a unique vacuum state~\cite{Antusch:2013toa}. A second possibility is to treat the $A_4$ discrete symmetry as an approximate global symmetry which is explicitly broken by quantum or gravitational corrections \cite{Riva:2010jm, Antusch:2013toa}. 
\footnote{The explicit breaking may lead to corrections to lepton mass matrix, providing another source of for non-zero $\theta_{13}$ \cite{Gelmini:2020bqg}.}
For instance, this breaking can be achieved by extending the discrete symmetry to the quark sector such that the symmetry is broken at the quantum level due to the QCD anomaly~\cite{Preskill:1991kd}.
However, the QCD anomaly alone is often insufficient to completely lift the vacuum degeneracy~\cite{Chigusa:2018hhl}. A more generic source of explicit breaking arises from gravitational effects, manifest as Planck-scale suppressed higher-dimension operators.
Explicit breaking splits the degeneracy of the vacua, creating an energy bias that drives the annihilation of the wall network. 
A potentially observable signature of this process is a stochastic gravitational wave (GW) background. The resulting signal links the wall annihilation temperature and the energy bias to the peak frequency of the GW spectrum. Furthermore, a distinctive feature of non-Abelian domain walls is that the presence of multiple vacua generates a rich spectrum of bias terms. This can lead to a pronounced multi-peak structure in the GW spectrum, which is quantitatively different from the signal predicted by simple $\mathbb{Z}_2$ domain walls~\cite{Gelmini:2020bqg,Fu:2024jhu}.

Finally, the non-Abelian discrete symmetry may arise as an effective subgroup resulting from the spontaneous breaking of a continuous gauge symmetry. For example, if a high-scale $SO(3)$ gauge symmetry is broken to $A_4$, $S_4$, or $A_5$ by high-dimensional Higgs representations, the usual domain wall problem associated with global symmetry breaking does not apply~\cite{Ovrut:1977cn,Etesi:1996urw}. A concrete example is presented in Ref.~\cite{King:2018fke}, where a two-step phase transition occurs: $SO(3) \to A_4$ followed by $A_4 \to Z_3, Z_2$. In the first step, the breaking of the gauge symmetry does not generate stable domain walls because the degenerate vacua are continuously connected in the gauge space. In the second step, the breaking of $A_4$ leads to degenerate $Z_3$- and $Z_2$-invariant vacua, generating $\mathrm{T}$- and $\mathrm{S}$-type domain walls, respectively. These walls store energy with barriers of order $\lambda v_\varphi^4$ or $\lambda v_\chi^4$ and eventually decay into light particles mediated by gauge bosons.

\subsection{SUSY Solutions}\label{sec:SUSY_solution}

In the exact SUSY limit, the scalar potential in Eq.~\eqref{eq:potential_SUSY} possesses five degenerate vacua: the trivial vacuum at $\varphi=0$ and the four non-trivial $\mathrm{T}^s$-type vacua defined in Eq.~\eqref{eq:Ts-vacua}. As discussed in \secref{sec:superfield}, the lowest-tension domain walls interpolate between the trivial vacuum and one of the $\mathrm{T}^s$ vacua, denoted by $\mybox{0}\mybox{w_i}$. If these SUSY domain walls were perfectly stable, their energy density would eventually dominate the Universe, leading to the standard domain wall overclosure problem. In realistic SUSY flavour models, however, this problem can be alleviated by including soft SUSY-breaking terms and a small explicit breaking of $A_4$.

The inclusion of $A_4$-invariant soft SUSY-breaking terms lifts the degeneracy between the trivial vacuum and the $\mathrm{T}^s$ vacua. The resulting energy difference acts as a volume pressure on the SUSY domain walls $\mybox{0}\mybox{w_i}$. As long as this vacuum pressure is small compared to the surface tension of the walls, the network evolves in a standard scaling regime. Once the volume pressure overcomes the surface tension (specifically, when it exceeds the Hubble friction term), the SUSY walls accelerate towards the trivial vacuum. Consequently, the network of $\mybox{0}\mybox{w_i}$ walls collapses, radiating gravitational waves with a peak frequency determined by the Hubble scale at the time of collapse and an amplitude controlled by the wall tension.

Below, we outline a step-by-step example of how the domain wall problem is resolved in a SUSY $A_4$ model:
\begin{itemize}
  \item $A_4$ is spontaneously broken to $Z_3$ at a scale much higher than the soft-SUSY breaking scale. Bubbles of the $w_i$ vacua and the $A_4$-invariant trivial vacuum coexist in the Universe, and walls form between $w_i$ and the trivial vacuum. Note that no walls form directly between $w_i$ and $w_j$ (for $i\neq j$) at this stage.

  \item Once the Hubble expansion reaches the epoch where the vacuum pressure induced by soft SUSY breaking overcomes the wall surface tension, the SUSY DWs collapse. At this point, the trivial vacuum becomes energetically disfavoured, and non-SUSY walls of the type $\mybox{w_i}\mybox{w_j}$ form. In the limit where the flavour symmetry breaking scale is much higher than the soft SUSY-breaking scale, these new walls resemble the ``two-peak'' solutions discussed in Section~\ref{sec:realtrip}, effectively acting as bound states of two SUSY walls. The tension of such a wall is approximately twice that of the original SUSY domain wall.

  \item Finally, additional explicit breaking terms of $A_4$ lead to the collapse of the non-SUSY DWs. This occurs at a scale where the vacuum pressure induced by the explicit breaking bias dominates the wall surface tension.
\end{itemize}
In this specific scenario, both the initial SUSY DWs and the subsequent non-SUSY DWs radiate GWs, producing a spectrum with peaks determined by the soft SUSY-breaking scale and the explicit $A_4$ breaking scale, respectively.

\section{Gravitational Waves from Multi-Scale Domain Wall Collapse}\label{sec:GW}
Gravitational waves are emitted by the collapsing domain wall network, giving rise to a stochastic background characterised by a broken power-law spectrum. The peak frequency, $f_{\rm peak}$, is directly correlated with the time at which the domain walls annihilate. In the simplest case, namely domain walls arising from the spontaneous breaking of a $Z_2$ symmetry, only two vacua exist in the system and the dynamics of the collapse is comparatively simple. In this case, the peak frequency is given by
\begin{equation}
f_{\rm peak} \;=\; a(t_{\rm ann})\, H(t_{\rm ann}) 
\simeq 3 \times 10^3 \, {\rm Hz} \,
\left(\frac{10}{g_*(T_{\rm ann})}\right)^{1/12}
\left(\frac{V_{\rm bias}}{\sigma \, {\rm TeV}} \right)^{1/2},
\end{equation}
which corresponds to a peak amplitude
\begin{equation}
  \Omega_{\rm gw} h^2\Big|_{\rm peak}
  \simeq 0.9 \times 10^{-67} \,
  \left(\frac{10}{g_*(T_{\rm ann})}\right)^{1/3} \,
  \left(\frac{\sigma}{{\rm TeV}^3}\right)^4
  \left(\frac{{\rm TeV}^4}{V_{\rm bias}}\right)^2.
\end{equation}
The spectrum at $f < f_{\rm peak}$ behaves as $\Omega_{\rm gw} h^2(f) \propto f^{3}$, as expected from causality arguments \cite{Caprini:2009fx,Hiramatsu:2013qaa,Saikawa:2017hiv}. For $f > f_{\rm peak}$, numerical simulations indicate that $\Omega_{\rm gw} h^2(f) \propto f^{-1}$ \cite{Hiramatsu:2013qaa}. Recent work has reconsidered the ultraviolet frequency dependence, separating the GW production during the scaling regime from that associated with the final collapse \cite{Ferreira:2023jbu,Li:2023yzq,Kitajima:2023kzu,Dankovsky:2024zvs,Ferreira:2024eru,Notari:2025kqq,Cyr:2025nzf,Babichev:2025stm, Blasi:2025tmn}. Although some of these studies find deviations from the simple $f^{-1}$ behaviour, we will, as a first approximation, adopt the $f^{-1}$ scaling for the high-frequency tail in the remainder of this work.
For domain walls associated with discrete groups larger than $Z_2$, more vacua arise and multiple energy biases between pairs of vacua are present, making the dynamics of the network substantially more complicated. For non-Abelian discrete symmetries this situation has been analysed in Refs.~\cite{Gelmini:2020bqg,Fu:2024jhu}, where it has been shown that the GW spectrum can develop a characteristic multi-peak structure due to the rich dynamics of domain wall collapse.

We argue that, in addition to this multi-peak structure, non-Abelian discrete flavour symmetries naturally realise another type of dynamics, namely \emph{multi-scale domain wall collapse}, which can imprint additional features on the GW spectrum. We first comment on why this behaviour is generic in  flavour models:
\begin{itemize}
  \item \textbf{Real scalars} In real scalar $A_4$ models, the cubic invariant exists but can be parametrically smaller than other terms in the potential. In this regime it can be treated as a soft breaking of an approximate $S_4$ symmetry at high scales. The symmetry breaking may then proceed in two steps,
  \begin{equation}
    S_4 \;\to\; A_4 \;\to\; \dots \, .
  \end{equation}
  Domain walls associated with the breaking of $S_4$ (for example the TI and TIII walls) are generated at the first step. These walls subsequently collapse when $S_4$ is broken down to $A_4$, leaving behind TII walls or composite configurations, as discussed previously. The remaining $A_4$ domain walls finally collapse when the soft breaking terms of $A_4$ become important.

  \item \textbf{SUSY Framework} In SUSY models, SUSY domain walls can be generated at early times, before SUSY breaking becomes relevant. These walls collapse once soft SUSY-breaking terms lift the degeneracy between the SUSY vacua. At lower energies, non-SUSY domain walls associated with the residual discrete flavour symmetry can form. These non-SUSY walls then undergo their own collapse via the usual mechanism driven by explicit breaking of the remaining symmetry. The two stages of wall formation and annihilation can occur at parametrically different scales.

  \item \textbf{Multiflavon scenarios} Realistic flavour models typically involve multiple flavon fields. For instance, in $A_4$ models one often introduces $\varphi^T$ and $\varphi^S$, as reviewed in the previous section. The VEV of $\varphi^T$ breaks $A_4 \to Z_3$ in the charged-lepton sector, while the VEV of $\varphi^S$ breaks $A_4 \to Z_2$ in the neutrino sector. Taken together, both $Z_2$ and $Z_3$ are broken and no residual flavour symmetry remains in the true vacuum. In the decoupling limit, however, $Z_3$-preserving VEVs of $\varphi^T$ and $Z_2$-preserving VEVs of $\varphi^S$ can coexist in the Universe. The resulting $\mathrm{T}$- and $\mathrm{S}$-type walls form two, in principle independent, networks. Their tensions can lie at different scales, depending on the respective potentials $V(\varphi^T)$ and $V(\varphi^S)$. Moreover, the difference between the two breaking patterns is precisely what is needed to generate non-trivial lepton flavour mixing. Bias terms in the two potentials can also differ significantly if the explicit breaking of $A_4$ feeds into $V(\varphi^T)$ and $V(\varphi^S)$ in different ways. Since the collapse time of a wall network depends on both its tension and its bias, the $\mathrm{T}$-type and $\mathrm{S}$-type walls can annihilate at parametrically different times.
\end{itemize}
In what follows, we treat each type of domain wall as evolving independently and producing its own $Z_2$-like GW spectrum. This approach is sufficient to obtain qualitative insight into the multi-peak structure generated by multi-scale domain wall collapse, different from the multi-peak structure from multiple degenerate vacua discussed in \cite{Fu:2024jhu}. Any quantitative prediction should be based on dedicated numerical simulations of the full coupled system, which we leave for future work.

\subsection{Gravitational Waves from Single-Flavon Models} 
\begin{figure}[t!]
\centering
\includegraphics[width = 0.5 \linewidth]{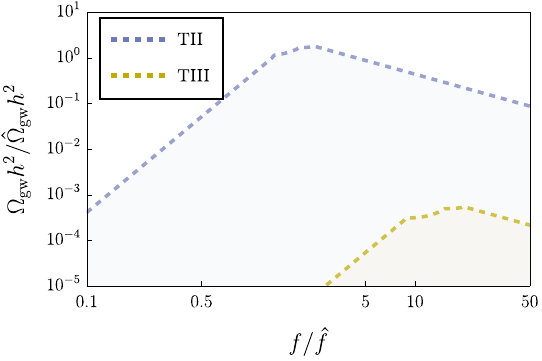}
  \caption{Spectrum of the gravitational wave background generated by collapsing T-type domain walls in a single-flavon model with real scalars.}
\label{fig:singleflavonexample}
\end{figure}
We consider the model described by \equaref{eq:potential}, with a single flavon that leads to $Z_3$-preserving vacua. We choose $a = -3\times10^{-5}$ and $\beta = -0.01$. For these parameter values, we have that the tension of the TI domain wall is considerably larger than the tension of TII and TIII domain walls. 
Therefore, the production of TI domain wall is exponentially suppressed during the Kibble mechanism, and only TII and TIII domain walls contribute significantly to the GW production. 

The cubic term in \equaref{eq:potential} acts as an effective bias for the TIII walls, which are realised as oreo-type walls with an energy profile as shown in \figref{fig:path_and_en}. An additional bias term must be introduced for the TII walls and, in general, can have a different microscopic origin. This naturally motivates a multi-scale pattern of biases, which in turn causes different classes of walls to annihilate at different epochs.
\figref{fig:singleflavonexample} shows an example of the GW signal from collapsing domain walls in the case of real scalars, although analogous considerations apply in the SUSY setup.
Even though multi-scale domain wall collapse is present in such frameworks, the GW signal from earlier-collapsing walls is typically overshadowed by that from walls collapsing later. The reason is that the different wall species have tensions of the same order, as we can see in \figref{fig:Ratio_Heatmap}, so the present-day GW amplitude is mainly controlled by the time of collapse: GWs produced earlier are more strongly redshifted and hence subdominant in the final spectrum.

\subsection{Gravitational Waves from Multi-Flavon Models}
For the multiflavon case, we consider the model described by \equaref{eq:model_1_vev}. The scalar potential is schematically given by \equaref{eq:potentialA4real} and we assume
\begin{equation}
  V(\phi^T, \phi^S) \ll V(\phi^T) + V(\phi^S)\,,
\end{equation}
that is, the cross-coupling terms are negligible at leading order. Such terms, when present, act as additional sources of bias. In this decoupling limit we can treat the T-type and S-type walls as evolving independently, which simplifies the discussion.

To avoid a stable domain wall network, it is necessary to introduce bias terms. In realistic models these can originate, for example, from Yukawa interactions with charged leptons and neutrinos via the Coleman–Weinberg potential and thermal corrections \cite{Gelmini:2020bqg, Zhang:2023nrs, Zeng:2025zjp}. Since we are interested in a generic, qualitative description of the GW signal from multiflavon domain walls, we parametrise the biases as
\begin{equation}
  V^S_{\rm bias} = \epsilon_S\, v_S^4\,,
  \qquad 
  V^T_{\rm bias} = \epsilon_T\, v_T^4\,.
\end{equation}
As discussed above, a multi-peak structure can be observed only if the two contributions are of comparable size, which requires $\epsilon_S \sim \epsilon_T$ up to $\mathcal{O}(1)$ factors. Motivated by the model under consideration, we take $v_S \gg v_T$. We focus on a scenario in which the GW signal from the $Z_3$- and $Z_2$-preserving sector is mainly generated by TII walls and SII walls respectively. This is justified for the values of $a$ and $\beta$ adopted here, namely $a= -3\times10^{-5}$ and $\beta = -0.01$, and by the considerations made earlier in this section.
\begin{figure}[t!]
  \centering
  \includegraphics[width=0.49\linewidth]{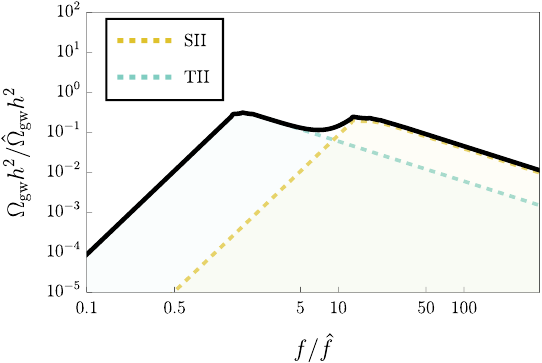}
  \includegraphics[width=0.49\linewidth]{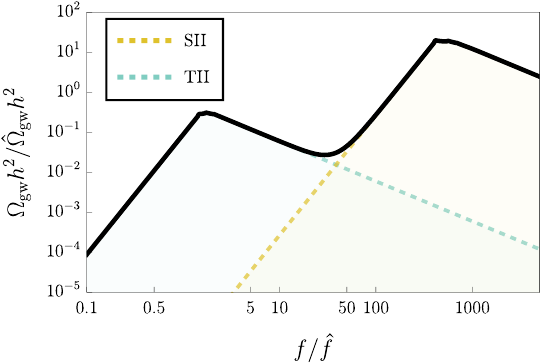}
  \caption{Spectrum of the gravitational wave background generated in a multiflavon scenario with real scalars. Realistic models with more than one flavon naturally give rise to hierarchies between the tensions and biases of different wall types, which can lead to a multi-peak structure in the GW spectrum. }
  \label{fig:multiflavonexample}
\end{figure}
In \figref{fig:multiflavonexample} we show two normalised GW spectra produced by multiflavon domain wall networks. For the first benchmark we take $\epsilon_S/\epsilon_T = 1$ and $v_S/v_T = 10$, while for the second we choose $\epsilon_S / \epsilon_T=0.1$ and $v_S/v_T = 10^3$. In both cases, the two peaks appear at significantly different frequencies while still having comparable amplitudes. Moreover, in computing the signal we took into consideration the presence of different bias for each kind of domain wall, as described in Ref.~\cite{Gelmini:2020bqg}. 

However, it is important to stress that the strict decoupling limit is an idealisation. Small non-decoupling effects are induced by loop corrections, for instance via Coleman-Weinberg potentials generated by Yukawa couplings to fermions, even if tree-level cross couplings between $\varphi^T$ and $\varphi^S$ are forbidden. These corrections can themselves generate additional biases. To suppress their impact one must assume that the explicit breaking of $A_4$ entering these loops is sufficiently small, or is arranged such that the induced biases remain subdominant.
\begin{figure}[t!]
  \centering
  \includegraphics[width=0.85\linewidth]{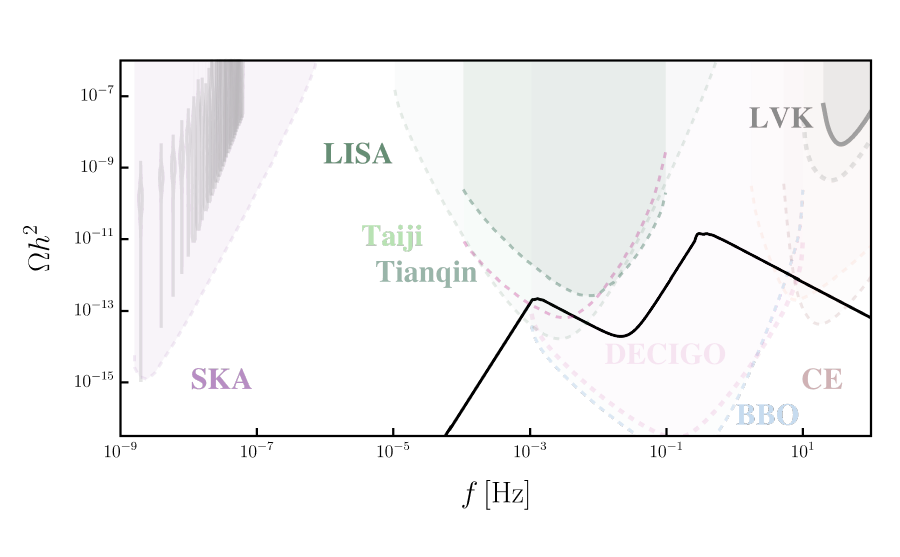}
  \caption{Domain walls from multiflavon can be potentially tested at next generation GW observatories. Here we show the GW signal generated by the second benchmark example, with $\sigma_S^{1/3} \sim 8.4 \times 10^6 \, \text{TeV} $, and $\sigma_T^{1/3}  \sim 8.4 \times 10^4 \, \text{TeV} $.  We show projections with dashed lines and current bound with solid lines. In grey it is shown the signal reported by NANOGrav collaboration. \cite{NANOGrav:2023gor} }
  \label{fig:testability}
\end{figure}
In our analytical treatment we further assume that each wall species produces a ``$Z_2$-like'' GW spectrum, as in the single-flavon case, and that interactions between different wall networks can be neglected. Relaxing these assumptions and including the full set of interactions could modify the detailed shape of the spectrum. Nevertheless, we expect the emergence of a qualitative multi-peak structure to be a robust feature of models with more than one flavon.
It is worth mentioning that this multi-peak spectrum is not exclusive to models with more than one flavon. For example, it has been argued recently that current-carrying domain walls from $Z_2$ symmetries can generate a similar GW spectrum \cite{Ghoshal:2025gci}.

We briefly mention the possibility of such kind of signal to be tested at current and next-generations GW observatories such as Pulsar Time Arrays, SKA, LIGO, LISA, Taiji, Tianqin, ET, CE, BBO and DECIGO  \cite{NANOGrav:2023gor, Zic:2023gta, EPTA:2023sfo, Xu:2023wog,Rana:2025ano, Janssen:2014dka, Li:2024rnk, Ruan:2018tsw, 2017arXiv170200786A, Punturo_2010, Reitze:2019iox, LIGOScientific:2025kry, Crowder:2005nr,Kawamura:2006up}. 
In \figref{fig:testability} we show, as an example, the second benchmark with $\sigma_S^{1/3} \sim 8.4 \times 10^6 \, \text{TeV} $, and $\sigma_T^{1/3} \sim  8.4 \times 10^4 \, \text{TeV} $. The bias terms for the T and S domain walls are respectively ${V^T_{\rm bias}}^{1/4} \sim 3.0 \, \text{TeV}$ and $ {V^S_{\rm bias}}^{1/4} \sim 1.4 \times 10^3 \, \text{TeV}$. This example shows how the peaks can be relatively well separated and how they  could, in principle, be observed in different GW observatories.
Finally, it is interesting to mention that DW could generate primordial black holes (PBH) \cite{Gouttenoire:2023gbn, Ferreira:2024eru}, and due to the different collapse times these could be produced at different mass ranges.

\section{Conclusion}
\label{sec:conclusion}
In this work we have carried out a systematic study of domain walls arising from the spontaneous breaking of an $A_4$ flavour symmetry and discussed the associated stochastic gravitational wave background.
We first constructed the most general renormalisable $A_4$-invariant scalar potential for a real triplet and classified all $Z_2$- and $Z_3$-preserving vacua. The cubic $A_4$ invariant splits the eight $\mathrm{T}$-type vacua into two sets with different depths, leading to both true and false vacua and consequently to metastable domain walls. We showed that certain $Z_3$ walls (TIII) are unstable and decay into a bound state of two lighter walls (TII-like), forming oscillating two-wall configurations. This behaviour has no analogue in simple $\mathbb{Z}_2$ models and is specific to the $A_4$ structure. We derived the corresponding wall profiles and tensions for these walls. 

We then extended the analysis to complex triplet flavons in an $A_4\times U(1)$ theory, where the additional global $U(1)$ gives rise to global cosmic strings in addition to non-Abelian domain walls. We classified the $\mathrm{S}^c$- and $\mathrm{T}^c$-type vacua, showed that static walls correspond to configurations with constant relative phases and reduced equations of motion equivalent to those of an $S_4$-like real mode and identified three classes of $\mathrm{T}'^c$ walls distinguished by their group-theoretic properties. The resulting defect network can involve both strings and walls and we identified regions of parameter space where some wall species are unstable against decay into lighter ones. Such string–wall systems provide a natural source of GWs whose spectrum is sensitive to both the non-Abelian and Abelian sectors. We point out that the existence of CP-violating domain walls between two CP-conjugated $T^{\prime c}$ vacua. Note that since the field space here is more than three dimensions, domain walls cannot be classified geometrically, and instead we develop an algorithm based on group transformations and CP transformation for domain wall classification. We also note that the non-Abelian vacuum structure studied here may support more complicated network configurations, including possible domain-wall junctions; a dedicated analysis of their formation and evolution is left for future work.

In the supersymmetric framework we considered $A_4$ triplet chiral superfields with $F$-term dominated potentials. We showed that the SUSY scalar potential admits an exactly degenerate trivial vacuum and four non-trivial $\mathrm{T}$-type vacua, and that the lowest-tension domain walls interpolate between the origin and each non-trivial vacuum. We specify SUSY domain walls as those which separate SUSY vacua. These walls satisfy first-order BPS-like equations, allowing analytic solutions and exact tensions. Introducing an $R$-symmetry and driving superfields, as in standard SUSY $A_4$ constructions, leaves the flavon potential effectively unchanged in the SUSY limit, while decoupling the driving fields along the wall profiles. Soft SUSY-breaking terms lift the degeneracy between the trivial and non-trivial vacua, providing a bias that triggers the annihilation of the SUSY walls to non-SUSY domain walls, $i.e.$, walls separating SUSY-breaking vacua. 

Through the above discussion for $A_4$ breaking via real scalars, complex scalars, and superfields, we observe different types of domain walls, respectively. Those are ``oreo''-type composite domain walls raised via cubic term of real scalars, CP-violating domain walls via CP-conjugate vacua of complex scalars, and SUSY non-Abelian domain walls distinghuiable from non-SUSY ones. 
 
Embedding these ingredients in explicit lepton flavour models requires multiple flavons.
For a toy model with two real triplet flavons $\phi^S$ and $\phi^T$, we showed how $Z_2$- and $Z_3$-preserving VEVs arise, leading to TBM mixing at leading order and to separate $\mathrm{S}$-type and $\mathrm{T}$-type wall networks with tensions of order $\mu_S v_S^2$ and $\mu_T v_T^2$. In complex flavon models with a global $U(1)_L$ (or its discrete or gauged variants), we explained how the same flavour structure can be realised while simultaneously generating global or local strings. In SUSY $A_4$ models with driving fields and $U(1)_R$, we highlighted how the alignment mechanism fixes the flavon directions.

Including explicit breaking terms provides a feasible way to solve the cosmological domain wall problem. In particular, soft SUSY breaking terms lift the trivial vacuum from the $A_4$-breaking vacua and thus the problem for SUSY domain walls is naturally solved. The explicit breaking of $A_4$ generates biases that drive wall collapsing, thereby alleviating the domain wall problem. We argue that multi-scale domain wall collapsing are widely predicted in $A_4$ flavour models. The multi-scale collapsing may happen in several different ways: 1) In real scalar models, approximate $S_4$ recover at high scale due to the smallness of cubic flavon couplings, and thus $S_4$ domain walls collapses first, and the $A_4$ walls collapse later; 2) In SUSY models, SUSY domain walls collapses after soft-SUSY breaking and non-SUSY walls collapse later; 3) In realistic flavon models when multiple flavons appear, those in the neutrino sector and charged-lepton sector can evolves independently in the decoupling limit and thus collapse at different scales.
These features imprint characteristic structures in the GW spectrum, including the possibility of multiple peaks.

Taken together, our results show that non-Abelian flavour symmetries such as $A_4$ naturally give rise to rich defect networks whose dynamics are tightly correlated with the structure of the flavour sector. The resulting stochastic GW backgrounds can display multiple peaks and characteristic features linked to the hierarchy of breaking scales (flavour, SUSY, and explicit breaking) and to the presence or absence of additional Abelian symmetries. This opens up the possibility that future GW measurements, in combination with precision flavour data, could provide novel probes of the symmetry structure underlying lepton masses and mixing. In the presence of multiple fields and many vacua, dynamics of domain wall evolution and collapsing can be more complicated than the $Z_2$ domain walls focused by most simulations. Thus, our discussion on sum of peaked GW spectra is qualitative. A natural next step is to perform dedicated simulations of $A_4$ domain wall and string networks, including both the real, complex and SUSY cases discussed here, in order to obtain quantitative GW predictions for specific benchmark models and to confront them with the sensitivity of current and forthcoming GW observatories.

\subsubsection*{Acknowledgements}
BF acknowledges Liaoning Natural Science Foundation No.~2025-BS-0086 and Guangdong Basic and Applied Basic Research Foundation No.~2025A1515011079.
SFK thanks IFIC, Valencia, for hospitality and acknowledges the STFC Consolidated Grant ST/X000583/1; his work was funded by a Leverhulme Emeritus Fellowship Grant.
L.M. is supported by the Generalitat Valenciana through the grant CIACIF/2023/91. L.M would like to thank the IPPP at Durham University for their hospitality during the completion of this work.
JT would like to thank the Quantum Field Theory Centre at the University of Southern Denmark
for their hospitality during the completion of this work. 
YLZ was supported by the National Natural Science Foundation of China (NSFC) under Grant Nos.~12535007, 12547104, and Zhejiang Provincial Natural Science Foundation of China under Grant No. LDQ24A050002. 


\appendix

\section{Complex scalar invariants}\label{app:complex_invariants} 
For completeness, we list the independent \(A_4\)-invariant operators constructed from a complex triplet \(\varphi=(\varphi_1,\varphi_2,\varphi_3)^T\) up to dimension four: 
\[
\begin{aligned}
\mathcal{I}_{11} &= |\varphi_{1}|^{2} + |\varphi_{2}|^{2} + |\varphi_{3}|^{2}, 
&\quad 
\mathcal{I}_{12} &= \varphi_{1}^{2} + \varphi_{2}^{2} + \varphi_{3}^{2}, \\[5pt]
\mathcal{I}_{21} &= |\varphi_{1}|^{2} |\varphi_{2}|^{2} + |\varphi_{2}|^{2} |\varphi_{3}|^{2} + |\varphi_{3}|^{2} |\varphi_{1}|^{2}, 
&\quad 
\mathcal{I}_{22} &= \varphi_{1}^{2} (\varphi_{2}^{2})^{*} + \varphi_{2}^{2} (\varphi_{3}^{2})^{*} + \varphi_{3}^{2} (\varphi_{1}^{2})^{*}, \\[5pt]
\mathcal{I}_{23} &= \varphi_{1}^{2} \varphi_{2}^{2} + \varphi_{2}^{2} \varphi_{3}^{2} + \varphi_{3}^{2} \varphi_{1}^{2}, 
&\quad 
\mathcal{I}_{24} &= \varphi_{1}^{2} |\varphi_{2}|^{2} + \varphi_{2}^{2} |\varphi_{3}|^{2} + \varphi_{3}^{2} |\varphi_{1}|^{2}, \\[5pt]
\mathcal{I}_{25} &= |\varphi_{1}|^{2} \varphi_{2}^{2} + |\varphi_{2}|^{2} \varphi_{3}^{2} + |\varphi_{3}|^{2} \varphi_{1}^{2}, \\[5pt]
\mathcal{I}_{31} &= \varphi_{1} \varphi_{2} \varphi_{3}, 
&\quad 
\mathcal{I}_{32} &= \tfrac{1}{3} \bigl( 
\varphi_{1} \varphi_{2} \varphi_{3}^{*} 
+ \varphi_{1} \varphi_{2}^{*} \varphi_{3} 
+ \varphi_{1}^{*} \varphi_{2} \varphi_{3} 
\bigr).
\end{aligned}
\]
These invariants, together with their hermitian conjugates, generate the general potential in \equaref{eq:potential_complex}.

\section{$\mathrm{S}^c$-type Domain Walls: Technical Details}
\label{app:Sdomain}

In this appendix we collect the explicit equations of motion and the Noether current analysis used in the discussion of $\mathrm{S}^c$-type domain walls.
With the normalisations
\[
\overline{\varphi}_i \equiv \frac{\sqrt{g_1}}{\mu}\,\varphi_i=\frac{\varphi_i}{v},
\qquad
\overline{z}\equiv \mu z,
\]
and the parametrisation
\[
\overline{\varphi}_1=\frac{\bar\phi_1}{\sqrt{2}}e^{i\alpha_1},
\qquad
\overline{\varphi}_2=\frac{\bar\phi_2}{\sqrt{2}}e^{i\alpha_2},
\]
the equations of motion, in polar form, reduce to
\be
\begin{aligned}
&\bar\phi_1'' = \bar\phi_1\left(\frac{\bar\phi_2^2}{\bar\phi_1^2+\bar\phi_2^2}\right)^2(\alpha_{12}')^2
+ \bar\phi_1\!\left[
1 - (\bar\phi_1^2+\bar\phi_2^2) - \beta\,\bar\phi_2^2 - \beta'\,\bar\phi_2^2\cos(2\alpha_{12})
\right], 
\\
&\bar\phi_2'' = \bar\phi_2\left(\frac{\bar\phi_1^2}{\bar\phi_1^2+\bar\phi_2^2}\right)^2(\alpha_{12}')^2
+ \bar\phi_2\!\left[
1 - (\bar\phi_1^2+\bar\phi_2^2) - \beta\,\bar\phi_1^2 - \beta'\,\bar\phi_1^2\cos(2\alpha_{12})
\right], 
\\
&\frac{d}{d\overline{z}}\!\left(
\frac{\bar\phi_1^2\,\bar\phi_2^2}{\bar\phi_1^2+\bar\phi_2^2}\,
\alpha_{12}'\right)
= \beta'\,\bar\phi_1^2\bar\phi_2^2\,\sin(2\alpha_{12})\,,
\end{aligned}
\ee
where primes denote derivatives with respect to $\overline{z}$ and
\[
\beta \equiv \frac{g_2}{g_1}, \qquad \beta' \equiv \frac{g_2'}{g_1}\,.
\]
Writing $\varphi_i=(h_i+i a_i)/\sqrt{2}$, the equations of motion become
\be
\begin{aligned}
h_1'' &= h_1\Big[-\mu^2 + g_1 (h_1^2+a_1^2+h_2^2+a_2^2) + g_2 (h_2^2+a_2^2) + g'_2 (h_2^2-a_2^2)\Big] +2g'_2 a_1 h_2 a_2, \\
a_1'' &= a_1\Big[-\mu^2 + g_1 (h_1^2+a_1^2+h_2^2+a_2^2) + g_2 (h_2^2+a_2^2) - g'_2 (h_2^2-a_2^2)\Big] +2g'_2 h_1 h_2 a_2 ,  \\
h_2'' &= h_2\Big[-\mu^2 + g_1 (h_1^2+a_1^2+h_2^2+a_2^2) + g_2 (h_1^2+a_1^2) + g'_2 (h_1^2-a_1^2)\Big] +2g'_2 a_2 h_1 a_1 , \\
a_2'' &= a_2\Big[-\mu^2 + g_1 (h_1^2+a_1^2+h_2^2+a_2^2) + g_2 (h_1^2+a_1^2) - g'_2 (h_1^2-a_1^2)\Big] +2g'_2 h_2 h_1 a_1 .
\label{eq:ah_eom}
\end{aligned}
\ee

The Noether current associated with the global $U(1)$ symmetry is
\be
J^\mu \;=\; i\sum_{i=1}^3\!\left(\varphi_i^{*}\,\partial^\mu\varphi_i
-\partial^\mu\varphi_i^{*}\,\varphi_i\right)
\;=\;\sum_{i=1}^3 \phi_i^{2}\,\partial^\mu \alpha_i\,.
\ee
Since only $\varphi_{1,2}$ vary across the wall, this reduces to
\[
J^\mu = \phi_1^2\,\partial^\mu\alpha_1 + \phi_2^2\,\partial^\mu\alpha_2.
\]
For a static planar wall $J^0=0$ and
\[
J^z = \phi_1^2\,\partial_z\alpha_1 + \phi_2^2\,\partial_z\alpha_2.
\]
In the normalised variables, we find
\[
\frac{d}{d\bar z}\Big(\bar\phi_1^{2}\,\alpha_1' + \bar\phi_2^{2}\,\alpha_2'\Big)=0
\;\;\Longrightarrow\;\;
\bar J^z \equiv \bar\phi_1^{2}\,\alpha_1' + \bar\phi_2^{2}\,\alpha_2' = \text{const.}
\]
For finite-energy static walls, $\bar J^z=0$. This immediately implies that the 
relative phase $\alpha_{12}$ must be constant across the wall wherever 
$\phi_1\phi_2\neq0$. This provides the current-conservation justification for the constant-phase assumption made in the main text. Another way to see this is that for a static domain wall to be stable, it must represent a configuration of minimum energy (tension) connecting two vacua. For a planar wall dependent on a coordinate $z$, the phase-dependent part of the wall tension is given by the energy functional:
\begin{equation}
  E_{\text{phase}} = \int dz \left[ \frac{1}{2} K(z) (\alpha_{12}')^2 + V_{\text{phase}}(\alpha_{12}) \right]
\end{equation}
where $\alpha_{12} \equiv \alpha_1 - \alpha_2$ is the relative phase, the prime denotes a derivative with respect to $z$, and
$$
K(z) = \frac{\phi_1^2 \phi_2^2}{\phi_1^2 + \phi_2^2} \ge 0, \quad V_{\text{phase}} = \frac{g_2'}{2} \phi_1^2 \phi_2^2 \cos(2\alpha_{12}).
$$
The total energy is a sum of two terms: a `kinetic' term, $\frac{1}{2}K(z)(\alpha_{12}')^2$, which penalises any spatial variation of the relative phase, and a potential term, $V_{\text{phase}}$, which depends on the value of the phase itself.
To minimise the integral $E_{\text{phase}}$, we must minimise the integrand at every point $z$ across the wall.
\begin{enumerate}
  \item \textbf{Kinetic Term:} Since $K(z) \ge 0$, the kinetic term is always non-negative. It is minimised to zero if and only if $\alpha_{12}'(z) = 0$.
  \item \textbf{Potential Term:} The potential term is minimised by choosing a constant value of $\alpha_{12}$ that makes $\cos(2\alpha_{12})$ most negative (if $g_2' > 0$) or most positive (if $g_2' < 0$). This corresponds to discrete values $\alpha_{12} = n\pi/2$ for some integer $n$.
\end{enumerate}
Any solution where the relative phase varies with $z$ ($i.e.$, $\alpha_{12}' \neq 0$) would incur a positive kinetic energy cost, leading to a higher total wall tension. Therefore, {the minimum-energy, and thus physically stable, domain wall solution must have a constant relative phase}, $\alpha_{12}'(z)=0$, across the entire interior of the wall where both fields $\phi_1$ and $\phi_2$ are non-zero.

\section{Solution to the EoMs in complex $A_4\times U(1)$ model \label{app:complex}}

To find the vacua of the complex $A_4\times U(1)$ model with non-zero expectation values for all three radial fields, we extremise the field components of the complex flavon with respect to both the magnitude of the field, $|\phi_i|$, and the phase differences, $\alpha_{ij}$. 
The EoMs for the phase differences are 
\begin{equation}
\begin{aligned}
\frac{\partial V}{\partial\alpha_{12}}&=
-g_{2}'\phi_{1}^{2}\Bigl(\phi_{2}^{2}\sin 2\alpha_{12}+\phi_{3}^{2} \sin2(\alpha_{12}+\alpha_{23})\Bigr)=0\,,\\[4pt]
\frac{\partial V}{\partial\alpha_{23}}&=
-g_{2}'\phi_{3}^{2}\Bigl(\phi_{2}^{2}\sin 2\alpha_{23}+\phi_{1}^{2}\sin 2(\alpha_{12}+\alpha_{23})\Bigr)=0\,.
\end{aligned}
\end{equation}
Solving these equations determines the possible phase alignments and, together with the EoMs of the radial fields \(\partial V/\partial\phi_{i}=0\), specifies the vacuum structure of the complex flavon potential.
There are three distinctive solutions of $\alpha$:
\begin{enumerate}[label=\textbf{Solution \arabic*:}, leftmargin=*]
  \item $\cos 2\alpha_{12} = \cos 2\alpha_{23} = 1$ \\ In this case the flavon phases are trivial and $I_2' = I_2$ with the potential reduced to that of the real $S_4$ theory with a shifted coupling $g_2 \rightarrow g_2 + g_2'$, $i.e.$
  \begin{eqnarray} \label{eq:case1}
    V(\phi) = -\frac{\mu^2}{2} I_1 + \frac{g_1}{4} I_1^2 + \frac{g_2+g_2'}{2} I_2 \,.
  \end{eqnarray}
  \item $\Big(\tan 2\alpha_{12},\,\tan 2\alpha_{23}\Big) =\pm\left( \dfrac{\sqrt{4I_2-I_1^2}}{-\phi_1^4-\phi_2^4+\phi_3^4},\, \dfrac{\sqrt{4I_2-I_1^2}}{\phi_1^4-\phi_2^4-\phi_3^4}\right)$ \\
  In the case, there is $I_2' = -\frac12 (\phi_1^4 + \phi_2^4 + \phi_3^4) = -\frac12 I_1^2 + I_2$ and the potential is reduced to 
  \begin{eqnarray} \label{eq:case2}
    V(\phi) = -\frac{\mu^2}{2} I_1 + \frac{g_1-g_2'}{4} I_1^2 + \frac{g_2+g_2'}{2} I_2 \,,
  \end{eqnarray}
  which is the potential of real $S_4$ model with shifted couplings $g_1 \rightarrow g_1 - g_2'$ and $g_2 \rightarrow g_2 + g_2'$.
  \item $\cos 2\alpha_{12} = \cos 2\alpha_{23} = -1$, or $\cos 2\alpha_{12}\cos 2\alpha_{23} = -1$\\
  It can be shown that 
  \begin{eqnarray}
    I_2 \geq I_2'\geq -\frac12 I_1^2 + I_2\,,
  \end{eqnarray}
  which means the value of $I_2'$ in this solution is always between the value of $I_2'$ in the first two solutions.
\end{enumerate} 
Therefore the potential is minimised when $\alpha_{12}$ and $\alpha_{23}$ satisfy solution 1 if $g_2'<0$ and when $\alpha_{12}$ and $\alpha_{23}$ satisfy solution 2 if $0<g_2'<-g_2$. 
In any case, the resulting potential of the radial fields are similar to the one of real $S_4$ scalar model. 
Therefore the solution of the radial fields are $\langle\phi_1\rangle = \langle\phi_2\rangle = \langle\phi_3\rangle = u$ with
\begin{eqnarray}
u=\begin{dcases}
  \frac{\mu}{\sqrt{3g_1 + 2g_2 + 2 g_2'}}\, & 
  \text{when } g_2'<0\,,\\
  \frac{\mu}{\sqrt{3g_1 + 2g_2 - g_2'}}\, & 
  \text{when } g_2'>0\,.
\end{dcases}
\end{eqnarray}
The corresponding values of the phases are 
\begin{eqnarray} 
\begin{pmatrix}
\langle\alpha_{12}\rangle \\
\langle\alpha_{23}\rangle
\end{pmatrix} = \begin{dcases}
  \left\{\begin{pmatrix} 0 \\ 0 \end{pmatrix},
  \begin{pmatrix} \pi \\ 0 \end{pmatrix},
  \begin{pmatrix} 0 \\ \pi \end{pmatrix},
  \begin{pmatrix} \pi \\ \pi \end{pmatrix} 
\right\}\, & 
  \text{when } g_2'<0\,,\\
  \left\{
  \pm\begin{pmatrix} \frac\pi3 \\[1.ex] \frac\pi3 \end{pmatrix},
  \pm\begin{pmatrix} \frac{2\pi}{3} \\[1.ex] \frac{2\pi}{3} \end{pmatrix},
  \pm\begin{pmatrix} \frac{\pi}{3} \\[1.ex] -\frac{2\pi}{3} \end{pmatrix},
  \pm\begin{pmatrix} \frac{2\pi}{3} \\[1.ex] -\frac{\pi}{3} \end{pmatrix}
  \right\}
\, & 
  \text{when } 0<g_2'<-g_2\,.
\end{dcases}
\end{eqnarray}
To find the distinct vacua, one must also solve for the $\partial V/\partial|\phi_{i}|=0$.
The corresponding vacuum energy is 
\begin{eqnarray}
V=\begin{dcases}
  -\frac{3\mu^4}{4(3g_1 + 2g_2 + 2 g_2')}\, & 
  \text{when } g_2'<0\,,\\
  -\frac{3\mu^4}{4(3g_1 + 2g_2 - g_2')}\, & 
  \text{when } g_2'>0\,.
\end{dcases}
\end{eqnarray}

In the case of $g_2'>0$, we end up with 8 different vacua: 
\begin{eqnarray}
  \left\{  
  \begin{pmatrix}
    1\\\pm\omega\\\omega^*
  \end{pmatrix},  
  \begin{pmatrix}
    1\\\pm\omega\\-\omega^*
  \end{pmatrix},
  \begin{pmatrix}
    1\\\pm\omega^*\\\omega
  \end{pmatrix},  
  \begin{pmatrix}
    1\\\pm\omega^*\\-\omega
  \end{pmatrix}
  \right\}e^{i\alpha}\,.
\end{eqnarray}

\section{$Z_2$ DWs in SUSY \label{app:Z2SUSY}}

In this appendix, we discuss $Z_2$ domain walls in the SUSY framework. Two toy models will be covered below: one involves only a $Z_2$-odd superfield and the other include one $Z_2$-odd superfield and one $Z_2$-even driving fields. DWs solutions will be calculated in two ways: the traditional 2ODE and 1ODE respect to F-terms.

\subsection{DWs with single superfield} 

We first consider the single-superfield case. A chiral superfield $\hat{\varphi}$, with its scalar component denoted as $\varphi$, transform as $\hat{\varphi} \to -\hat{\varphi}$ under the $Z_2$ symmetry. We consider the following superpotential
\begin{eqnarray}
  W = M \hat{\varphi}^2 + \frac{-1}{4 \Lambda} \hat{\varphi}^4
\end{eqnarray}
Here, in order to obtain a non-trivial VEV, we included a non-renormalisable term. The mass-dimensional parameter $M$ and $\Lambda$ can be kept real without loss of generality. It is achieved as follows: the relative phase between $M$ and $1/\Lambda$ can be absorbed via the phase redefinition of the superfield $\hat{\varphi}$ and the overall phase is rotated away via the 
$F$-term of $\hat\varphi$ at $\theta = 0$ is given by
\begin{eqnarray} \label{eq:F_term}
  F = \frac{\partial W}{\partial \hat{\varphi}}\Big|_{\theta=0} = M \varphi + \frac{-1}{\Lambda} \varphi^3
\end{eqnarray}
The potential of the scalar $\varphi$ is derived via $F$-term as
\begin{eqnarray}
  V = F \bar{F} + V_{\rm soft}
  = M^2 \varphi \varphi^* + \frac{1}{\Lambda^2} \varphi^3 \varphi^{*3}
  +\frac{-M}{\Lambda} (\varphi^3 \varphi^* + \varphi \varphi^{*3})
  + V_{\rm soft}
\end{eqnarray}
where $\varphi^*$ is the complex conjugate of $\varphi$. $V_{\rm soft}$ is the soft breaking terms as a consequence of SUSY breaking. For NP much higher than the SUSY breaking scale, this effect should be sufficient small and thus we can ignore its effect. In this case, $V \geqslant 0$ always hold. Once the flat direction $F = 0$ is achieved at some value of $\varphi$, $V$ gains the global minimum. 

\begin{figure}
  \centering
  \includegraphics[width=0.7\linewidth]{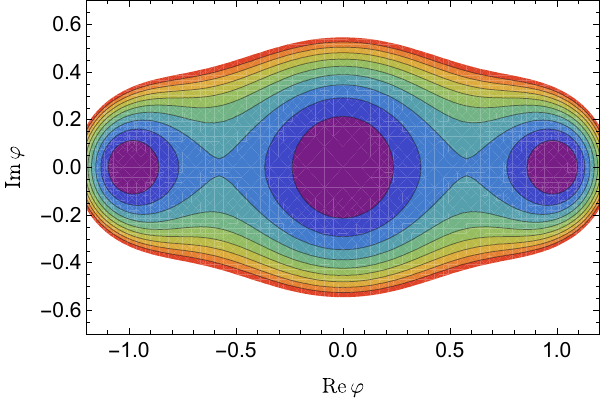}
  \caption{Contour plot of $Z_2$-invariant scalar potential in SUSY.}
  \label{fig:contour_Z2}
\end{figure}

Given the $F$-term in \equaref{eq:F_term}, the flat direction is solved at
\begin{eqnarray}
  v_0 = 0 \,,\quad 
  v_{1,2} = \pm \sqrt{M \Lambda} \,. 
\end{eqnarray}
The first solution is a trivial solution and the second two solutions break $Z_2$. Contour of potential and vacua located in the field space are shown in \figref{fig:contour_Z2}. 

Different from the domain wall in $Z_2$-invariant non-SUSY theory, here we have no domain wall formed between the two $Z_2$-breaking vacua. However, we have domain walls separating one $Z_2$-preserving and one $Z_2$-breaking vacua, 
\begin{eqnarray}
\mybox{v_0}\mybox{v_1},~ \mybox{v_0}\mybox{v_2}
\end{eqnarray}
Below, we give domain wall solutions in two ways. One follows the traditional ways, solving the 2nd-order ODE (2ODE) and the other is to solve 1st-order ODE (1ODE) derived in the SUSY framework.

{\bf DW solution in traditional 2ODE}. 
The 2ODE to be solved is 
\begin{eqnarray} \label{eq:2ODE}
  \frac{d^2 \varphi}{d z^2} = \frac{\partial V}{d \varphi^*} =
  M^2 \varphi + \frac{3}{\Lambda^2} \varphi^3 \varphi^{*2}
  +\frac{-M}{\Lambda} (\varphi^3 + 3 \varphi \varphi^{*2})
\end{eqnarray}
Boundary conditions (BCs) are given by 
\begin{eqnarray} 
  \varphi|_{z\to +\infty} = + \sqrt{M \Lambda},\quad 
  \varphi|_{z\to -\infty} = 0 \,.
  \label{eq:BC}
\end{eqnarray}
Once the scalar profile of a domain wall solution is obtained, the tension of domain wall is calculated via
\begin{eqnarray}
  \sigma &=& \int_{-\infty}^{+\infty} dz \, \varepsilon(z) \nonumber \\
  \varepsilon(z) &=& \frac{d \varphi(z)}{dz} \frac{d \varphi^*(z)}{dz} + \Delta V (\varphi(z), \varphi^*(z)) \label{eq:tension}
\end{eqnarray}
where $\Delta V (\varphi(z), \varphi^*(z)) = V (\varphi(z), \varphi^*(z)) - V_{\rm min}$. 
The first and second term of $\varepsilon$ represent the gradient energy and potential energy, respectively. Both are not negative at any place. A zero tension is obtained only if $d \varphi(z)/dz = 0 $ and $V(\varphi(z)) = V_{\rm min}$. This is inconsistent with a domain wall solution. Thus, the tension of a domain wall is always positive. 

We solve the ODE below. As the equation and BCs do not evolve any phase, we guess the solution should be real. Thus, we will consider only the solution of $|\varphi|$. We perform the following parameterisation,
\begin{eqnarray}
  |\varphi| = \sqrt{M \Lambda} f \,,\quad
  z = \frac{x}{M}\,,
\end{eqnarray}
\equaref{eq:2ODE} is simplified to be
\begin{eqnarray}
  f''(x) = f + 3 f^5 - 4 f^3
\end{eqnarray}
Multiplying $2 f'$ on both side, 
\begin{eqnarray}
  \frac{d}{dx}[f'(x)]^2 = \frac{d}{dx}f^2 (1- f^2)^2
\end{eqnarray}
The solution is given by
\begin{eqnarray} \label{eq:df}
  f'(x) = f (1-f^2)
\end{eqnarray}
It satisfies BCs in \equaref{eq:BC}. The above equation is further solved to be a square root of the sigmoid function
\begin{eqnarray}
  f(x) = \frac{1}{\sqrt{1+ e^{-2x}}} \,.
\end{eqnarray}
Taking the solution back to \equaref{eq:tension}, we obtain
\begin{eqnarray}
  \sigma = \frac{1}{2} M^2 \Lambda \,.
\end{eqnarray}

{\bf DW solution in 1ODE in SUSY}.
In SUSY, a sufficient condition of \equaref{eq:2ODE} is an 1ODE 
\begin{eqnarray} \label{eq:1ODE}
  \frac{d \varphi}{dz} = e^{i\theta} \bar{F}
\end{eqnarray}
We also fix the BC at $z\to \infty$ as $v_2$
\begin{eqnarray} \label{eq:BC_1st_ODE}
  \text{BC}: \varphi|_{z \to + \infty} = + \sqrt{M \Lambda}
\end{eqnarray}
As a 1ODE, the solution of \equaref{eq:1ODE} leaves only one free parameter, which can be fixed by the BC in \equaref{eq:BC_1st_ODE}. And the value of $\varphi$ at $z \to -\infty$ should be treated as an outcome. As a comparison, a solution of 2ODE in \equaref{eq:2ODE} has two free parameters, requiring two BCs to fix. Namely, we include BCs at both $z \to \pm \infty$ in \equaref{eq:BC}. 

Given a domain wall solution satisfying the 1ODE, the tension is calculated very simply
\begin{eqnarray}
  \sigma = \int_{-\infty}^{+\infty} dz \, \Big( \frac{d \varphi}{dz} \frac{d \varphi^*}{dz} + F \bar{F} \Big) 
  = \int_{-\infty}^{+\infty} dz \, \Big( \frac{d \varphi}{dz} \frac{\partial W}{\partial \varphi} e^{-i\theta} + \frac{d \varphi^*}{dz} \frac{\partial \bar{W}}{\partial \varphi^*} e^{i\theta} \Big)
\end{eqnarray}
On the right hand side, the first term is integrated to be $\Delta W e^{-i\theta}$ with $\Delta W = W|_{\varphi \to +\infty} - W|_{\varphi \to -\infty}$. This term should be real and positive from its initial setup. Thus, $\theta$ represent the phase of $\Delta W$. Similarly, the second term is integrated to be $\Delta \bar{W} e^{i\theta}$ and thus $-\theta$ is the phase of $\Delta W$. Finally, we arrive at 
\begin{eqnarray}
  \sigma = 2 |\Delta W|. 
\end{eqnarray}

We can use this formula to check which domain wall solution does not satisfy the 1ODE. It is straightforward to check that $|\Delta W|$ for type-1 domain wall is zero. To obtain a zero tension, we need require gradient energy and potential energy at any place to be zero, c.f., \equaref{eq:tension}. This is impossible for a non-trivial soliton solution. Thus, type-I domain wall cannot be solved via the 1ODE. 

On the other hand, we obtain the tension of type-2 domain wall as $\sigma = \frac{1}{2} M^2 \Lambda$, which is consistent to that obtained via 2ODE. It is worth noting that \equaref{eq:1ODE} gives also \equaref{eq:df} after some parametrisation. Here we will not repeat. 

\subsection{DWs in R symmetry\label{app:Z2SUSY_R}}

In flavour model building, the driving-field method is widely used. This approach assumes an $U(1)_R$ symmetry with the superspace coordinate $\theta$ taking a charge $+1$ and the driving field taking a charge $+2$. An $U(1)_R$-invariant action requires the superpotential takes charge $+2$. With this setup, the action is invariant under $U(1)_R$, and the driving field always appears linearly in the superpotential. In the following, we show a toy model with flavour symmetry assumed to be $Z_2$. We introduce two superfields, the flavon superfield $\varphi$ is parity-odd in $Z_2$, and the driving field $\chi$ is blind in $Z_2$. The superpotential is given by
\begin{eqnarray}
  W = ({\rm g} \hat{\varphi} \hat{\varphi} -\mu^2) \hat{\chi}
\end{eqnarray}
where both $\mu$ and ${\rm g}$ are real and positive. 
$F$-terms are written out to be
\begin{eqnarray}
  && F_\chi (\varphi) = {\rm g} \varphi \varphi -\mu^2 \nonumber\\
  && F_\varphi (\varphi, \chi) = 2 {\rm g} \varphi \chi
\end{eqnarray}
The flat direction gives $\chi = 0$ and $\varphi = \pm \mu /\sqrt{\rm g}$. 
We fix them as BCs at $z \to \pm \infty$ for 
DW solution. 
\begin{eqnarray}
  \varphi|_{z \to +\infty} = + \frac{\mu}{\sqrt{\rm g}}\,, && \chi|_{z \to +\infty} = 0 \nonumber\\
  \varphi|_{z \to -\infty} = - \frac{\mu}{\sqrt{\rm g}}\,, && \chi|_{z \to -\infty} = 0
\end{eqnarray} 

We first calculate the DW solution via the traditional 2ODE. The potential of $\varphi$, ignoring the driving field, 
is given by
\begin{eqnarray}
  V(\varphi,\varphi^*) = ({\rm g} \varphi \varphi - \mu^2) ({\rm g} \varphi^* \varphi^* - \mu^2)
\end{eqnarray}
The DW solution is simply given by the tanh function,
\begin{eqnarray}
  \varphi = \varphi^* = \frac{\mu}{\sqrt{\rm g}} \tanh (\sqrt{2 \rm g}\mu \,z)
\end{eqnarray}

We further check the solution via the corresponding 1ODEs. The latter are given by
\begin{eqnarray}
  \frac{d \varphi}{dz} &=& F^*_\varphi (\varphi^*, \chi^*)\,, \nonumber\\
  \frac{d \chi}{dz} &=& F^*_\chi(\varphi^*) \,.
\end{eqnarray}
Here, the global phase has been fixed to be zero as no phase varying between VEVs. 
Again, we ignoring the phase and parametrise $\varphi(z) = \frac{\mu}{\sqrt{g}} f(z)$ and $\chi(z) = \frac{\mu}{\sqrt{\rm g}} h(z)$. We obtain
\begin{eqnarray}
  \frac{dh}{df} = \frac{f^2-1}{2fh}
\end{eqnarray}
The solution is given by
\begin{eqnarray}
  1-f^2 = C \exp(\pm h^2)
\end{eqnarray}
This solution is inconsistent with BCs. That means 1ODE cannot be used find the domain wall solution. Indeed, there is one 1ODE the domain wall solution satisfies,
\begin{eqnarray}
  \frac{d \varphi}{dz} &=& F^*_\chi(\varphi^*) \,.
\end{eqnarray}

\bibliographystyle{JHEP}
\bibliography{Ref}

\providecommand{\href}[2]{#2}\begingroup\raggedright\begin{thebibliography}{10}

\bibitem{Ma:2001dn}
E.~Ma and G.~Rajasekaran, {{Softly broken A(4) symmetry for nearly degenerate
  neutrino masses}}, \href{https://doi.org/10.1103/PhysRevD.64.113012}{{Phys.
  Rev. D} {\bfseries 64} (2001) 113012}
  [\href{https://arxiv.org/abs/hep-ph/0106291}{{\ttfamily hep-ph/0106291}}].

\bibitem{Altarelli:2005yx}
G.~Altarelli and F.~Feruglio, {{Tri-bimaximal neutrino mixing, A(4) and the
  modular symmetry}},
  \href{https://doi.org/10.1016/j.nuclphysb.2006.02.015}{{Nucl. Phys. B}
  {\bfseries 741} (2006) 215}
  [\href{https://arxiv.org/abs/hep-ph/0512103}{{\ttfamily hep-ph/0512103}}].

\bibitem{King:2013eh}
S.F.~King and C.~Luhn, {{Neutrino Mass and Mixing with Discrete Symmetry}},
  \href{https://doi.org/10.1088/0034-4885/76/5/056201}{{Rept. Prog. Phys.}
  {\bfseries 76} (2013) 056201}
  [\href{https://arxiv.org/abs/1301.1340}{{\ttfamily 1301.1340}}].

\bibitem{Harrison:2002er}
P.F.~Harrison, D.H.~Perkins and W.G.~Scott, {{Tri-bimaximal mixing and the
  neutrino oscillation data}},
  \href{https://doi.org/10.1016/S0370-2693(02)01336-9}{{Phys. Lett. B}
  {\bfseries 530} (2002) 167}
  [\href{https://arxiv.org/abs/hep-ph/0202074}{{\ttfamily hep-ph/0202074}}].

\bibitem{Harrison:2002kp}
P.F.~Harrison and W.G.~Scott, {{Symmetries and generalizations of tri -
  bimaximal neutrino mixing}},
  \href{https://doi.org/10.1016/S0370-2693(02)01753-7}{{Phys. Lett. B}
  {\bfseries 535} (2002) 163}
  [\href{https://arxiv.org/abs/hep-ph/0203209}{{\ttfamily hep-ph/0203209}}].

\bibitem{Xing:2002sw}
Z.-z.~Xing, {{Nearly tri bimaximal neutrino mixing and CP violation}},
  \href{https://doi.org/10.1016/S0370-2693(02)01649-0}{{Phys. Lett. B}
  {\bfseries 533} (2002) 85}
  [\href{https://arxiv.org/abs/hep-ph/0204049}{{\ttfamily hep-ph/0204049}}].

\bibitem{Harrison:2002et}
P.F.~Harrison and W.G.~Scott, {{mu - tau reflection symmetry in lepton mixing
  and neutrino oscillations}},
  \href{https://doi.org/10.1016/S0370-2693(02)02772-7}{{Phys. Lett. B}
  {\bfseries 547} (2002) 219}
  [\href{https://arxiv.org/abs/hep-ph/0210197}{{\ttfamily hep-ph/0210197}}].

\bibitem{Harrison:2003aw}
P.F.~Harrison and W.G.~Scott, {{Permutation symmetry, tri - bimaximal neutrino
  mixing and the S3 group characters}},
  \href{https://doi.org/10.1016/S0370-2693(03)00183-7}{{Phys. Lett. B}
  {\bfseries 557} (2003) 76}
  [\href{https://arxiv.org/abs/hep-ph/0302025}{{\ttfamily hep-ph/0302025}}].

\bibitem{Lam:2008rs}
C.S.~Lam, {{Determining Horizontal Symmetry from Neutrino Mixing}},
  \href{https://doi.org/10.1103/PhysRevLett.101.121602}{{Phys. Rev. Lett.}
  {\bfseries 101} (2008) 121602}
  [\href{https://arxiv.org/abs/0804.2622}{{\ttfamily 0804.2622}}].

\bibitem{Tsumura:2009yf}
K.~Tsumura and L.~Velasco-Sevilla, {{Phenomenology of flavon fields at the
  LHC}}, \href{https://doi.org/10.1103/PhysRevD.81.036012}{{Phys. Rev. D}
  {\bfseries 81} (2010) 036012}
  [\href{https://arxiv.org/abs/0911.2149}{{\ttfamily 0911.2149}}].

\bibitem{Berger:2014gga}
E.L.~Berger, S.B.~Giddings, H.~Wang and H.~Zhang, {{Higgs-flavon mixing and LHC
  phenomenology in a simplified model of broken flavor symmetry}},
  \href{https://doi.org/10.1103/PhysRevD.90.076004}{{Phys. Rev. D} {\bfseries
  90} (2014) 076004} [\href{https://arxiv.org/abs/1406.6054}{{\ttfamily
  1406.6054}}].

\bibitem{Arroyo-Urena:2018mvl}
M.A.~Arroyo-Ure{\~n}a, J.L.~D{\'\i}az-Cruz, G.~Tavares-Velasco, A.~Bola{\~n}os
  and G.~Hern{\'a}ndez-Tom{\'e}, {{Searching for lepton flavor violating flavon
  decays at hadron colliders}},
  \href{https://doi.org/10.1103/PhysRevD.98.015008}{{Phys. Rev. D} {\bfseries
  98} (2018) 015008} [\href{https://arxiv.org/abs/1801.00839}{{\ttfamily
  1801.00839}}].

\bibitem{Bauer:2016rxs}
M.~Bauer, T.~Schell and T.~Plehn, {{Hunting the Flavon}},
  \href{https://doi.org/10.1103/PhysRevD.94.056003}{{Phys. Rev. D} {\bfseries
  94} (2016) 056003} [\href{https://arxiv.org/abs/1603.06950}{{\ttfamily
  1603.06950}}].

\bibitem{Heinrich:2018nip}
L.~Heinrich, H.~Schulz, J.~Turner and Y.-L.~Zhou, {{Constraining A$_{4}$
  leptonic flavour model parameters at colliders and beyond}},
  \href{https://doi.org/10.1007/JHEP04(2019)144}{{JHEP} {\bfseries 04} (2019)
  144} [\href{https://arxiv.org/abs/1810.05648}{{\ttfamily 1810.05648}}].

\bibitem{MEG:2016leq}
{\scshape MEG} collaboration, {{Search for the lepton flavour violating decay
  $\mu ^+ \rightarrow \mathrm {e}^+ \gamma $ with the full dataset of the MEG
  experiment}}, \href{https://doi.org/10.1140/epjc/s10052-016-4271-x}{{Eur.
  Phys. J. C} {\bfseries 76} (2016) 434}
  [\href{https://arxiv.org/abs/1605.05081}{{\ttfamily 1605.05081}}].

\bibitem{Vilenkin:1984ib}
A.~Vilenkin, {{Cosmic Strings and Domain Walls}},
  \href{https://doi.org/10.1016/0370-1573(85)90033-X}{{Phys. Rept.} {\bfseries
  121} (1985) 263}.

\bibitem{Kibble:1976sj}
T.W.B.~Kibble, {{Topology of Cosmic Domains and Strings}},
  \href{https://doi.org/10.1088/0305-4470/9/8/029}{{J. Phys. A} {\bfseries 9}
  (1976) 1387}.

\bibitem{Zeldovich:1974uw}
Y.B.~Zeldovich, I.Y.~Kobzarev and L.B.~Okun, {{Cosmological Consequences of the
  Spontaneous Breakdown of Discrete Symmetry}}, {{Zh. Eksp. Teor. Fiz.}
  {\bfseries 67} (1974) 3}.

\bibitem{Saikawa:2017hiv}
K.~Saikawa, {{A review of gravitational waves from cosmic domain walls}},
  \href{https://doi.org/10.3390/universe3020040}{{Universe} {\bfseries 3}
  (2017) 40} [\href{https://arxiv.org/abs/1703.02576}{{\ttfamily 1703.02576}}].

\bibitem{Gelmini:2020bqg}
G.B.~Gelmini, S.~Pascoli, E.~Vitagliano and Y.-L.~Zhou, {{Gravitational wave
  signatures from discrete flavor symmetries}},
  \href{https://doi.org/10.1088/1475-7516/2021/02/032}{{JCAP} {\bfseries 02}
  (2021) 032} [\href{https://arxiv.org/abs/2009.01903}{{\ttfamily
  2009.01903}}].

\bibitem{Fu:2024jhu}
B.~Fu, S.F.~King, L.~Marsili, S.~Pascoli, J.~Turner and Y.-L.~Zhou,
  {{Non-Abelian Domain Walls and Gravitational Waves}},
  \href{https://arxiv.org/abs/2409.16359}{{\ttfamily 2409.16359}}.

\bibitem{Jueid:2023cgp}
A.~Jueid, M.A.~Loualidi, S.~Nasri and M.A.~Ouahid, {{Cosmological domain walls
  from the breaking of S4 flavor symmetry}},
  \href{https://doi.org/10.1103/PhysRevD.109.055048}{{Phys. Rev. D} {\bfseries
  109} (2024) 055048} [\href{https://arxiv.org/abs/2312.04388}{{\ttfamily
  2312.04388}}].

\bibitem{Bogomolny:1975de}
E.B.~Bogomolny, {{Stability of Classical Solutions}}, {{Sov. J. Nucl. Phys.}
  {\bfseries 24} (1976) 449}.

\bibitem{Prasad:1975kr}
M.K.~Prasad and C.M.~Sommerfield, {{An Exact Classical Solution for the 't
  Hooft Monopole and the Julia-Zee Dyon}},
  \href{https://doi.org/10.1103/PhysRevLett.35.760}{{Phys. Rev. Lett.}
  {\bfseries 35} (1975) 760}.

\bibitem{Chibisov:1997rc}
B.~Chibisov and M.A.~Shifman, {{BPS saturated walls in supersymmetric
  theories}}, \href{https://doi.org/10.1103/PhysRevD.58.109901}{{Phys. Rev. D}
  {\bfseries 56} (1997) 7990}
  [\href{https://arxiv.org/abs/hep-th/9706141}{{\ttfamily hep-th/9706141}}].

\bibitem{Kubotani:1991kw}
H.~Kubotani, {{The domain wall network of explicitly broken O(N) model}},
  \href{https://doi.org/10.1143/PTP.87.387}{{Prog. Theor. Phys.} {\bfseries 87}
  (1992) 387}.

\bibitem{Vilenkin:2000jqa}
A.~Vilenkin and E.P.S.~Shellard, {{Cosmic Strings and Other Topological
  Defects}}, Cambridge University Press (7, 2000).

\bibitem{Pascoli:2016eld}
S.~Pascoli and Y.-L.~Zhou, {{The role of flavon cross couplings in leptonic
  flavour mixing}}, \href{https://doi.org/10.1007/JHEP06(2016)073}{{JHEP}
  {\bfseries 06} (2016) 073}
  [\href{https://arxiv.org/abs/1604.00925}{{\ttfamily 1604.00925}}].

\bibitem{Wu:2022tpe}
Y.~Wu, K.-P.~Xie and Y.-L.~Zhou, {{Classification of Abelian domain walls}},
  \href{https://doi.org/10.1103/PhysRevD.106.075019}{{Phys. Rev. D} {\bfseries
  106} (2022) 075019} [\href{https://arxiv.org/abs/2205.11529}{{\ttfamily
  2205.11529}}].

\bibitem{Hiramatsu:2012sc}
T.~Hiramatsu, M.~Kawasaki, K.~Saikawa and T.~Sekiguchi, {{Axion cosmology with
  long-lived domain walls}},
  \href{https://doi.org/10.1088/1475-7516/2013/01/001}{{JCAP} {\bfseries 01}
  (2013) 001} [\href{https://arxiv.org/abs/1207.3166}{{\ttfamily 1207.3166}}].

\bibitem{Wainwright:2011kj}
C.L.~Wainwright, {{CosmoTransitions: Computing Cosmological Phase Transition
  Temperatures and Bubble Profiles with Multiple Fields}},
  \href{https://doi.org/10.1016/j.cpc.2012.04.004}{{Comput. Phys. Commun.}
  {\bfseries 183} (2012) 2006}
  [\href{https://arxiv.org/abs/1109.4189}{{\ttfamily 1109.4189}}].

\bibitem{Bazzocchi:2007na}
F.~Bazzocchi, S.~Kaneko and S.~Morisi, {{A SUSY A(4) model for fermion masses
  and mixings}}, \href{https://doi.org/10.1088/1126-6708/2008/03/063}{{JHEP}
  {\bfseries 03} (2008) 063} [\href{https://arxiv.org/abs/0707.3032}{{\ttfamily
  0707.3032}}].

\bibitem{Ding:2013bpa}
G.-J.~Ding, S.F.~King and A.J.~Stuart, {{Generalised CP and $A_4$ Family
  Symmetry}}, \href{https://doi.org/10.1007/JHEP12(2013)006}{{JHEP} {\bfseries
  12} (2013) 006} [\href{https://arxiv.org/abs/1307.4212}{{\ttfamily
  1307.4212}}].

\bibitem{Altarelli:2010gt}
G.~Altarelli and F.~Feruglio, {{Discrete Flavor Symmetries and Models of
  Neutrino Mixing}}, \href{https://doi.org/10.1103/RevModPhys.82.2701}{{Rev.
  Mod. Phys.} {\bfseries 82} (2010) 2701}
  [\href{https://arxiv.org/abs/1002.0211}{{\ttfamily 1002.0211}}].

\bibitem{JUNO:2025gmd}
{\scshape JUNO} collaboration, {{First measurement of reactor neutrino
  oscillations at JUNO}},  \href{https://arxiv.org/abs/2511.14593}{{\ttfamily
  2511.14593}}.

\bibitem{Antusch:2008gw}
S.~Antusch, S.F.~King, M.~Malinsky, L.~Velasco-Sevilla and I.~Zavala, {{Flavon
  Inflation}}, \href{https://doi.org/10.1016/j.physletb.2008.07.051}{{Phys.
  Lett. B} {\bfseries 666} (2008) 176}
  [\href{https://arxiv.org/abs/0805.0325}{{\ttfamily 0805.0325}}].

\bibitem{Antusch:2013toa}
S.~Antusch and D.~Nolde, {{Matter inflation with $A_4$ flavour symmetry
  breaking}}, \href{https://doi.org/10.1088/1475-7516/2013/10/028}{{JCAP}
  {\bfseries 10} (2013) 028} [\href{https://arxiv.org/abs/1306.3501}{{\ttfamily
  1306.3501}}].

\bibitem{Riva:2010jm}
F.~Riva, {{Low-Scale Leptogenesis and the Domain Wall Problem in Models with
  Discrete Flavor Symmetries}},
  \href{https://doi.org/10.1016/j.physletb.2010.05.073}{{Phys. Lett. B}
  {\bfseries 690} (2010) 443}
  [\href{https://arxiv.org/abs/1004.1177}{{\ttfamily 1004.1177}}].

\bibitem{Preskill:1991kd}
J.~Preskill, S.P.~Trivedi, F.~Wilczek and M.B.~Wise, {{Cosmology and broken
  discrete symmetry}},
  \href{https://doi.org/10.1016/0550-3213(91)90241-O}{{Nucl. Phys. B}
  {\bfseries 363} (1991) 207}.

\bibitem{Chigusa:2018hhl}
S.~Chigusa and K.~Nakayama, {{Anomalous Discrete Flavor Symmetry and Domain
  Wall Problem}}, \href{https://doi.org/10.1016/j.physletb.2018.11.027}{{Phys.
  Lett. B} {\bfseries 788} (2019) 249}
  [\href{https://arxiv.org/abs/1808.09601}{{\ttfamily 1808.09601}}].

\bibitem{Ovrut:1977cn}
B.A.~Ovrut, {{Isotropy Subgroups of SO(3) and Higgs Potentials}},
  \href{https://doi.org/10.1063/1.523660}{{J. Math. Phys.} {\bfseries 19}
  (1978) 418}.

\bibitem{Etesi:1996urw}
G.~Etesi, {{Spontaneous symmetry breaking in SO(3) gauge theory to discrete
  subgroups}}, \href{https://doi.org/10.1063/1.531470}{{J. Math. Phys.}
  {\bfseries 37} (1996) 1596}
  [\href{https://arxiv.org/abs/hep-th/9706029}{{\ttfamily hep-th/9706029}}].

\bibitem{King:2018fke}
S.F.~King and Y.-L.~Zhou, {{Spontaneous breaking of $SO(3)$ to finite family
  symmetries with supersymmetry - an $A_4$ model}},
  \href{https://doi.org/10.1007/JHEP11(2018)173}{{JHEP} {\bfseries 11} (2018)
  173} [\href{https://arxiv.org/abs/1809.10292}{{\ttfamily 1809.10292}}].

\bibitem{Caprini:2009fx}
C.~Caprini, R.~Durrer, T.~Konstandin and G.~Servant, {{General Properties of
  the Gravitational Wave Spectrum from Phase Transitions}},
  \href{https://doi.org/10.1103/PhysRevD.79.083519}{{Phys. Rev. D} {\bfseries
  79} (2009) 083519} [\href{https://arxiv.org/abs/0901.1661}{{\ttfamily
  0901.1661}}].

\bibitem{Hiramatsu:2013qaa}
T.~Hiramatsu, M.~Kawasaki and K.~Saikawa, {{On the estimation of gravitational
  wave spectrum from cosmic domain walls}},
  \href{https://doi.org/10.1088/1475-7516/2014/02/031}{{JCAP} {\bfseries 02}
  (2014) 031} [\href{https://arxiv.org/abs/1309.5001}{{\ttfamily 1309.5001}}].

\bibitem{Ferreira:2023jbu}
R.Z.~Ferreira, S.~Gasparotto, T.~Hiramatsu, I.~Obata and O.~Pujolas, {{Axionic
  defects in the CMB: birefringence and gravitational waves}},
  \href{https://doi.org/10.1088/1475-7516/2024/05/066}{{JCAP} {\bfseries 05}
  (2024) 066} [\href{https://arxiv.org/abs/2312.14104}{{\ttfamily
  2312.14104}}].

\bibitem{Li:2023yzq}
Y.~Li, Y.~Jia and L.~Bian, {{Numerical simulation of domain wall and
  first-order phase transition in an expanding universe}},
  \href{https://doi.org/10.1088/1475-7516/2025/02/038}{{JCAP} {\bfseries 02}
  (2025) 038} [\href{https://arxiv.org/abs/2304.05220}{{\ttfamily
  2304.05220}}].

\bibitem{Kitajima:2023kzu}
N.~Kitajima, J.~Lee, F.~Takahashi and W.~Yin, {{Stability of domain walls with
  inflationary fluctuations under potential bias, and gravitational wave
  signatures}}, \href{https://doi.org/10.1088/1475-7516/2025/07/053}{{JCAP}
  {\bfseries 07} (2025) 053}
  [\href{https://arxiv.org/abs/2311.14590}{{\ttfamily 2311.14590}}].

\bibitem{Dankovsky:2024zvs}
I.~Dankovsky, E.~Babichev, D.~Gorbunov, S.~Ramazanov and A.~Vikman,
  {{Revisiting evolution of domain walls and their gravitational radiation with
  CosmoLattice}}, \href{https://doi.org/10.1088/1475-7516/2024/09/047}{{JCAP}
  {\bfseries 09} (2024) 047}
  [\href{https://arxiv.org/abs/2406.17053}{{\ttfamily 2406.17053}}].

\bibitem{Ferreira:2024eru}
R.Z.~Ferreira, A.~Notari, O.~Pujol{\`a}s and F.~Rompineve, {{Collapsing domain
  wall networks: impact on pulsar timing arrays and primordial black holes}},
  \href{https://doi.org/10.1088/1475-7516/2024/06/020}{{JCAP} {\bfseries 06}
  (2024) 020} [\href{https://arxiv.org/abs/2401.14331}{{\ttfamily
  2401.14331}}].

\bibitem{Notari:2025kqq}
A.~Notari, F.~Rompineve and F.~Torrenti, {{The spectrum of gravitational waves
  from annihilating domain walls}},
  \href{https://doi.org/10.1088/1475-7516/2025/07/049}{{JCAP} {\bfseries 07}
  (2025) 049} [\href{https://arxiv.org/abs/2504.03636}{{\ttfamily
  2504.03636}}].

\bibitem{Cyr:2025nzf}
B.~Cyr, S.~Cotterill and R.~Battye, {{Near-Peak Spectrum of Gravitational Waves
  from Collapsing Domain Walls}},
  \href{https://arxiv.org/abs/2504.02076}{{\ttfamily 2504.02076}}.

\bibitem{Babichev:2025stm}
E.~Babichev, I.~Dankovsky, D.~Gorbunov, S.~Ramazanov and A.~Vikman, {{Biased
  domain walls: faster annihilation, weaker gravitational waves}},
  \href{https://doi.org/10.1088/1475-7516/2025/10/103}{{JCAP} {\bfseries 10}
  (2025) 103} [\href{https://arxiv.org/abs/2504.07902}{{\ttfamily
  2504.07902}}].

\bibitem{Blasi:2025tmn}
S.~Blasi, A.~Mariotti, A.~Rase and M.~Vanvlasselaer, {{Domain walls in the
  scaling regime: Equal Time Correlator and Gravitational Waves}},
  \href{https://arxiv.org/abs/2511.16649}{{\ttfamily 2511.16649}}.

\bibitem{Zhang:2023nrs}
Z.~Zhang, C.~Cai, Y.-H.~Su, S.~Wang, Z.-H.~Yu and H.-H.~Zhang, {{Nano-Hertz
  gravitational waves from collapsing domain walls associated with freeze-in
  dark matter in light of pulsar timing array observations}},
  \href{https://doi.org/10.1103/PhysRevD.108.095037}{{Phys. Rev. D} {\bfseries
  108} (2023) 095037} [\href{https://arxiv.org/abs/2307.11495}{{\ttfamily
  2307.11495}}].

\bibitem{Zeng:2025zjp}
Q.-Q.~Zeng, X.~He, Z.-H.~Yu and J.~Zheng, {{Collapsing domain walls with
  Z2-violating coupling to thermalized fermions and their impact on
  gravitational wave detections}},
  \href{https://doi.org/10.1103/cdsj-bmvx}{{Phys. Rev. D} {\bfseries 111}
  (2025) 115017} [\href{https://arxiv.org/abs/2501.10059}{{\ttfamily
  2501.10059}}].

\bibitem{NANOGrav:2023gor}
{\scshape NANOGrav} collaboration, {{The NANOGrav 15 yr Data Set: Evidence for
  a Gravitational-wave Background}},
  \href{https://doi.org/10.3847/2041-8213/acdac6}{{Astrophys. J. Lett.}
  {\bfseries 951} (2023) L8}
  [\href{https://arxiv.org/abs/2306.16213}{{\ttfamily 2306.16213}}].

\bibitem{Ghoshal:2025gci}
A.~Ghoshal and Y.~Hamada, {{Gravitational waves created by current-carrying
  domain walls}}, \href{https://doi.org/10.1103/24w5-rflt}{{Phys. Rev. D}
  {\bfseries 112} (2025) 115019}.

\bibitem{Zic:2023gta}
A.~Zic et~al., {{The Parkes Pulsar Timing Array third data release}},
  \href{https://doi.org/10.1017/pasa.2023.36}{{Publ. Astron. Soc. Austral.}
  {\bfseries 40} (2023) e049}
  [\href{https://arxiv.org/abs/2306.16230}{{\ttfamily 2306.16230}}].

\bibitem{EPTA:2023sfo}
{\scshape EPTA} collaboration, {{The second data release from the European
  Pulsar Timing Array - I. The dataset and timing analysis}},
  \href{https://doi.org/10.1051/0004-6361/202346841}{{Astron. Astrophys.}
  {\bfseries 678} (2023) A48}
  [\href{https://arxiv.org/abs/2306.16224}{{\ttfamily 2306.16224}}].

\bibitem{Xu:2023wog}
H.~Xu et~al., {{Searching for the Nano-Hertz Stochastic Gravitational Wave
  Background with the Chinese Pulsar Timing Array Data Release I}},
  \href{https://doi.org/10.1088/1674-4527/acdfa5}{{Res. Astron. Astrophys.}
  {\bfseries 23} (2023) 075024}
  [\href{https://arxiv.org/abs/2306.16216}{{\ttfamily 2306.16216}}].

\bibitem{Rana:2025ano}
P.~Rana et~al., {{The Indian Pulsar Timing Array Data Release 2: I. Dataset and
  Timing Analysis}},  \href{https://arxiv.org/abs/2506.16769}{{\ttfamily
  2506.16769}}.

\bibitem{Janssen:2014dka}
G.~Janssen et~al., {{Gravitational wave astronomy with the SKA}},
  \href{https://doi.org/10.22323/1.215.0037}{{PoS} {\bfseries AASKA14} (2015)
  037} [\href{https://arxiv.org/abs/1501.00127}{{\ttfamily 1501.00127}}].

\bibitem{Li:2024rnk}
E.-K.~Li et~al., {{Gravitational wave astronomy with TianQin}},
  \href{https://doi.org/10.1088/1361-6633/adc9be}{{Rept. Prog. Phys.}
  {\bfseries 88} (2025) 056901}
  [\href{https://arxiv.org/abs/2409.19665}{{\ttfamily 2409.19665}}].

\bibitem{Ruan:2018tsw}
W.-H.~Ruan, Z.-K.~Guo, R.-G.~Cai and Y.-Z.~Zhang, {{Taiji program:
  Gravitational-wave sources}},
  \href{https://doi.org/10.1142/S0217751X2050075X}{{Int. J. Mod. Phys. A}
  {\bfseries 35} (2020) 2050075}
  [\href{https://arxiv.org/abs/1807.09495}{{\ttfamily 1807.09495}}].

\bibitem{2017arXiv170200786A}
P.~{Amaro-Seoane}, H.~{Audley}, S.~{Babak}, J.~{Baker}, E.~{Barausse},
  P.~{Bender} et~al., {{Laser Interferometer Space Antenna}},
  \href{https://doi.org/10.48550/arXiv.1702.00786}{{arXiv e-prints} (2017)
  arXiv:1702.00786} [\href{https://arxiv.org/abs/1702.00786}{{\ttfamily
  1702.00786}}].

\bibitem{Punturo_2010}
M.~Punturo, M.~Abernathy, F.~Acernese, B.~Allen, N.~Andersson, K.~Arun et~al.,
  {The einstein telescope: a third-generation gravitational wave observatory},
  \href{https://doi.org/10.1088/0264-9381/27/19/194002}{{Classical and Quantum
  Gravity} {\bfseries 27} (2010) 194002}.

\bibitem{Reitze:2019iox}
D.~Reitze et~al., {{Cosmic Explorer: The U.S. Contribution to
  Gravitational-Wave Astronomy beyond LIGO}}, {{Bull. Am. Astron. Soc.}
  {\bfseries 51} (2019) 035}
  [\href{https://arxiv.org/abs/1907.04833}{{\ttfamily 1907.04833}}].

\bibitem{LIGOScientific:2025kry}
{\scshape LIGO Scientific, VIRGO, KAGRA} collaboration, {{Cosmological and High
  Energy Physics implications from gravitational-wave background searches in
  LIGO-Virgo-KAGRA's O1-O4a runs}},
  \href{https://arxiv.org/abs/2510.26848}{{\ttfamily 2510.26848}}.

\bibitem{Crowder:2005nr}
J.~Crowder and N.J.~Cornish, {{Beyond LISA: Exploring future gravitational wave
  missions}}, \href{https://doi.org/10.1103/PhysRevD.72.083005}{{Phys. Rev. D}
  {\bfseries 72} (2005) 083005}
  [\href{https://arxiv.org/abs/gr-qc/0506015}{{\ttfamily gr-qc/0506015}}].

\bibitem{Kawamura:2006up}
S.~Kawamura et~al., {{The Japanese space gravitational wave antenna DECIGO}},
  \href{https://doi.org/10.1088/0264-9381/23/8/S17}{{Class. Quant. Grav.}
  {\bfseries 23} (2006) S125}.

\bibitem{Gouttenoire:2023gbn}
Y.~Gouttenoire and E.~Vitagliano, {{Primordial black holes and wormholes from
  domain wall networks}},
  \href{https://doi.org/10.1103/PhysRevD.109.123507}{{Phys. Rev. D} {\bfseries
  109} (2024) 123507} [\href{https://arxiv.org/abs/2311.07670}{{\ttfamily
  2311.07670}}].

\end{thebibliography}\endgroup

\end{document}